\documentclass[aps,twocolumn,amsmath,amsfonts,nofootinbib,preprintnumbers,
  superscriptaddress,secnumarabic,notitlepage]{revtex4-1}
\usepackage[utf8]{inputenc}
\usepackage{color,graphicx,hyperref,amsmath,amssymb,multirow,slashed,nicefrac,xspace}
\hypersetup{colorlinks=true,linkcolor=blue,citecolor=magenta,filecolor=magenta,
  urlcolor=cyan}
\newcommand{\be}{\begin{equation}}
\newcommand{\ee}{\end{equation}}
\def\bsp#1\esp{\begin{split}#1\end{split}}
\newcommand{\ie}{\textit{i.e.}}

\def\ov{\overline}
\def\sss{\scriptscriptstyle}
\def\mh{m_{\sss h}}
\def\mw{m_{\sss W}}
\def\mz{m_{\sss Z}}
\def\gs{g_{\sss s}}
\def\cw{c_{\sss W}}
\def\sw{s_{\sss W}}

\def\lagr{\mathcal{L}}

\def\beq{\begin{equation}}
\def\eeq{\end{equation}}
\def\nn{\nonumber}
\def\ov{\overline}

\def\frules{{\sc FeynRules}\xspace}
\def\lilith{{\sc Lilith}\xspace}
\def\hbounds{{\sc HiggsBounds}\xspace}
\def\hsignals{{\sc HiggsSignals}\xspace}
\def\MadGraph{{\sc MG5\_aMC}\xspace}
\def\momegas{{\sc MicrOMEGAs}\xspace}

\begin{document}

\title{Heavy dark matter through the dilaton portal}

\author{Benjamin Fuks}
\email{fuks@lpthe.jussieu.fr}
\affiliation{Sorbonne Universit\'e, CNRS, Laboratoire de Physique Th\'eorique et Hautes \'Energies, LPTHE, F-75005 Paris, France}
\affiliation{Institut Universitaire de France, 103 boulevard Saint-Michel, 75005 Paris, France}
\author{Mark D. Goodsell}
\email{goodsell@lpthe.jussieu.fr}
\affiliation{Sorbonne Universit\'e, CNRS, Laboratoire de Physique Th\'eorique et Hautes \'Energies, LPTHE, F-75005 Paris, France}
\author{Dong Woo Kang}
\email{dongwookang@kias.re.kr}
\affiliation{School of Physics, Korea Institute for Advanced Study, Seoul 02455, Korea}
\author{Pyungwon Ko}
\email{pko@kias.re.kr}
\affiliation{School of Physics, Korea Institute for Advanced Study, Seoul 02455, Korea}
\author{Seung J.~ Lee}
\email{sjjlee@korea.ac.kr}
\affiliation{Department of Physics, Korea University, Seoul 136-713, Korea}
\author{Manuel Utsch}
\email{utsch@lpthe.jussieu.fr}
\affiliation{Sorbonne Universit\'e, CNRS, Laboratoire de Physique Th\'eorique et Hautes \'Energies, LPTHE, F-75005 Paris, France}

\date{\today}

\begin{abstract} 
We re-examine current and future constraints on a heavy dilaton coupled to a
simple dark sector consisting of a Majorana fermion or a St\"uckelberg vector
field. We include three different treatments of dilaton-Higgs mixing, paying
particular attention to a gauge-invariant formulation of the model. Moreover, we
also invite readers to re-examine effective field theories of vector dark matter, which we show are missing important terms. Along with the latest Higgs coupling data,
heavy scalar search results, and dark matter density/direct detection
constraints, we study the LHC bounds on the model and estimate the prospects of
dark matter production at the future HL-LHC and 100 TeV FCC colliders. We
additionally compute novel perturbative unitarity constraints involving vector
dark matter, dilaton and gluon scattering.
\end{abstract}

\maketitle


\section{Introduction}
\label{introduction}

Any theory can be made scale invariant by coupling it to a dilaton. The scale invariance can then be softly broken, giving the dilaton a mass and self-interactions, and this becomes a popular proposal \cite{Bardeen:1985sm,Buchmuller:1987uc,Buchmuller:1988cj,Rattazzi:2000hs,Csaki:2000zn,Dominici:2002jv,Csaki:2007ns,Goldberger:2008zz,Bai:2009ms,Grzadkowski:2012ng,Chacko:2012sy,Bellazzini:2012vz,Ahmed:2015uqt,Ahmed:2019csf} for solving the hierarchy problem of the Standard Model (SM). Such SM plus dilaton theories can either be thought of as fundamental, or as the low-energy limit of composite theories, where the dilaton becomes the pseudo-Goldstone boson associated with the spontaneous breaking of scale invariance. It therefore couples to the SM fields through the trace of the energy-momentum tensor.

The dilaton portal is also an extremely economic way of coupling the SM to a dark matter particle: a massive dark matter field automatically couples to the dilaton, so that there is no need to add any additional interactions with the SM. Such models are very economical in terms of new parameters: we have effectively just the dilaton mass, dark matter mass, and the dilaton decay constant/symmetry breaking scale as extra degrees of freedom relative to the SM.

Models of dilaton portal dark matter have also been well studied in the literature \cite{Bai:2009ms,Jung:2014zga,Blum:2014jca,Efrati:2014aea,Kim:2016jbz}. In this paper, we will study both the fermionic and vector dark matter cases in detail, pointing out along the way the connection between the dilaton portal and models of vector dark matter based on effective field theory. We shall consider both the vanilla dilaton scenario without mixing with the Higgs boson, and also (in section \ref{sec:HiggsMixing}) two different formulations of the theory once mixing is included: the theory is not uniquely defined, and the simplest way to include mixing is not gauge invariant; along the way we also write a gauge-invariant formulation of the Lagrangian. Our aim is to revisit and update the constraints on the main parameters of the model, via dark matter constraints, the latest Higgs-like particle searches for the dilaton itself, collider constraints from direct searches for the dark matter particle at the LHC, and perturbative unitarity of scattering amplitudes.  We then provide projections for future colliders. In particular, we shall restrict our attention to a heavy dilaton above about $300$ GeV (see ref.~\cite{Ahmed:2019csf} for a detailed recent examination of constraints on the dilaton in the low-mass window) where diboson searches push the dilaton to large masses and weak couplings. However, a vector dark matter candidate is produced much more copiously at the LHC than a fermion candidate, and thus this model is more promising for future searches.

The rest of this paper is organised as follows. In section~\ref{sec:th}, we
introduce our theoretical framework, discuss the problematics of the
Higgs-dilaton mixing and address the issue of perturbative unitarity.
Section~\ref{SEC:HiggsDilMixing} is dedicated to the estimation of the
constraints on the Higgs-dilaton mixing and section~\ref{sec:DM} to the bounds
that can be imposed from dark matter direct detection and LHC searches.
Section~\ref{SEC:FCC} finally focuses on future collider prospects. We conclude
and summarise our work in section~\ref{sec:conclusion}. In addition, more
extensive details on perturbative unitarity constraints are presented in the
appendix.


\section{The dilaton portal to dark matter}\label{sec:th}

\subsection{Theoretical framework}
We consider an extension of the SM that contains a dark matter
candidate, taken to be a Majorana fermion or a neutral vector boson,
without any tree-level interaction with the Standard Model particles. Assuming
spontaneously broken scale invariance at a high scale $f$, the particle
content of the model includes a light dilaton $\sigma$ of mass $m_\sigma$ that couples to all
other fields through the trace of the energy-momentum
tensor~\cite{Bardeen:1985sm,Buchmuller:1987uc,Buchmuller:1988cj,Csaki:2007ns,Goldberger:2008zz,Chacko:2012sy,
Bellazzini:2012vz}.
As pointed out in the introduction, the dark matter candidate couples to the dilaton via its mass term,
and then the dilaton
acts as a mediator for
dark matter interactions with the Standard Model particles. The dilaton,
its self-interactions and its interactions with the SM sector arise
from a procedure\footnote{For a detailed explanation see, for example, the work of ref.~\cite{Goldberger:2008zz}.} that converts a theory into a
scale-invariant one. We briefly summarise this procedure below, in the
context of the model studied in this work.

We consider a generic Lagrangian ${\cal L}_0$ that we write in terms of
operators $\mathcal{O}_i$ of classical scaling dimension $d_i=[\mathcal{O}_i]$
and coefficients $g_i$ depending on the renormalisation scale $\mu$,
\begin{equation}
  \mathcal{L}_0=\sum\limits_i g_i(\mu)\mathcal{O}_i(x)\,.
\end{equation}
Scale-invariance requires that the above Lagrangian contains only dimensionless,
scale-independent couplings, or equivalently that it is invariant under a scale
transformation of parameter $\lambda$,
\begin{equation}
  x^\mu\to e^\lambda x^\mu\,,\quad
  \mathcal{O}_i(x) \to e^{\lambda d_i}\mathcal{O}_i(e^\lambda x)\,,\quad
  \mu\to e^{-\lambda}\mu\,.
\end{equation}
However, those rules yield a non-vanishing 
Lagrangian variation 
$\delta \mathcal{L}$ equal
to the trace of the energy-momentum tensor $T^\mu{}_\nu$,
\be
  \delta {\cal L} = \sum\limits_i \Big[ (d_i-4){\cal O}_i
     - \mu\frac{\partial g_i}{\partial\mu} \Big]  = T^\mu{}_\mu \,.
\ee

Scale invariance can then be established by introducing a conformal
compensator field $\chi(x)$, which is a scalar field of dimension $[\chi]=1$,
and by enforcing the replacement of the couplings $g_i$~\cite{Chacko:2012sy,
Bellazzini:2012vz}
\begin{equation}
  g_i(\mu)\quad\to\quad g_i\left(\mu\frac{\chi}{f}\right)\left(\frac{\chi}{f}\right)^{4-d_i}\,.\label{eq_replacement_gi}
\end{equation}
Such an introduction of the appropriate powers of the compensator field allows
us to retrieve scale invariance after imposing the scale transformation
law of the compensator
\begin{equation}
  \chi(x)\to e^\lambda\chi(e^\lambda x) \,.
\end{equation}
Along with the additional field $\chi$, the replacement in
eq.~\eqref{eq_replacement_gi} also induces a new parameter, the cut-off scale
 $f=\langle\chi\rangle$. This is an artifact of the breaking of the
scale symmetry, \ie\ the vacuum expectation value of $\chi$. Therefore, an
appropriate parametrisation involving the associated Goldstone boson $\sigma$,
namely the dilaton field, would be based on the field
redefinition~\cite{Blum:2014jca,Goldberger:2008zz}
\begin{equation}
  \chi(x) = f  + \sigma(x)\,.\label{chi_parameterization}
\end{equation}
Inserting these modifications into the Lagrangian leads to
\begin{equation}
  \begin{split}
    \hspace{-.5cm}\mathcal{L}&= \sum\limits_ig_i\Big(\mu (1 + \sigma(x)/f)\Big)
       \ \Big(1 + \sigma(x)/f\Big)^{4-d_i} \mathcal{O}_i(x)\\
 &= {\cal L}_0 \!+\! \sum\limits_i\frac{\sigma(x)}{f}\Big[
     g_i(\mu)(4-d_i)\mathcal{O}_i(x) \!+\!
     \beta(g_i)\mathcal{O}_i(x)\Big]\\ &\qquad
     + \frac{\sigma^2(x)}{2f^2}\sum\limits_i\Big[ (4-d_i)(3-d_i) g_i(\mu)
     {\cal O}_i(x) \Big]\\ & \qquad +\dots \,,
  \end{split}
\end{equation}
where we restricted ourselves to the leading terms in $\nicefrac{\sigma}{f}$
explicitly, and the ellipsis stands for any (potentially-relevant) higher-order
contributions.

Considering the broken electroweak  phase, the interaction terms of the dilaton
with the SM sector are given, including all the Lagrangian terms of
dimension 6 or below, by
\begin{equation}
  \begin{split}
    \hspace{-.35cm}\mathcal{L}_\sigma =&
    \frac{\sigma}{f}\bigg[
      2\mw^2 W_\mu^+W^{-\mu} \!+\! \mz^2Z_\mu Z^\mu \!-\! \mh^2h^2
      \!-\! \sum\limits_{\psi}m_\psi\ov{\psi}\psi\\
    &\quad -\frac{\mh^2}{2v}\Big[ h^3 + hG^0G^0 + 2 h G^+G^-\Big]\\
    &\quad +\frac{gv}{2}\Big[ \partial^\mu G^- W_\mu^+
         + \partial^\mu G^+ W_\mu^-
         + \frac{1}{\cw} \partial^\mu G^0 Z_\mu\Big]\\
    &\quad + g \mw h W_\mu^+ W^{\mu-} + \frac{g}{2\cw} \mz h Z_\mu Z^\mu\\
    &\quad + ig'\mw \big(G^- W_\mu^+-G^+ W_\mu^-\big)
     \big(\cw A^\mu-\sw Z^\mu\big)\\
    &\quad
      +\frac{11\alpha_{\text{EM}}}{24\pi} F_{\mu\nu}F^{\mu\nu}
      -\frac{7\alpha_s}{8\pi} G^a_{\mu\nu}G^{a\mu\nu}\bigg] \\
    & +\frac{\sigma^2}{2 f^2}
      \Big[2 \mw^2 W_\mu^+W^{-\mu} \!+\! \mz^2Z_\mu Z^\mu \!-\! \mh^2h^2\Big]\,,
  \end{split}
\label{eq:lag}\end{equation}
where the summation over $\psi$ refers to all the SM fermionic
mass-eigenstates of mass $m_\psi$, and $m_h$, $m_W$ and $m_Z$ stand for the
Higgs boson, $W$-boson and $Z$-boson masses. Moreover, $v$ denotes the vacuum
expectation value of the Standard Model Higgs field, $\cw$ and $\sw$ the cosine
and sine of the electroweak mixing angle, and $g$, $g'$ and $\gs$ the weak,
hypercharge and strong coupling constants. In our expressions, we have included
the interactions between the physical Higgs ($h$) and electroweak ($A_\mu$,
$W_\mu$, $Z_\mu$) bosons and the three Goldstone bosons ($G^0$, $G^\pm$).
For the corresponding expression in the unbroken electroweak phase, we refer to
ref.~\cite{Jung:2014zga} that additionally includes a complete Higgs-dilaton
mixing analysis.

The interactions of the dilaton with the two considered dark matter candidates,
namely a Majorana fermion $\Psi_X$ and a real vector field $X_\mu$, read
\begin{equation}
  \hspace{-.2cm}\mathcal{L}_\sigma^{\text{DM}} \!=\!
  \begin{cases}
    -\frac{\sigma}{2f}m_\Psi\overline{\Psi}_X\Psi_X&\text{(Majorana fermion)}\,,\\
    \Big(\frac{\sigma}{f}\!+\!\frac{\sigma^2}{2 f^2}\Big)
       m_V^2 X_\mu X^\mu&\text{(vector boson)} \,,
  \end{cases}
\end{equation}
where we denote the mass of the dark matter state by $m_\Psi$ and $m_V$ in the
fermion and vector case respectively. In order
to ensure the stability of the dark matter particle, our setup assumes a $\mathbb{Z}_2$
symmetry.

At this level, the only purpose of introducing the dilaton was to make the
existing terms of the Lagrangian scale-invariant. We should however also include
a kinetic and a mass term for the dilaton, as well as its potential
self-interaction terms. We rely on a dilaton potential that is constructed under
the assumption that the conformally-invariant field theory, for which our model
represents an effective theory, is explicitly broken due to the addition of an
operator with a scaling dimension $\Delta_{\mathcal{O}}\neq 4$.
This yields a potential $V(\chi)$~\cite{Goldberger:2008zz,Rattazzi:2000hs},
\begin{equation}
  V(\chi)=\chi^4\sum\limits_{n=0}^\infty c_n(\Delta_{\mathcal{O}})\left(\frac{\chi}{f}\right)^{n(\Delta_{\mathcal{O}}-4)}\,,
\end{equation}
that we add to the effective Lagrangian. At the minimum of the potential at
which $\langle\chi\rangle=f$, the coefficients $c_n$ can be related to the
the parameters of the underlying conformal field theory\footnote{Note that the exact details of the underlying theory is beyond the scope of this paper, and in our analysis we take an effective theory approach.}. This potential is supposed to be at the origin
of the dilaton mass $m_\sigma$, which translates into the condition
\be
  m_\sigma^2 =
    \frac{{\rm d}^2V(\chi)}{{\rm d}\chi^2}\bigg|_{\langle\chi\rangle=f}>0\,.
\ee
With the assumption that $|\Delta_{\mathcal{O}}-4|\ll 1$, it is possible to
expand the potential in $|\Delta_{\mathcal{O}}-4|$ such that the explicit
$\Delta_{\mathcal{O}}$-dependence of the different coefficients $c_n$
disappears~\cite{Goldberger:2008zz},
\begin{equation}
  V(\chi)=\frac{1}{16}\frac{m^2}{f^2}\chi^4\left[
    4\ln\frac{\chi}{f}-1\right]+\mathcal{O}(|\Delta_{\mathcal{O}}-4|^2)\,.
\end{equation}
By applying the parametrisation of eq.~\eqref{chi_parameterization},
one obtains, after adding the dilaton kinetic term~\cite{Blum:2014jca},
\begin{equation}
  \mathcal{L}_\sigma^{\rm self} \!=\!
    \frac12\partial_\mu\sigma\partial^\mu\sigma
    \!-\! \frac{m_\sigma^2}{2}\sigma^2
    \!-\! \frac{5}{6}\frac{m_\sigma^2}{f}\sigma^3
    \!-\! \frac{11}{24}\frac{m_\sigma^2}{f^2}\sigma^4+\dots\,,
\label{EQ:dilpot}\end{equation}
where the dots stand for higher-dimensional interactions.

\subsection{Higgs-Dilaton mixing}
\label{sec:HiggsMixing}

When two physical neutral scalars are present in the theory (the Higgs field $h$ and the
dilaton field $\sigma$), they could in principle mix, unless it is forbidden
by some symmetry. If the mixing is allowed, its origins can be found through a
UV completion of our effective setup. This option was first studied in
ref.~\cite{Efrati:2014aea}, and will be re-investigated in the light of the most
recent experimental data in section \ref{SEC:HiggsDilMixing}.

Keeping in mind the effective approach adopted in this work, we will  not speculate about any UV-physics that drives the source, and
hence the amount, of mixing. Instead, we will study the possibility of a
non-vanishing Higgs-dilaton mixing by introducing a mixing angle $\alpha$ as an
additional parameter. We will \emph{relabel} our original flavour states as $h_0, \sigma_0$ with mass parameters $m_{h,0}, m_{\sigma, 0}$ and relate them to the new mass eigenstates $h, \sigma$ via the rotation
\begin{equation}
  \begin{pmatrix}
    h\\
    \sigma
  \end{pmatrix}\!=\!
  \begin{pmatrix}
    \cos\alpha&\sin\alpha\\
    -\sin\alpha&\cos\alpha
  \end{pmatrix}
  \begin{pmatrix}
    h_0\\
    \sigma_0
  \end{pmatrix}\!\equiv\!
  \begin{pmatrix}
    c_\alpha&s_\alpha\\
   -s_\alpha&c_\alpha
  \end{pmatrix}
  \begin{pmatrix}
    h_0\\
    \sigma_0
  \end{pmatrix} \,.
  \label{EQ:MixingRotation}
\end{equation}
 We then assume that the (lighter, by assumption) scalar $h$ can be identified as
the experimentally confirmed scalar of mass of about 125~GeV, so that it is
mostly compatible with the SM Higgs boson. In contrast, the heavier
scalar field $\sigma$ is mostly dilaton-like.

The impact of the non-zero mixing of the Higgs boson and the dilaton can be
inferred from the similarity between the couplings of the $h_0$ and $\sigma_0$ states
with the remaining particle content in the zero-mixing case.
Without mixing and in unitary gauge (\ie\ ignoring Goldstone bosons), there is a corresponding dilaton coupling
for every Higgs boson coupling. This is not surprising since these dilaton couplings originate from the presence of a dimensionful coupling in the SM Lagrangian after electroweak symmetry breaking,
where the dimensionful quantity in the coupling constants is actually the Higgs vacuum expectation value $v$.
In all of these cases, the coupling constants appearing in the dilaton
interaction vertices therefore differ by a factor of $r_f=v/f$ for
each dilaton participating in the interaction, when compared with the
corresponding Higgs-boson interaction. In principle, dilaton couplings involving
the Higgs or Goldstone bosons (leaving the unitary gauge) should be
discussed as well, but they cannot be related to any SM counterpart by
factors of $v/f$. We refer to the discussion below for what concerns
those multi-scalar interactions.

For the dilaton and Higgs Yukawa couplings to the SM fermions, this
yields the interaction Lagrangian ${\cal L}_\sigma^\psi$,
\begin{equation}\begin{split}
    \mathcal{L}_\sigma^\psi &=
      \sum\limits_\psi\frac{m_\psi}{v}\Big(h_0+r_f\sigma_0\Big)\ov{\psi}\psi\\
    &=\sum\limits_\psi\frac{m_\psi}{v}\Big[(c_\alpha \!+\! r_fs_\alpha)h +
     (r_fc_\alpha\!-\!s_\alpha)\sigma\Big]\ \ov{\psi}\psi\,.
\end{split}\label{EQ:MixingFermionCouplings}\end{equation}
Similarly, the massive gauge boson interactions read, at leading order in the
scalar fields,
\begin{equation}\begin{split}
    \mathcal{L}_\sigma^V&=
      \Big[(c_\alpha\!+\!r_fs_\alpha)h + (r_fc_\alpha\!-\!s_\alpha)\sigma \Big]
     \hspace{1cm}\\
    &\qquad \times \bigg[ \frac{2m_W^2}{v} W_\mu^+W^{-\mu}
       +\frac{m_Z^2}{v} Z_\mu Z^\mu \bigg]\,.
\end{split}\label{EQ:MixingWZCouplings}\end{equation}
The $W$- and $Z$-boson couplings to a pair of scalar fields $h$ and/or $\sigma$ are
obtained analogously.

These couplings are potentially relevant for processes addressing both scalar
production and decays.
For example, in scalar production at hadron colliders through gluon fusion, the
leading-order contribution involves triangle diagrams featuring a loop of
quarks (the top quark one being the most relevant by virtue of its largest
mass), so that the associated predictions are affected by the modifications of
eq.~\eqref{EQ:MixingFermionCouplings}. In case of the light scalar, a factor of
$(c_\alpha+r_fs_\alpha)$ is introduced into the amplitude compared to the
zero-mixing case. On the contrary, for the production of the heavy scalar, the
extra factor is given by $(-s_\alpha+r_fc_\alpha)$.
This feature is obviously also present for any other production mode at
colliders, like associated production ($Vh$ or $V\sigma$) or vector-boson
fusion, that involve the coupling of the scalar to the $W$- and $Z$-bosons.

On the other hand, the dark matter mass term is at the origin of a dilaton coupling which obviously does not have an analogue for the Higgs boson due to the
absence of the dark matter in the SM and of a Higgs portal relating
the dark and the visible sector.
Depending on the type of dark matter, the Lagrangian in terms of the scalars $h$ and $\sigma$ gives
\begin{equation}
  \mathcal{L}_\sigma^{\text{DM}}=-\frac{m_\Psi}{2f}s_\alpha h\ov{\Psi}_X\Psi_X-\frac{m_\Psi}{2f}c_\alpha \sigma\ov{\Psi}_X\Psi_X
\end{equation}
in the Majorana case and
\begin{equation}
  \mathcal{L}_\sigma^{\text{DM}} = 
      \frac{m_V^2}{f}   (s_\alpha h\!+\!c_\alpha \sigma)X_\mu X^\mu 
    + \frac{m_V^2}{2f^2}(s_\alpha h\!+\!c_\alpha \sigma)^2X_\mu X^\mu
\end{equation}
in the vector case. In the latter case, the trilinear couplings are the most
relevant ones for Higgs measurements and heavy scalar searches, but
the quartic couplings are very important for dark matter searches when $m_V \gg
m_\sigma$. Consequently, mixing leads to a coupling of the SM-like Higgs state
to the dark-matter candidate that would not exist without mixing.

The same applies to the tree-level couplings of the dilaton to the massless gauge-bosons, which are not induced by dimensionful couplings
but by the scale-dependence of the electromagnetic and the strong coupling.
Here the contribution to the Lagrangian in the mass eigenbasis $\mathcal{L}_{VVS}$ reads
\begin{equation}
  \begin{split}
    \mathcal{L}_{VVS} = &\ 
     \frac{11\alpha_{\text{EM}}}{24\pi v}
       \Big(r_fs_\alpha h+r_fc_\alpha \sigma\Big)\ F_{\mu\nu}F^{\mu\nu}\\
     &\ -\frac{7\alpha_s}{8\pi v}
       \Big(r_fs_\alpha h+r_fc_\alpha \sigma\Big)\ G_{\mu\nu}^a G_a^{\mu\nu}\,.
  \end{split}
\label{EQ:PhotonGluonCouplings}\end{equation}
Without mixing, the dilaton coupling to photons or gluons must take into account
both a tree-level and a one-loop contribution at leading order in the
electromagnetic or strong coupling, while the Higgs boson interacts with these
bosons only beyond tree-level. Eq.~\eqref{EQ:PhotonGluonCouplings} shows that in
the mixing case, the tree level contributions enter into the amplitudes
for \emph{both} the light and the heavy scalar with a factor of $\sin\alpha$ or $\cos\alpha$ respectively.

Furthermore, there are also couplings which involve some combination of the two scalars. 
These result
from the Higgs and the dilaton potentials, where couplings between the
two scalars $h$ and $\sigma$ 
are only present when the mixing is non-zero. On the other hand, the dimensionful couplings of the Higgs potential also give rise to interaction terms involving both the
dilaton and the Higgs boson, which are present even in the zero-mixing case.
These trilinear couplings are phenomenologically very important, because
they allow the dilaton to decay into a pair of Higgs bosons. Unfortunately, in
our model they are \emph{not uniquely defined}. Starting from the Lagrangian
introduced in the previous subsection, the cubic couplings come from the terms
\begin{align}
\mathcal{L}_{\phi^3} =& - \frac{1}{2v} m_{h,0}^2 h_0^3 - \xi \frac{m_{\sigma, 0}^2}{f} \sigma_0^3 - \frac{m_{h_0}^2}{f} h_0^2 \sigma_0 \,.
\end{align}
Here $\xi$ is a model-dependent dilaton self-coupling, that we have fixed to 5/6
in eq.~\eqref{EQ:dilpot}. The trilinear couplings between the two mass
eigenstates are then found by substituting, in the above equation, the mixing
relation of eq.~\eqref{EQ:MixingRotation}.

Note that the parameter $m_{h,0} \equiv \sqrt{2\lambda_{SM} v^2}$ is \emph{not} equal
to the mass of the lightest scalar. The physical masses are instead determined
by the diagonalisation of the dilaton-Higgs mass matrix extracted from the
bilinear terms of the scalar potential,
\begin{align}
\mathcal{L}_{\phi^2} =&\ - \frac{1}{2} m_{h,0}^2 h_0^2 - \frac{1}{2} m_{\sigma,0}^2 \sigma_0^2 - m_{h \sigma}^2 h_0 \sigma_0
\label{EQ:MinimalMixing}\end{align}
where
\be\bsp
  m_{h,0}^2      =&\ m_h^2 c_\alpha^2 + m_\sigma^2 s_\alpha^2 \,,\quad
  m_{\sigma,0}^2 = m_h^2 s_\alpha^2 + m_\sigma^2 c_\alpha^2 \,, \\
  m_{h \sigma}^2 =&\ - c_\alpha s_\alpha (m_\sigma^2 - m_h^2) \,.
\esp\ee

Note also that in the work of ref.~\cite{Efrati:2014aea} the parameter $m_{h \sigma}^2$ is not explicitly discussed. However, it is clear that if it is present, then the full Lagrangian of the theory cannot be written in a gauge-invariant way: we will have to give up precision calculations for the model and it is harder to make a connection with a UV-completion. On the other hand, if we want to write the lowest-dimension effective operator which can yield such a mass and preserve the gauge symmetry, then we must write the SM degrees of freedom in terms of a doublet $H \supset H^0 = \frac{1}{\sqrt{2}} [(v + h_0) + i G^0],$ and it follows that the physics that generates a dilaton mass should also generate a new coupling
\begin{align}
\mathcal{L} \supset& -\frac{m_{h \sigma }^2}{v} \sigma |H|^2 \supset -m_{h \sigma }^2 h_0 \sigma_0 - \frac{1}{2} \frac{m_{h \sigma }^2}{v} h_0^2 \sigma_0 \,.
\label{EQ:GaugeInvariantMixing}\end{align}
This new trilinear coupling has dramatic consequences for the phenomenology: it
allows for unsuppressed decays of the dilaton into two Higgs bosons. Therefore
we will investigate both scenarios: the ``minimal mixing scenario'' of
eq.~\eqref{EQ:MinimalMixing} and the ``gauge invariant mixing scenario'' of
eq.~\eqref{EQ:GaugeInvariantMixing}.

\subsection{Fermionic and vector dark matter through the dilaton portal}

The dilaton portal is particularly interesting and simple for dark matter models
because the dark matter coupling is unique (being determined only by the dark
matter mass). The most commonly considered and perhaps simplest dark matter candidate in a dilaton portal model is a Majorana fermion $\Psi$ or a real scalar $S$, with a $\mathbb{Z}_2$ symmetry.  
Note that in ref.~\cite{Efrati:2014aea} a \emph{Dirac} fermion was considered, which has instead a continuous unbroken global $U(1)$ symmetry (although this was not explicitly stated). The phenomenology is however very similar up to some factors of $2$.

There is little to add to the analysis of dark matter constraints performed in
this earlier work (and for the case of zero mixing, in the works of
ref.~\cite{Bai:2009ms,Jung:2014zga,Blum:2014jca}). However, we update those
results by providing limits from recent monojet and multijet searches in the
fermionic case in section~\ref{SEC:LHC}, and projections for a future 100~TeV
collider in section~\ref{SEC:FCC}.

A vector boson as the dark matter particle has attracted some interest as an alternative to the more common fermion or scalar candidates. The particular challenge for the vector case is to allow it to remain stable; a $\mathbb{Z}_2$ symmetry forbids conventional gauge interactions. We must either accept this \cite{Cheng:2003ju,Birkedal:2006fz,Bai:2009ms,Lebedev:2011iq,Arcadi:2020jqf} or we can stabilise the vector through another symmetry such as a custodial one \cite{Hambye:2008bq}. Here we shall be considering the minimal model, which can arise from a St\"uckelberg $U(1)$ field coupled to the dilaton. This automatically features
a $\mathbb{Z}_2$ symmetry provided there is no matter charged under it; and the
symmetry also forbids any kinetic mixing term involving the dark vector field, through
which it could decay.

Since our dilaton must be rather heavy and couples to all the SM particles
including the Higgs boson, we could consider it as generating effective Higgs
portal interactions. Indeed, integrating it out generates the effective
Lagrangian ${\cal L}_{\rm eff}$,
\be\bsp
  \mathcal{L}_{\rm eff} = &\ \frac{1}{2f^2 m_\sigma^2} \bigg[
     \sum_\psi m_\psi \ov{\psi} \psi \!-\! m_V^2 X_\mu X^\mu \!-\!
     \mz^2 Z_\mu Z^\mu\\
    &\quad  -2 \mw^2 W^+_\mu W^{-\mu} +\mh^2 h^2 \bigg]^2 \ ,
\esp\ee
after neglecting any higher-dimensional operators. This can be compared with the
standard Higgs-portal Lagrangian for vector dark matter ${\cal L}_{\rm VHPDM}$
(which has become a popular benchmark scenario) \cite{Lebedev:2011iq,
Arcadi:2020jqf},
\be
\mathcal{L}_{\rm VHPDM} = \frac{\lambda_{hv}}{8} h^2 X_\mu X^\mu + \frac{1}{2} m_V^2 X_\mu X^\mu + \frac{\lambda}{4} (X_\mu X^\mu)^2\,,
\ee
that exhibits the same classes of terms. 
These types of vector dark matter effective Lagragians including a mass term generated by
the St\"uckelberg mechanism exhibit a violation of unitarity, in particular at
high energy colliders and even for the standard Higgs portal. The Higgs
invisible decay width hence diverges when $m_V \rightarrow 0$. One simple way to
cure these problems consists of introducing a Higgs mechanism in the dark
sector~\cite{Baek:2012se}. Then, the Higgs invisible decay width becomes
finite~\cite{Baek:2014jga} and unitarity is restored~\cite{Ko:2016xwd,
Kamon:2017yfx,Dutta:2017sod}.  In the following, we will not suffer this complication.

In the dilaton case, the generic
coefficients $\lambda_{hv}$, $m_V$ and $\lambda$ are expressed in terms
of $m_V$ and $f$. While in $\mathcal{L}_{\rm VHPDM}$ the vector self-coupling
$\lambda$ is rather unimportant, ${\cal L}_{\rm eff}$ instead features
additional dimensionless quartic terms $ (X_\mu X^\mu) (Z_\mu Z^\mu) $ and
$(X_\mu X^\mu) (W^+_\mu W^{-\mu})$ that are \emph{crucial} for the phenomenology
of the model. They indeed provide the principal annihilation channels for the
dark matter. We therefore see no reason why these couplings should be neglected
in a generic vector dark matter effective field theory, and invite the reader to
reconsider benchmark scenarios that omit them.

Another interesting feature of a vector coupling to the dilaton is that a heavy dilaton will \emph{predominantly} decay into vectors. The partial width of the dilaton into dark matter becomes
\begin{align}
  \Gamma (\sigma \rightarrow X X ) = & \frac{m_\sigma^3}{32\pi f^2 } \bigg[ 1 - 4 \frac{m_V^2}{m_\sigma^2} + 3 \frac{m_V^4}{m_\sigma^4}\bigg] \sqrt{1 - \frac{4 m_V^2}{m_\sigma^2}}  \nn\\
     \underset{m_\sigma \gg m_V}{\rightarrow}&  \frac{m_\sigma^3}{32\pi f^2},
\end{align}
which is independent of the vector mass. When we fully consider the decay
channels into $Z$ and $W$ bosons, we find that the total width of the dilaton
can be well approximated by
\begin{align}
\Gamma \underset{m_\sigma \gg m_V, m_Z}{\simeq}&  \frac{m_\sigma^3}{8\pi f^2},
\label{EQ:TOTALWIDTH}\end{align}
and the branching ratio of the dilaton into dark matter then becomes roughly
0.25. This is much larger than for the scalar or fermionic cases, where the
dilaton branching ratio into dark matter tends to zero for small dark matter
masses. Consequently, vector dark matter production at colliders could be
potentially enhanced via a dilaton resonance.

This width also provides a limit on the size of the dilaton mass from the requirement that it be a narrow state. However, as we shall see, this is actually a weaker constraint than perturbative unitarity of scattering amplitudes.

\subsection{Perturbative Partial Wave Unitarity}

Since our theory contains non-renormalisable operators, it must have a cutoff
comparable to the scale $f$. We expect this to be manifest, even at tree level,
as perturbative unitarity constraints on two-body scattering amplitudes. In
particular, there are some couplings that could in principle be large compared
to $f$ since they come with additional massive factors -- such as the masses of
vector bosons, and particularly the dark matter (in the vector dark matter
scenario). We have therefore calculated the constraints originating from
imposing the unitarity of two-particle scattering when vector and dilaton fields
are involved, as well as the one stemming from the scattering of gluons into
heavy vectors via a
dilaton exchange. The details are given in appendix~\ref{APP:UNITARITY} that
also includes the derivation of useful approximate formulae, such as an upper
bound on the dilaton mass from self-scattering in
eq.~\eqref{EQ:UnitarityMaxMsig} of
\be m_\sigma \lesssim 2 f\, , \ee
which gives $\Gamma/m_\sigma \lesssim 1/2\pi$.

In particular, the scattering of gluons into vectors via a dilaton places a constraint on the \emph{maximum scattering energy} permissible in our theory. This will be relevant for the LHC and future collider constraints; and also for all other unitarity constraints (which necessarily also involve choosing a scattering momentum).

As shown in appendix~\ref{APP:gluonunitarity}, unitarity is violated if
\be
s > \frac{8\pi^2 f^2}{7 \alpha_s}\,.
\label{EQ:gluonslimit}\ee
In principle, when the process centre-of-mass energy $s$ is above this value ($30$ TeV for $f=3$ TeV), the cross section calculation in our theory is not reliable. However, in collider processes, this is of course not the centre-of-mass energy
of the proton-proton collision that is relevant, but the partonic one $\hat{s}$.
We have instead
\be
\hat{s} \equiv x_{1}x_{2} s < \frac{8\pi^{2} f^{2}}{7\alpha_{s}} \equiv \hat{s}_{\rm unitarity}\,,
\ee
or equivalently
\be
    x_{1}x_{2} < 0.08 \left( \frac{f}{3~\rm{TeV}} \right)^{2} \frac{(100 ~\rm{TeV})^{2}}{s}.
\ee
Here $x_{1,2}$ are the momentum fractions carried by the intial-state gluons,
the relevant processes being induced by gluon fusion. So
provided that the  gluon parton distribution function (PDF) is negligible for
\mbox{$x_{i} > \mathcal{O}(10^{-1})$}, calculations for a $100$ TeV collider are
safe. We consider processes where a single dilaton is resonantly produced. The
relevant scale therefore consists of its mass $m_\sigma$, and taking into
account the dilaton mass range investigated in this work, the typical
gluon-gluon parton collision scale is \mbox{$x_{1}x_{2}\sim 10^{-4}-10^{-3}$}.
This therefore guarantees the suppression of any growth in the
cross section coming from high-energy subprocesses.

On the other hand, the above constraint is very important when considering the
unitarity of (vector) dark matter scattering. While we typically want to take the high-energy limit for unitarity calculations to simplify matters and avoid resonances, we have seen that it is not possible for our model. In particular, the scattering of dilaton pairs into longitudinal vectors exhibits (only) a logarithmic growth with energy, so that the cutoff of eq.~\eqref{EQ:gluonslimit} also reduces the power of other unitarity constraints. To be safe for $f=2$ TeV we should take $\sqrt{s}< 20$ TeV and for $f=1$ TeV we should take $\sqrt{s}< 10$ TeV.

Throughout the rest of the paper we shall use the above limit as a guide for
fixing the \emph{cutoff} on our theory. This only appears in the other unitarity constraints, arising from vector and dilaton scattering, which we scan all momentum values up to a maximum \emph{vector} centre-of-mass momentum $p_{V,\mathrm{max}}$ where $p_V = \sqrt{s/4 - m_V^2}$. Since these other constraints grow logarithmically with $p_{V,\mathrm{max}} $ this is both necessary (to give finite results) and conservative.

\section{Constraints on the Higgs-dilaton mixing\label{SEC:HiggsDilMixing}}
\label{SEC:MIXING}

In order to assess the viable regions of the model parameter space, we use
input from experimental Higgs-boson measurements and heavy scalar searches, as
well as constraints on the magnitude of the new physics contributions to the
Peskin-Takeuchi parameters $S$ and $T$~\cite{Peskin:1991sw}.

For the Higgs and heavy scalar set of constraints, we enforce that predictions
for the signal strengths of various scattering processes (\ie\ the ratio of
total rates to their Standard Model counterparts) is consistent with
observations (within uncertainties).
In this context, it is assumed that the light or heavy scalar is resonantly
produced. It is therefore allowed to factorise the scattering cross sections
into the scalar production cross section times the branching ratio of the
considered decay mode.

The Higgs and heavy scalar constraints are determined using
\lilith-2.2.0~\cite{Kraml:2019sis} and \hsignals-2.4.0~\cite{Bechtle:2014ewa}
for the SM-like Higgs measurements, and \hbounds-5.5.0~\cite{Bechtle:2020pkv}
for the heavy scalar searches. The input for these codes is generated using a
specially written main program for \momegas-5.0.8~\cite{Belanger:2018ccd} using
model files generated with \frules-2.3.36~\cite{Alloul:2013bka}. The Higgs and
heavy scalar search limits require the computation of ratios of scalar couplings
to the ones of a SM-like Higgs particle of the same mass, both for the lighter
(observed) state and the heavier one. These are estimated using a modified
version of the routines embedded into \hbounds, that calculate tree-level scalar
decays into quarks with their masses evaluated at a fixed running scale of
100~GeV, as well as quark-loop-induced couplings to photons and gluons
(including some higher-order QCD factors) which we supplement with the
higher-order operators given by eq.~\eqref{EQ:PhotonGluonCouplings}. We have
verified that in the limit of small dilaton masses, the loop-induced operators
exactly cancel against the dilatonic ones for the dilaton state, as they should.

Predictions for the electroweak parameters $S$ and $T$ are obtained by rescaling
Standard Model results, which we compare with the experimental values extracted
from the electroweak precision fits~\cite{Haller:2018nnx} according to
formul\ae\ taken from ref.~\cite{Bernon:2015hsa}. For the calculation of the
contributions from the $h$ and $\sigma$ scalars ($\delta S$ and $\delta T$), we
use the SM Higgs-boson
contribution $X_{\rm SM}$ (for $X=S,T$)~\cite{Hagiwara:1994pw}, after replacing
the Higgs-boson mass by the corresponding scalar mass and modifying the
couplings as explained in section~\ref{sec:HiggsMixing}. This leads to
\be\bsp
  \delta X= &\ \big[(r_fs_\alpha+c_\alpha)^2-1\big]X_{\rm SM}(m_h)\\
    &\ \ +(r_fc_\alpha-s_\alpha)^2 X_{\rm SM}(m_\sigma)\,.
\esp\ee

\subsection{Constraints from the light SM-like Higgs boson}

Since a mass-mixing term between the dilaton and the SM-like Higgs scalar must
violate the electroweak symmetry, it is natural to assume that this mixing
should be small. However, in the absence of complete top-down constraints, we should consider all possible bottom-up values for the mixing angle $\alpha$. The allowed amount of mixing, depending on the parameter $f$ (or $r_f$), was examined
in ref.~\cite{Efrati:2014aea} for heavy scalar masses of 200, 600 and 900~GeV
with the Dirac dark matter mass being fixed to 300~GeV. While the bounds from heavy scalar searches obviously are strongly dependent on the mass of the heavy scalar (having a rather weak dependence on the dark matter mass with some provisos), the constraints on the light Higgs couplings are independent of both the dilaton and the dark matter masses (provided that the dark matter is not lighter
than half the Higgs-boson mass, opening up invisible decays). The constraints
from the $S$ and $T$ parameters are also only relatively weakly dependent on
the dilaton mass and have hence not drastically altered since 2014. The possible
values are therefore restricted to $|\sin\alpha|\ll 1$ or $r_f\ll 1$ (or both),
whereas the combination with electroweak precision measurements imposes an
additional upper bound of approximately 0.4 for $|\sin\alpha|$ and $r_f$.

\begin{figure}\centering 
  \includegraphics[width=\columnwidth]{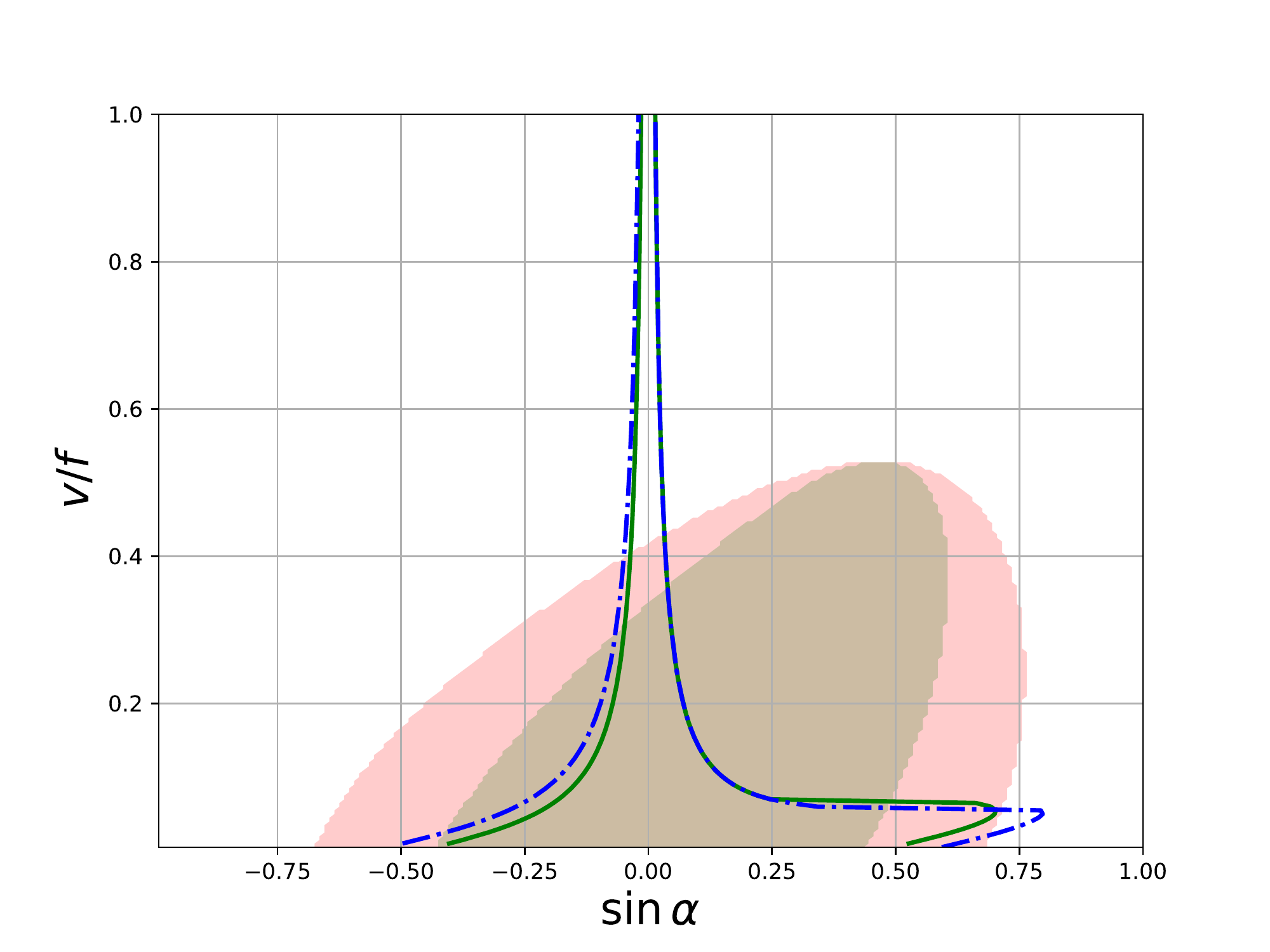}
  \caption{Constraints on the mixing angle between the Higgs and dilaton against
   the ratio $v/f$, computed using \lilith\ (solid green contour) and \hsignals\
   (blue dot-dashed contour). We also show the allowed regions from constraints
   originating from the electroweak $S$ and $T$ parameters for dilaton masses of
   3000~GeV (inner, light brown shaded region) and 500~GeV (outer, light red
   shaded region).}\label{FIG:LilithHiggsSignals}
\end{figure}

Figure~\ref{FIG:LilithHiggsSignals} shows the preferred parameter space regions
in the $(\sin\alpha, r_f)$ plane, after considering $99\%$ exclusion bounds
stemming from the most recent experimental input from Higgs measurements (as
implemented in \lilith~\cite{Kraml:2019sis} and
\hsignals~\cite{Bechtle:2014ewa}), as well as values for the $S$ and $T$
parameters extracted from the most up-to-date
electroweak fits~\cite{Haller:2018nnx}. The dark matter mass has been chosen
sufficiently high such that none of the scalars can decay invisibly, in order to
obtain bounds on $\sin\alpha$ and $r_f$ which are not influenced by dark matter.
The bounds originating from the electroweak precision fit are presented for
heavy scalar masses of 500 and 3000~GeV.

We have found out that a large amount of mixing is still permitted by data.
Larger mixing however modifies the constraints related to both heavy Higgs
searches (see below) and the dark matter sector. For the latter, the effect is
to more strongly couple the dark matter to the Higgs boson, which greatly
strengthens constraints from direct detection since the Higgs couples much more
strongly to light quarks than the dilaton does.

Our results additionally allow for a comparison between the \lilith\ and
\hsignals\ programs, the former requiring slightly more input (the decays
needing to be specified) while the latter calculates the decay table based on
coupling ratios. The results agree at an excellent level and do not
substantially differ from older bounds~\cite{Efrati:2014aea}.

\subsection{Heavy scalar searches}

Heavy scalar searches provide rather powerful complementary information, and
indeed have been substantially strengthened during the second run of the LHC.
These constitute some of the main new results of this paper. Here we probe the allowed parameter space for $f$ and $m_\sigma$ for fixed mixing angles. The results depend crucially on the treatment of the dilaton-Higgs mixing term, as we describe below.

\begin{figure*}\centering
  \includegraphics[width=0.49\textwidth]{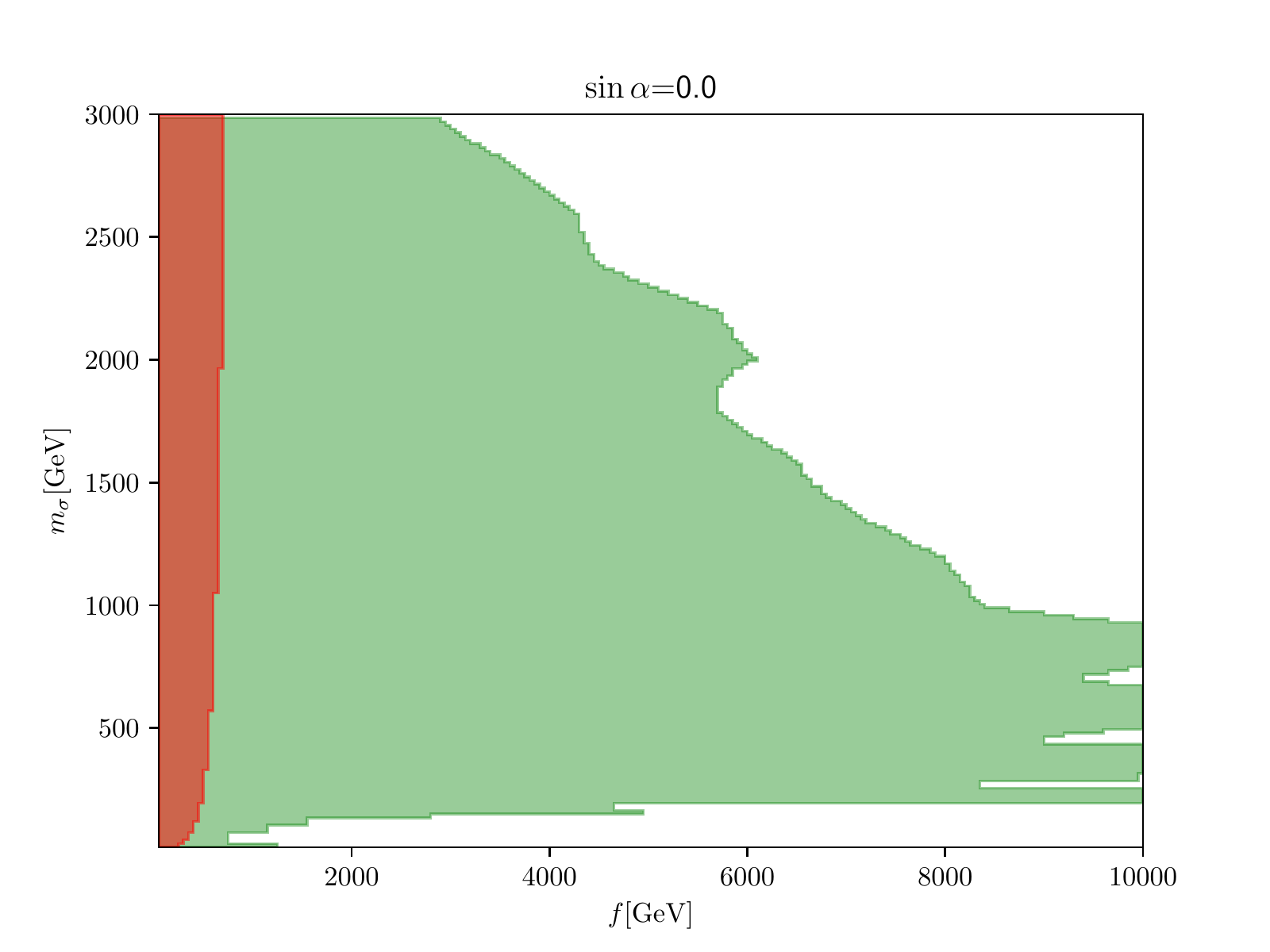}
  \includegraphics[width=0.49\textwidth]{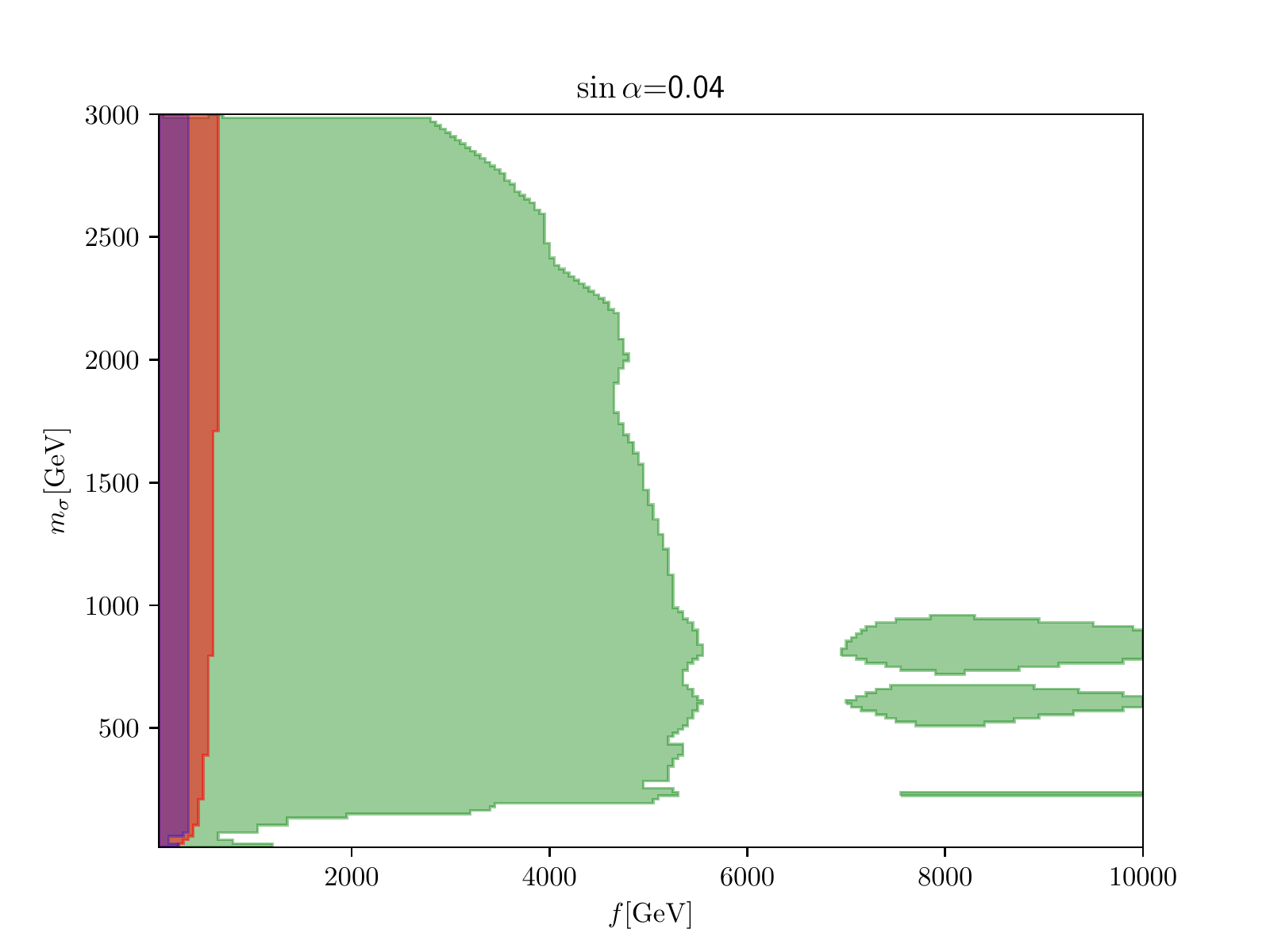}
  \includegraphics[width=0.49\textwidth]{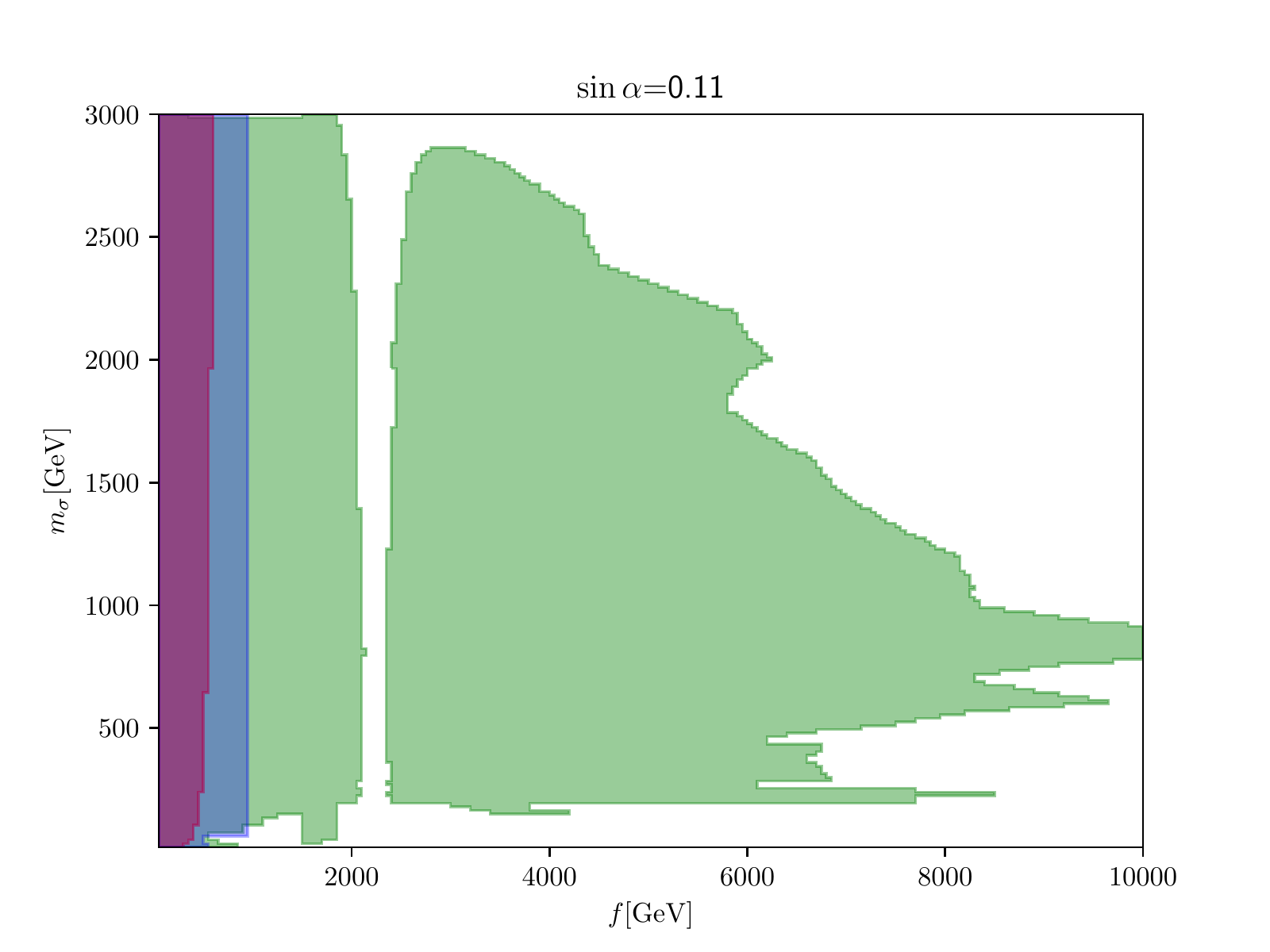}
  \includegraphics[width=0.49\textwidth]{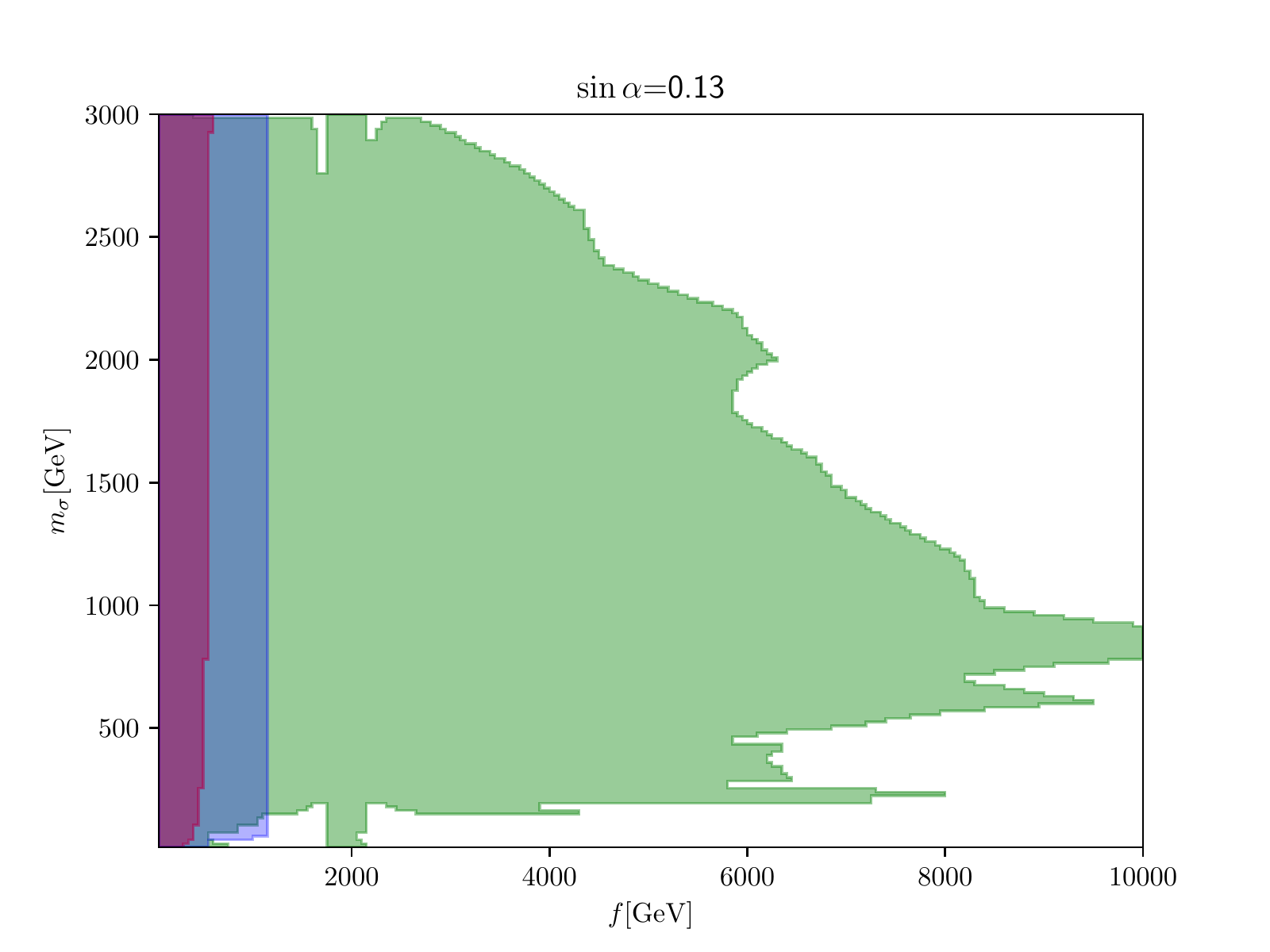}
  \caption{Excluded regions of the model parameter space, presented in the $(f,
   m_\sigma)$ plane for a selection of mixing angles, with the `minimal mixing
   term' treatment of the Higgs-dilaton mixing. We include 99\% confidence level
   exclusions from Higgs measurements (blue) and the electroweak $S$ and $T$
   parameters (red/purple), as well as 95\% confidence level bounds from heavy
   scalar searches (green). The parameter
   scans were done using \lilith\ and \hbounds. Our findings demonstrate that
   for sufficiently low positive mixing angles, a gap in the exclusion emerges
   around the value $f$ that satisfies $\nicefrac{v}{f}=\tan\alpha$. In this
   gap, the couplings of the heavy scalar to the Standard Model fermions and
   massive gauge bosons are close or equal to zero, so that searches for heavy
   scalars turn out to be insensitive and arbitrary heavy scalar masses are
   allowed.}\label{FIG:HeavyScalar}
\end{figure*}

We consider a heavy dilaton with a mass larger than 300~GeV. Throughout most of
the parameter space the main constraint therefore comes from diboson searches
since the dilaton predominantly decays to heavy vector bosons. However, there
are two notable special cases.

The first is when the dilaton decay constant is not large compared to the
dilaton mass. As we have seen, the dilaton can become a wide resonance, for
large $m_\sigma$. Even though unitarity constrains
$\Gamma/m_\sigma \lesssim 1/2\pi$, the width may be large enough that the standard narrow resonance searches do not apply. We shall avoid this issue by restricting our attention to a narrow dilaton.

The second special case, as already pointed out in ref.~\cite{Efrati:2014aea}, is when $f$ and $\alpha$ fulfill the condition
\begin{equation}
  \left(-s_\alpha+r_fc_\alpha\right)=0\quad\Leftrightarrow\quad r_f=\tan\alpha\,.
\end{equation}
Here the couplings of the heavy scalar $\sigma$ to fermions and heavy vector
bosons in eqs.~\eqref{EQ:MixingFermionCouplings} and
\eqref{EQ:MixingWZCouplings} vanish, \emph{independently of $m_\sigma$}.
This does not mean that the heavy scalar completely decouples from the Standard Model: the tree-level couplings to photons and especially gluons do not carry the factor $(-s_\alpha+r_fc_\alpha)$ and are therefore non-vanishing (although the contribution from fermion and vector loops vanishes). Moreover, since the dark matter particle does not obtain its mass from the Higgs mechanism, its coupling to the heavy scalar is also non-vanishing at this point. Therefore, in a `magic window' around $r_f = \tan \alpha $ the most stringent constraints on the heavy scalar disappear, yet it can still be produced by gluon fusion and decay significantly to dark matter.

There are still constraints even exactly at this magic value, notably from
diphoton and di-Higgs decays. The details depend crucially on the treatment of the mixing mass term, as described in section \ref{sec:HiggsMixing}. If we take the minimal approach to the dilaton-Higgs mixing, and do not introduce additional interactions, then for small values of the mixing angle we do see this `magic window' appear. The width of this window in terms of values of $f$ depends on the mixing and mass $m_\sigma$, and so does the \emph{depth}: for sufficiently large mixing the window actually disappears. The appearance of a wide window at small $\sin \alpha = 0.04$ and large $f=6$ TeV, and its eventual disappearance around $\sin \alpha = 0.13$ are illustrated in figure~\ref{FIG:HeavyScalar}.
At larger mixing angles, the constraints from the light Higgs measurements and
the $S/T$ parameters moreover dominate.

 \begin{figure*}\centering
  \includegraphics[width=0.49\textwidth]{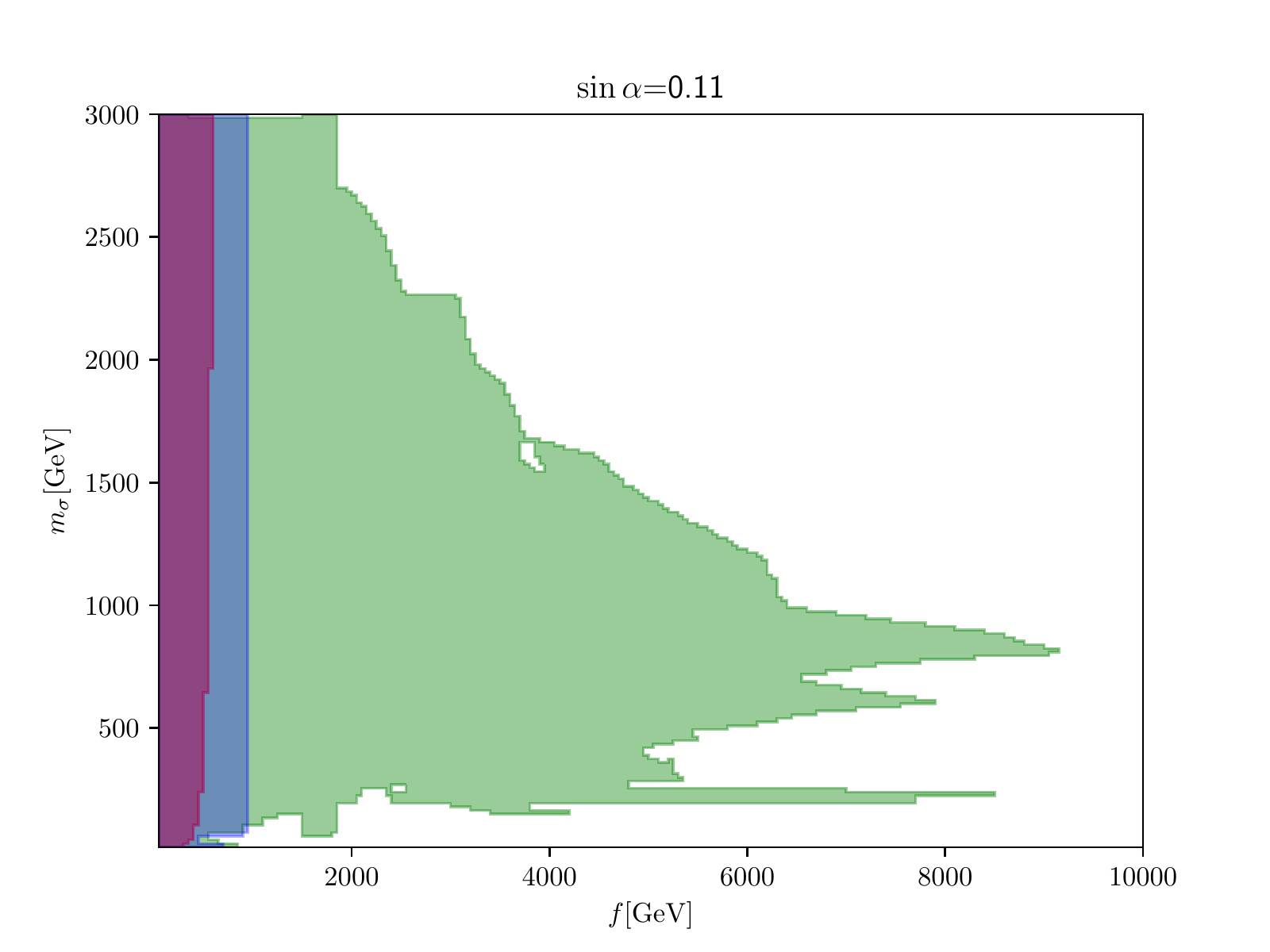}
  \includegraphics[width=0.49\textwidth]{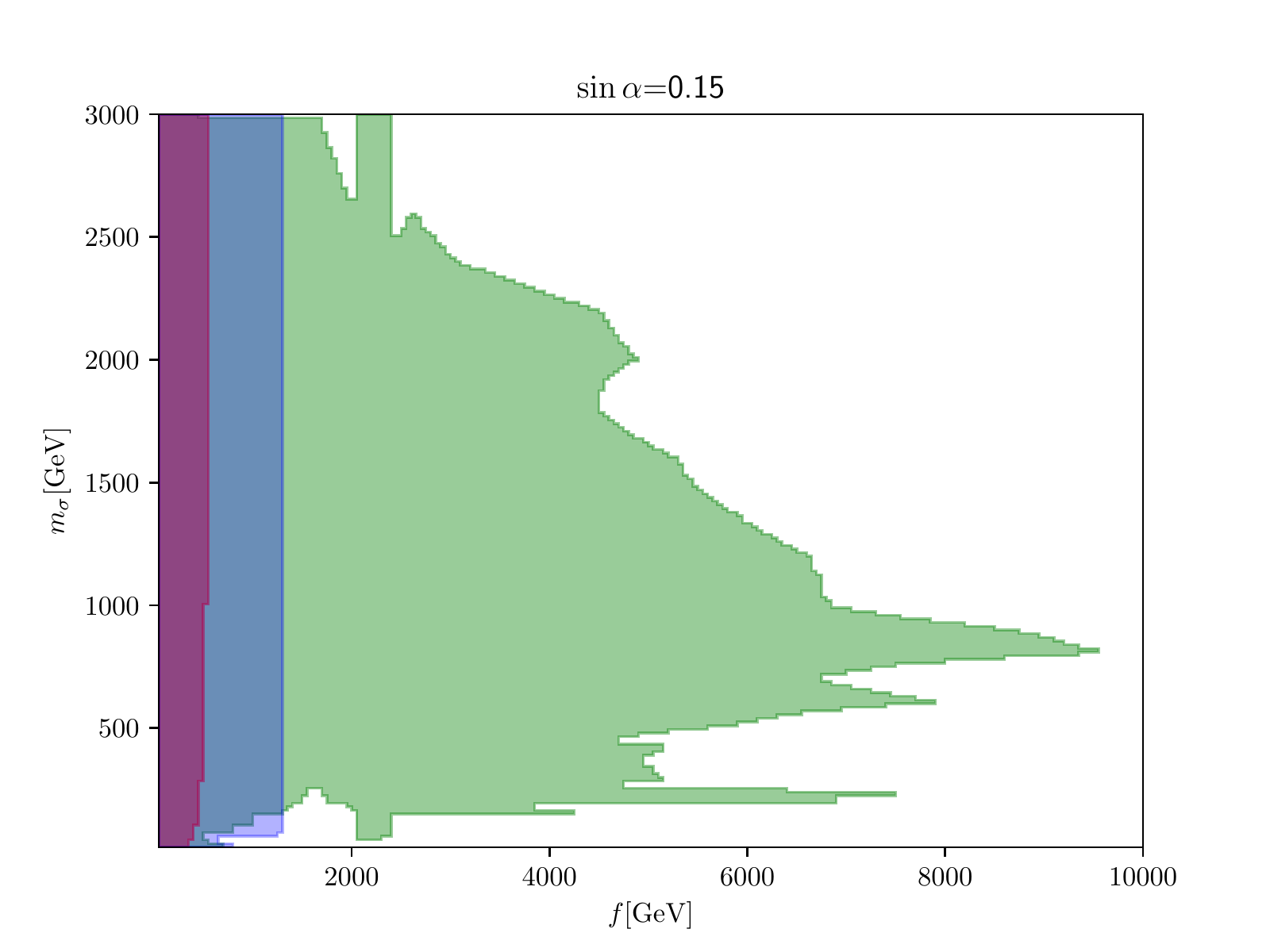}
  \caption{Constraints on the parameter space of $f$ and $m_\sigma$ for $\sin \alpha = 0.11$ (left) and $\sin \alpha =0.15$ (right) with a `gauge invariant' treatment of the dilaton-Higgs mixing. The colour coding is the same as for figure~\ref{FIG:HeavyScalar}. }\label{FIG:HeavyScalarGI}
\end{figure*}
 
On the other hand, if we introduce an additional term in the Lagrangian to
restore gauge invariance, then our results are dramatically modified. In
particular, the `magic window' around $v/f = \tan \alpha$ that was found with
the minimal treatment of the mixing disappears. In the `minimal mixing' case and
for large dilaton mass, the trilinear $\sigma h^2$ coupling $\lambda_{\sigma h h}$ becomes
\begin{align}
\lambda_{\sigma h h} \underset{v/f = s_\alpha/c_\alpha}{\longrightarrow} & \frac{m_\sigma^2}{v} c_\alpha^2 s_\alpha^3 (\xi -2) - 2 \frac{m_h^2}{v} c_\alpha^4 s_\alpha + ...
\end{align}
where the ellipsis denotes additional terms suppressed by powers of $s_\alpha$. On the other hand, in the `gauge invariant mixing' case it becomes
\begin{align}
\lambda_{\sigma hh} \underset{v/f = s_\alpha/c_\alpha}{\longrightarrow} & - c_\alpha^4 s_\alpha \frac{m_\sigma^2}{v} +...
\end{align}
which is dramatically enhanced compared to the previous value, by a factor of order $m_\sigma^2/m_h^2$. This leads to dilaton decays into two Higgs fields dominating for much of the parameter space, and has the effect of \emph{completely erasing the `magic window'}. We show the constraints in this case  in figure~\ref{FIG:HeavyScalarGI}, which shows current heavy scalar searches wiping out all of the interesting parameter space.

\section{Underground searches for dark matter}
\label{sec:DM}

We now turn to considering the constraints on our dilaton dark matter model
originating from the relic density, direct detection, and the LHC. Combined with
unitarity constraints, we will show which part of the parameter space remains
viable, and to what extent the different searches are complementary.

\subsection{Collider constraints}
\label{SEC:LHC}

\begin{figure*}
  \centering
  \includegraphics[width=.48\textwidth]{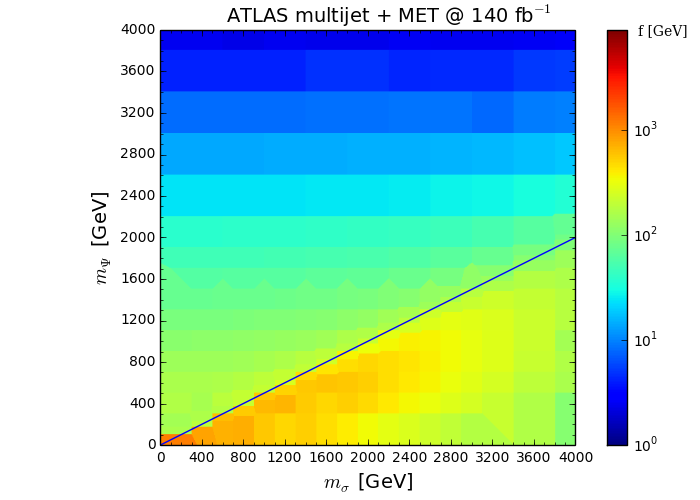}
  \includegraphics[width=.48\textwidth]{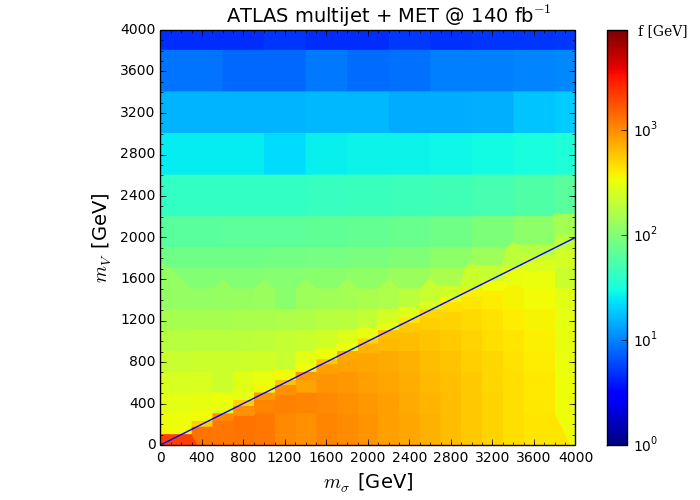}
  \caption{
    Constraints on dilaton-induced dark matter for fermionic dark matter (left)
    and vector dark matter (right), presented in the $(m_\sigma, m_V)$
    plane. For each mass configuration, we evaluate the maximum value of the
    cut-off scale that can be probed by using 140~fb$^{-1}$ of LHC data and the
    ATLAS analysis of ref.~\cite{ATLAS:2019vcq}.
  \label{fig:LHC_limits}}
\end{figure*}

In order to assess the constraints originating from dark matter searches at the
LHC, we used the implementation in the \frules\ package~\cite{Alloul:2013bka} of
the Lagrangian in eq.~\eqref{eq:lag} described in section \ref{SEC:MIXING}, to generate a UFO
library~\cite{Degrande:2011ua}. We then employ \MadGraph~\cite{Alwall:2014hca}
to generate hard-scattering events relevant for the production of a pair of dark
matter particles together with jets,
\begin{equation}
 p p \to X X j\ ,
 \label{monojet}
\end{equation}
where $X$ generically denotes the dark matter particle (a
Majorana fermion or a vector boson). In our simulations, we convolute
leading-order (LO) matrix elements with the LO set of NNPDF~3.0
parton densities~\cite{Ball:2014uwa}. Moreover, for every dilaton/dark matter
mass configuration, we evaluate the dilaton width with the
{\sc MadSpin}~\cite{Artoisenet:2012st} and {\sc MadWidth}~\cite{Alwall:2014bza}
packages and make sure that the dilaton is narrow (see
eq.~\eqref{EQ:TOTALWIDTH}). In other words, we enforce the $m_\sigma/f$ ratio to
be small enough.

The above hard process gives rise to
a monojet or a multijet plus missing transverse energy (MET) collider signature, that
is targeted by numerous dark matter searches undertaken by the ATLAS and CMS
collaborations~\cite{Sirunyan:2017jix,ATLAS:2019vcq}. As those searches usually
select events featuring at least one highly-energetic central jet, we impose,
at the event generator level, that the transverse momentum of the jet satisfies
$p_T > 100$~GeV and that its pseudo-rapidity fulfills $|\eta|<5$.

The simulation of the QCD environment relevant for proton-proton collisions is
achieved by matching our hard-scattering events with parton showering and
hadronisation as modeled in the {\sc Pythia}~8
program~\cite{Sjostrand:2014zea}.
The LHC sensitivity to the model is then estimated by re-interpreting the
results of the ATLAS-CONF-2019-040 analysis~\cite{ATLAS:2019vcq} that probes
dark matter models through a luminosity of 140~fb$^{-1}$ of LHC data at a
centre-of-mass energy of 13~TeV. This analysis targets multijet events featuring
a monojet-like topology, {\it i.e.} it requires events to exhibit a large amount
of missing transverse energy, a large number of jets with at least one of them
being very hard. As a consequence, it gives a great handle on the model
considered in this work and dark matter models in general.

Starting from Monte Carlo simulations of the dilaton-induced dark matter signal,
we make use of the {\sc MadAnalysis}~5 program~\cite{Conte:2012fm,
Conte:2018vmg} to automatically simulate the response of the ATLAS detector
through a tune of the {\sc Delphes}~3 package~\cite{deFavereau:2013fsa},
that internally relies on the {\sc FastJet}~software~\cite{Cacciari:2011ma}
for event reconstruction on the basis of the anti-$k_T$ jet
algorithm~\cite{Cacciari:2008gp}. We then estimate, still within the
{\sc MadAnalysis}~5 framework, the efficiencies of the different signal regions 
of the ATLAS-CONF-2019-040 analysis as it has been implemented in the 
Public Analysis Database~\cite{Dumont:2014tja}. The sensitivity of the LHC run~2
to the signal is extracted through the CL$_s$ method~\cite{Read:2002hq}.

In figure~\ref{fig:LHC_limits}, we present the obtained constraints for various
dark matter and dilaton mass configurations, both for Majorana fermion (left
panel) and vector (right panel) dark matter. For each mass configuration, we
evaluate the maximum value of the cut-off scale that is excluded by the ATLAS
analysis under consideration.

This shows that the LHC has no sensitivity to
scenarios in which the dilaton cannot be produced on-shell and then decay into
a pair of dark matter particles (which corresponds to the parameter space region
lying above the blue line). In contrast, when $2 m_X \lesssim m_\sigma$ (with
$m_X$ generically denoting the dark matter mass), 
cut-off scales $f$ around the TeV scale can be reached, which closes a part of
the small light dilaton window visible allowed by Higgs data (see the upper left
panel of figure~\ref{FIG:HeavyScalar}). Moreover, as expected from the spin
nature, constraints are tighter in the vector dark matter case than in
the fermionic one.

Enforcing a naive scaling of both the background and the
signal~\cite{Araz:2019otb}, we have verified that the future high-luminosity
operation of the LHC will not substantially affect those conclusions, even with 3000~fb$^{-1}$
of data. Multijet plus missing transverse energy LHC analysis targeting a
monojet-like topology within a multijet environment are indeed already limited
by the systematics~\cite{Banerjee:2017wxi} so that mild improvements can only be
expected with a larger amount of luminosity. Moreover, in the present case, the
limit depends on the fourth power of the cut-off scale, so that a noticeable
improvement at the level of the bounds on $f$ would require a huge improvement
at the analysis level.

\subsection{Vector dark matter at zero mixing}

The dilaton portal for dark matter is very simple, and, as described above, rather special. Its phenomenology in the case of no mixing between the Higgs and the dilaton is therefore rather straightforward: we have two regimes, either near the $s$-channel resonance where $m_V \simeq m_\sigma/2$,
or in the parameter space region where $m_V \gg m_\sigma$ in which there is a
combination of $t$-channel annihilations of the dark matter to dilaton pairs and $s$-channel annihilation. However, in the previous sections we saw that heavy Higgs searches force $m_\sigma > 3$ TeV for $f < 3$ TeV at zero mixing, and the bound on $m_\sigma $ is roughly given by $3000\, \mathrm{GeV}/f.$ Clearly this will limit the possibilities for detecting dilaton-induced vector dark matter at colliders or via direct detection.

\begin{figure}\centering
  \includegraphics[width=\columnwidth]{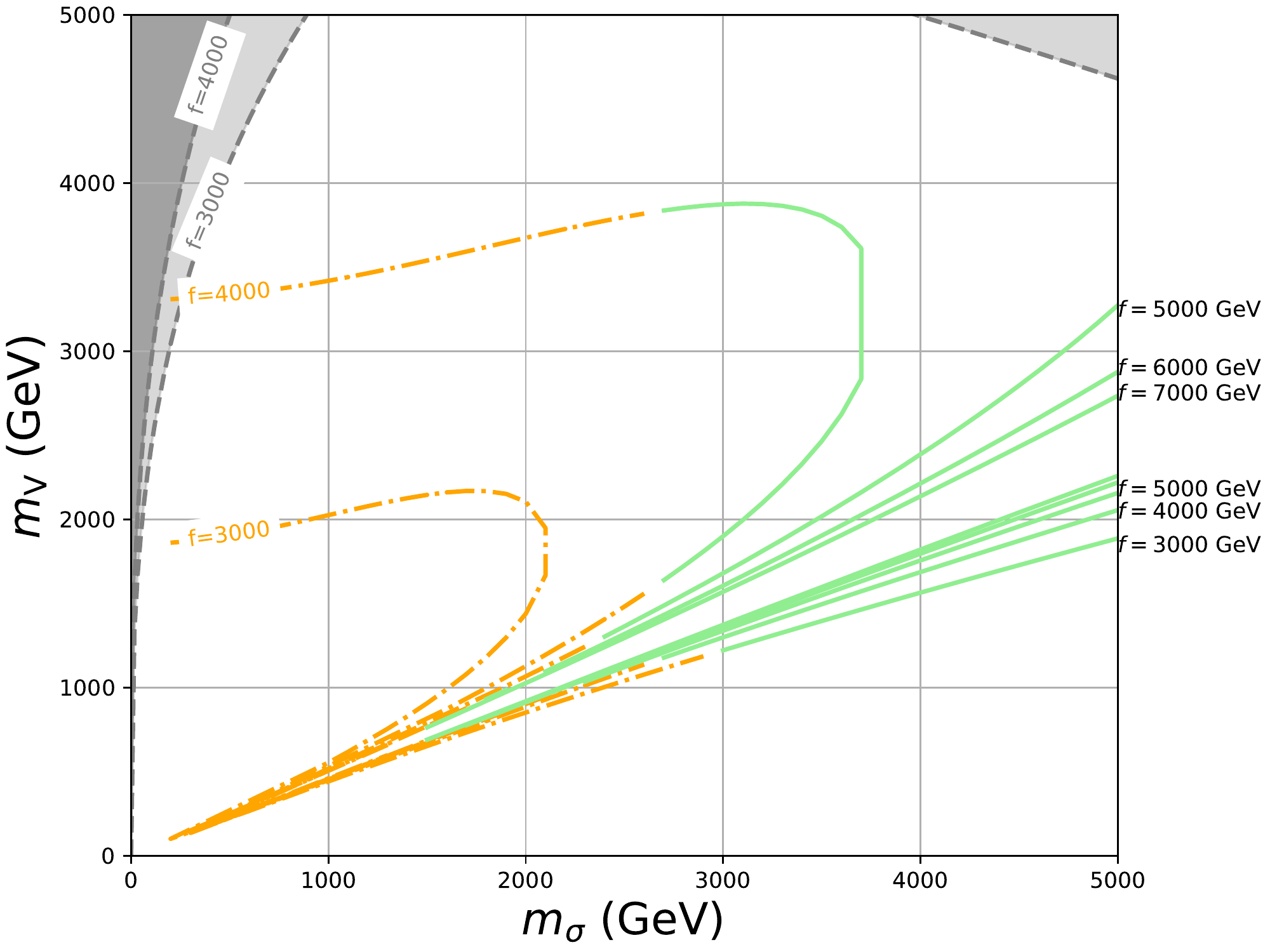}
  \caption{Dilaton and vector dark matter masses that saturate the observed dark
   matter relic density $\Omega h^2 =0.12$, for differing values of $f$ and no
   mixing between the dilaton and the Higgs boson. The curves are green and
   solid when the points are not excluded by any observations; they are orange
   and dot-dashed when excluded by heavy Higgs searches. The dark matter is
   \emph{underdense} (and thus allowed if there is another source of dark
   matter) between the two curves related to a given $f$ value, and
   \emph{overdense} (hence excluded unless there is some mechanism to dilute the
   dark matter density) outside them. For the values $f=3000, 4000$~GeV the
   underdense region therefore extends above the curves towards $m_V \rightarrow
   \infty$. Unitarity constraints are shown as the shaded grey regions.}
\label{FIG:VectorDM}
\end{figure}

We show in figure~\ref{FIG:VectorDM} the contours of relic density matching the
Planck results~\cite{Aghanim:2018eyx} for fixed $f$ values in the $(m_\sigma,
m_V)$ plane, for values of $f$ upwards of 3~TeV. We see that the sensitivity of
heavy scalar searches is not noticeably weakened due to the invisible
dilaton decays into dark matter, and so the most promising regions for detection
corresponds to a resonant configuration. This thus leads to $m_\sigma>2700$~GeV
(rather than 3000~GeV without the presence of dark matter) for $f=4000$~GeV, but where unfortunately $\sigma_{\rm proton, spin-independent} \simeq 10^{-50} \, \mathrm{cm}^2,$ well beyond the reach of current and near-future direct-detection experiments. Moreover, the LHC and HL-LHC searches described in the previous section do not limit the parameter space in the figure at all. Alternatively we can say that this model has a large unexcluded viable parameter space.

We also present limits from unitarity as described in
appendix~\ref{APP:UNITARITY}, which appear at the edges of
figure~\ref{FIG:VectorDM}. The constraints come from dark-matter scattering at
low momentum via a dilaton exchange, and at small dilaton masses. This could
also be interpreted as the regime where Sommerfeld enhancement should be taken
into account in the calculations. In the top right corner of the figure,
high-energy scattering constraints become visible, where we have taken a maximum
centre-of-mass vector momentum of $p_{V,\mathrm{max}} = 20$ TeV. 

\begin{figure}\centering
  \includegraphics[width=\columnwidth]{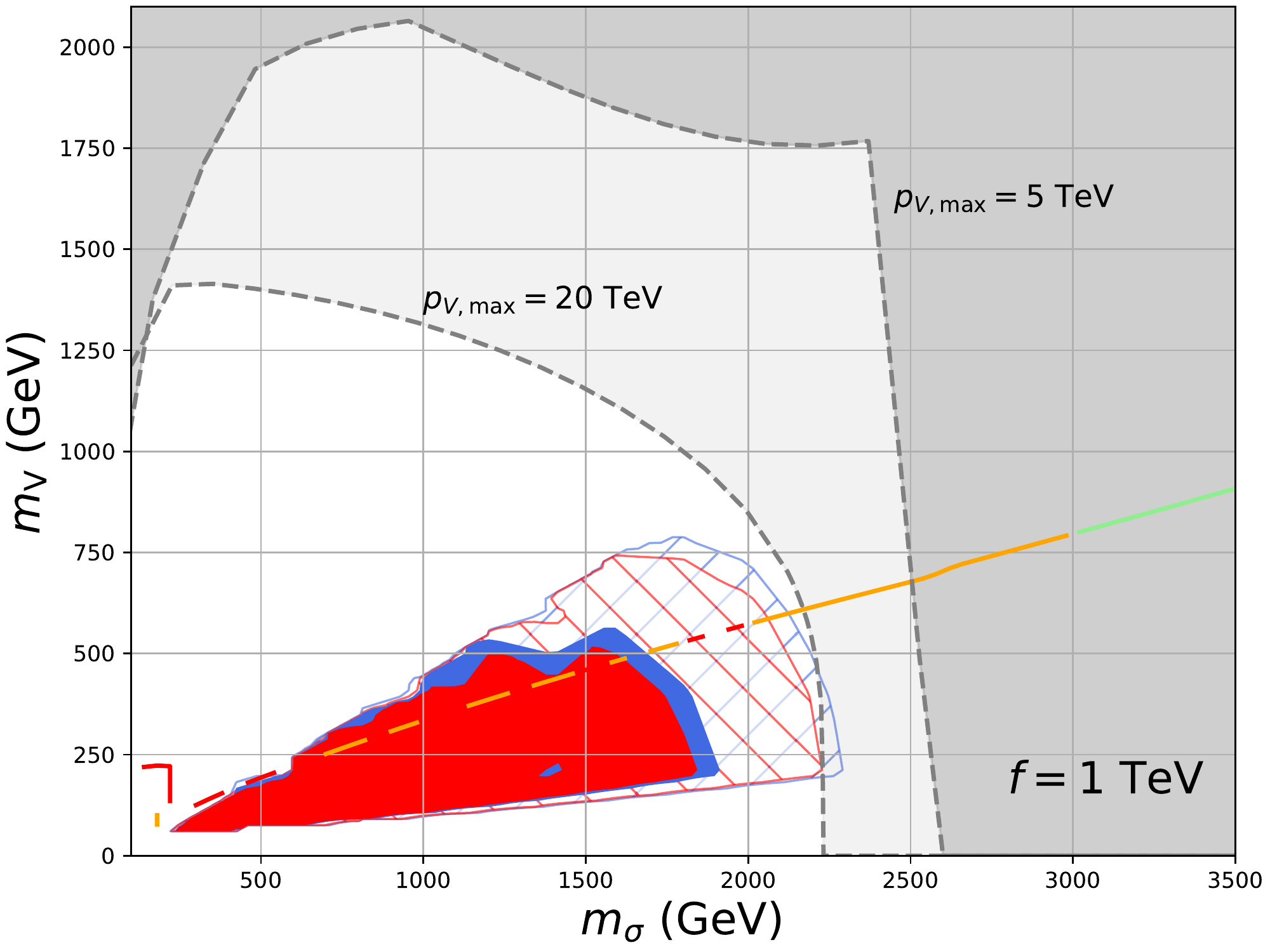}  
  \includegraphics[width=\columnwidth]{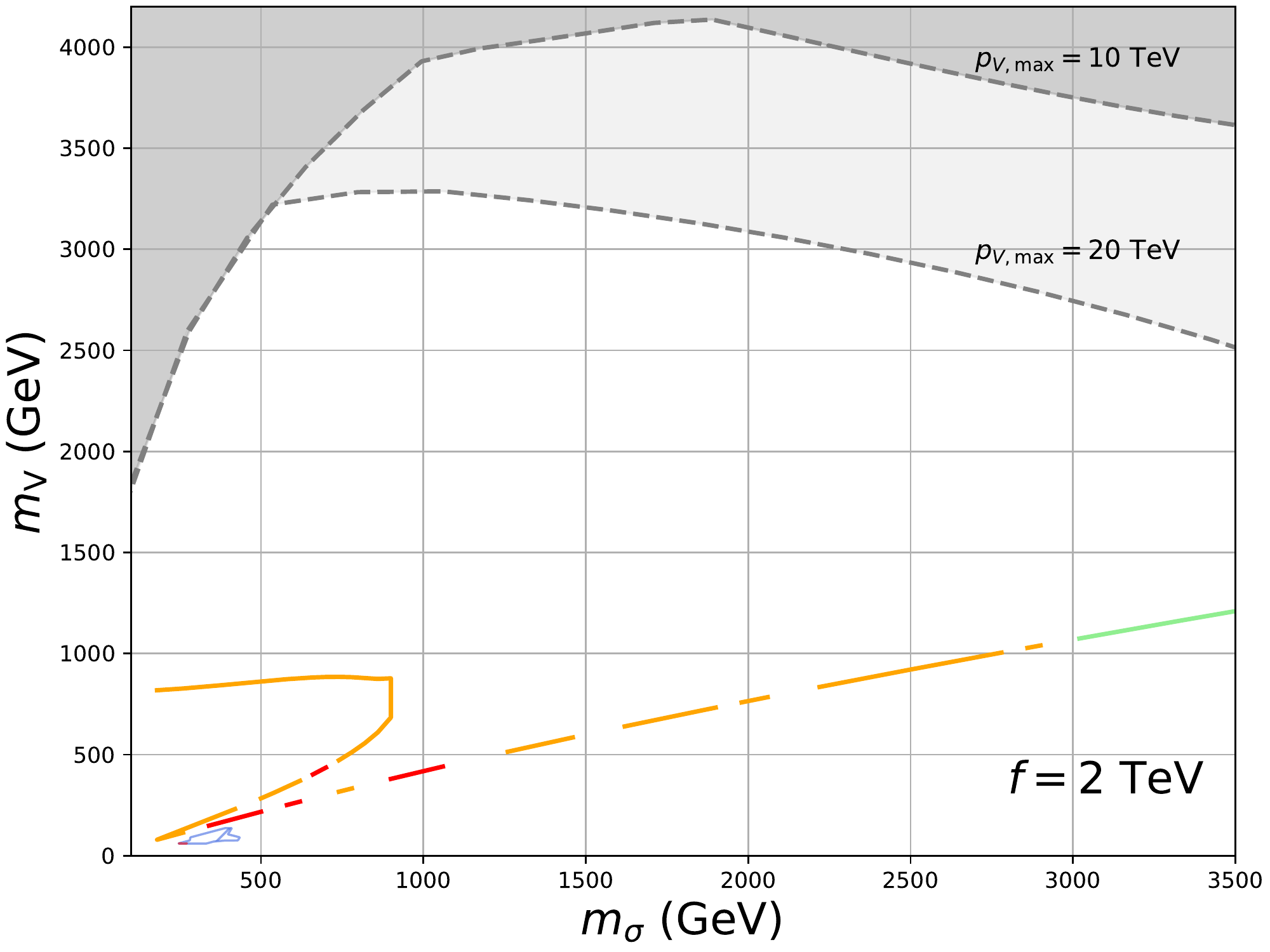}
  \caption{Dark matter \emph{curve} as in figure~\ref{FIG:VectorDM}, but for
  \mbox{$f=1000$}~GeV (upper) and 2000~GeV (lower); the red portion of the curve
  corresponds to regions excluded by both heavy Higgs searches \emph{and} dark
  matter direct detection experiments. The solid and hatched \emph{regions} show
  (current) LHC and (future) HL-LHC exclusions from dark-matter-inspired
  collider searches: the solid blue region is the future exclusion reach at the
  HL-LHC, after accounting for LO signal cross sections, and the hatched blue
  region shows the same constraint but with a signal enhanced by a
  $K$-factor of $2$. Similarly, the red solid and hatched regions depict the
  current LHC exclusion without and with a $K$-factor of $2$. Unitarity
   constraints are again shown as shaded grey regions. On the upper figure, the
   entire parameter space of the model is excluded by a combination of heavy
   Higgs searches and unitarity. On the lower panel, the monojet/multijet+MET
   searches are barely visible, and some viable parameter space exists above the
   reach of heavy Higgs searches. }
\label{FIG:lowf}
\end{figure}

It is legitimate to ask what happens at smaller values of $f$: can there remain some viable parameter space? For $f=1$ TeV, the constraints on the gluon scattering momentum force us to impose a maximum cutoff of \mbox{$s=10$}~TeV, leading to $p_{V,\mathrm{max}} = 5$ TeV.
In the upper panel of figure~\ref{FIG:lowf}, we see that this excludes $m_\sigma
< 2.5$~TeV. However, if we are more aggressive and allow for a higher cutoff,
the parameter space can be further shrunk. In that figure, we also show the
sensitivity of the collider searches for dark matter, although \emph{the entire
$(m_\sigma, m_V)$ plane is excluded by a combination of heavy Higgs search
results, dark matter and unitarity constraints}.

On the other hand, as shown in the lower panel of figure~\ref{FIG:lowf}, the
unitarity constraints give us an upper bound on the dark matter mass for
$f=2$~TeV. In our results, we naively limit the considered parameter space
regions to $m_\sigma > 3$~TeV and $m_V\in [1.1, 2.5]$~TeV. The reason is that at
$m_\sigma = 3$~TeV the dilaton is rather \emph{wide}, with $\Gamma_\sigma\simeq
250$~GeV. This means that the dark matter density constraint does not longer
result in a funnel, as dark matter is underdense everywhere above the shown
curve. Clearly, the choice $f=2$~TeV is therefore rather borderline in terms of
whether we trust the results of the numerical calculations.

\subsection{Collider and vector dark matter constraints at non-zero mixing}

The hope of detecting dark matter greatly improves once we allow for
dilaton-Higgs boson mixing: the
dark matter acquires a coupling to the Higgs, and so interacts much more strongly with nuclei (in principle the model can then accommodate a Higgs portal, which we shall not consider as being very fine-tuned). Moreover, we also have the possibility of sitting in the `magic window' where $f = v/\tan \alpha$, which should also maximise the reach of dark matter collider searches.

In the `minimal mixing' case, this would seem to be the ideal situation: the
dilaton still has barely suppressed gluon and dark vector couplings, but its
couplings to SM bosons and fermions vanish. This means that the dilaton decays
only to the dark vector and the SM Higgs boson. Potentially, then, monojet and multijet + MET searches could probe some interesting part of the parameter space of the model.
Since the collider searches depend so strongly on $f$, to have the best sensitivity we should look for the lowest possible value. We saw earlier that in the `minimal mixing' case at the magic window, for $m_\sigma > 300$ GeV, the minimum value of $f$ that survives all constraints was for $\sin \alpha =0.11$,
giving $f = 2.2$ TeV. 

\begin{figure}\centering
  \includegraphics[width=\columnwidth]{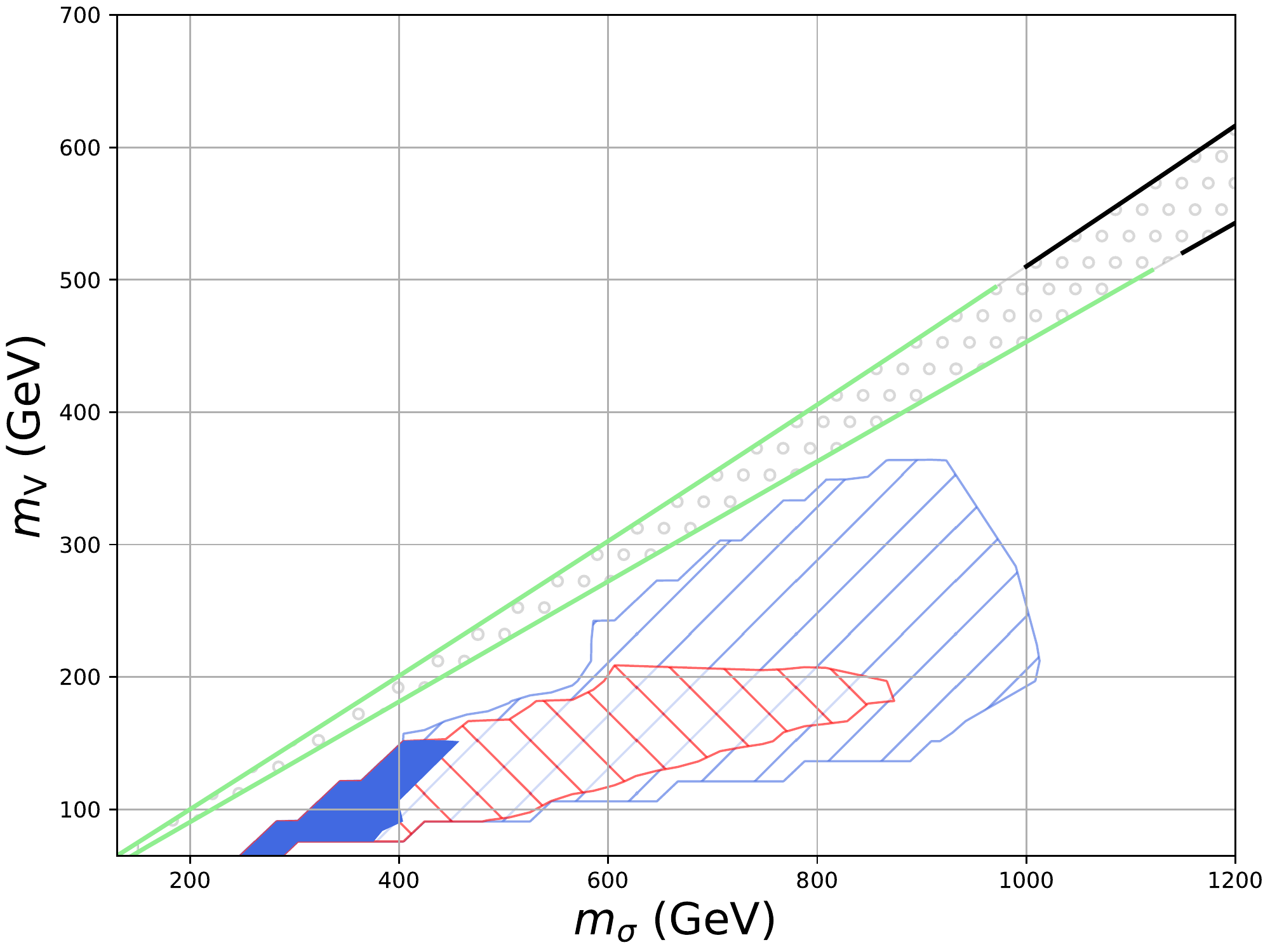}
  \includegraphics[width=\columnwidth]{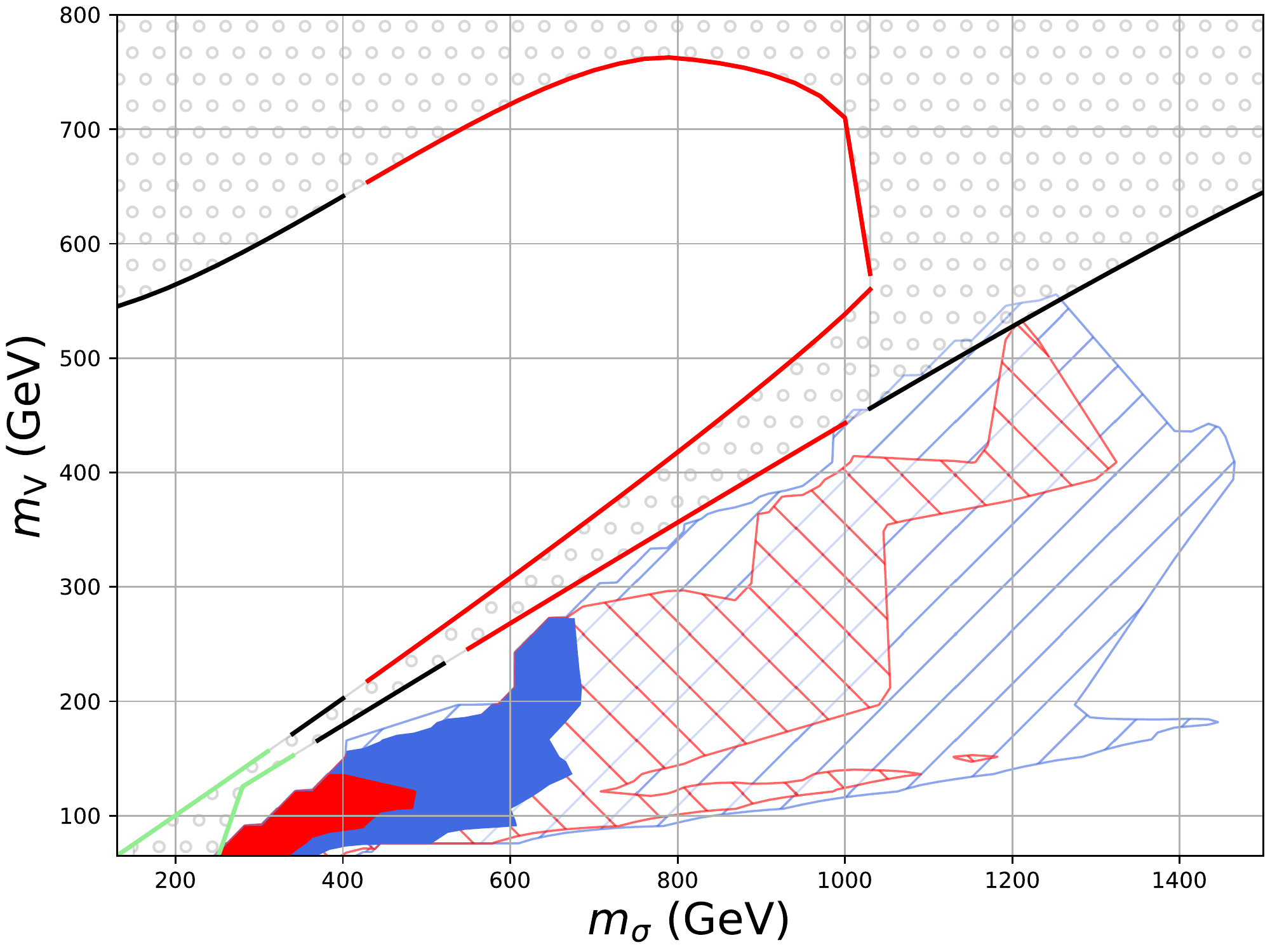}
  \caption{Combined dark matter, Higgs and collider constraints on the
    considered model for $\sin \alpha = 0.11$ (upper) and 0.15 (lower) in the
    `magic window' where $f = v/\tan \alpha$ and under the assumption of a
    `minimal Higgs-dilaton mixing' treatment. The solid lines show the curves
    where the dark matter density matches the Planck limit of $\Omega h^2=0.12$
    with the circular shading between them showing the \emph{underdense}
    regions. The solid line is green and orange for allowed and excluded by
    heavy Higgs searches, and black when excluded by dark matter direct
    detection. The solid and hatched regions show the current and future
    exclusions from dark matter collider searches: the solid blue region is the
    future exclusion reach at the HL-LHC after accounting for LO signal cross
    sections and the hatched blue region is the same constraint but with the
    signal enhanced by $2\sigma$ according to the uncertainty on its total rate.
    The red solid and hatched region represent the corresponding constraints
    at the end of the LHC run~2.
    \label{FIG:sin011}}
\end{figure}

For the `minimal mixing' case we show the results for both relic density, direct
detection and direct production at the LHC and HL-LHC in
figure~\ref{FIG:sin011}, for mixing angles of $\sin \alpha = 0.11$ and $\sin
\alpha = 0.15$. The relic density and direct detection cross sections were
computed using \momegas\ and compared with the limits summarised in
ref.~\cite{Schumann:2019eaa}. The LHC limits and projections were inferred from the results in section \ref{SEC:LHC} by recomputing the production cross sections (for the $pp \rightarrow VVj$ process with the same cuts on the hard jet) for the mixing case.
Due to the dilaton widths and the relative coupling changes, we could not
naively rescale the cross sections. However, the same cutflows/limits on the
total rate can be used.
We observe that the (HL-)LHC searches do not overlap with any of the viable
regions of the parameter space, so that for $\sin \alpha =0.11$ we could have a
viable dark matter model for $m_\sigma < 1$ TeV.

\begin{figure}\centering
\includegraphics[width=\columnwidth]{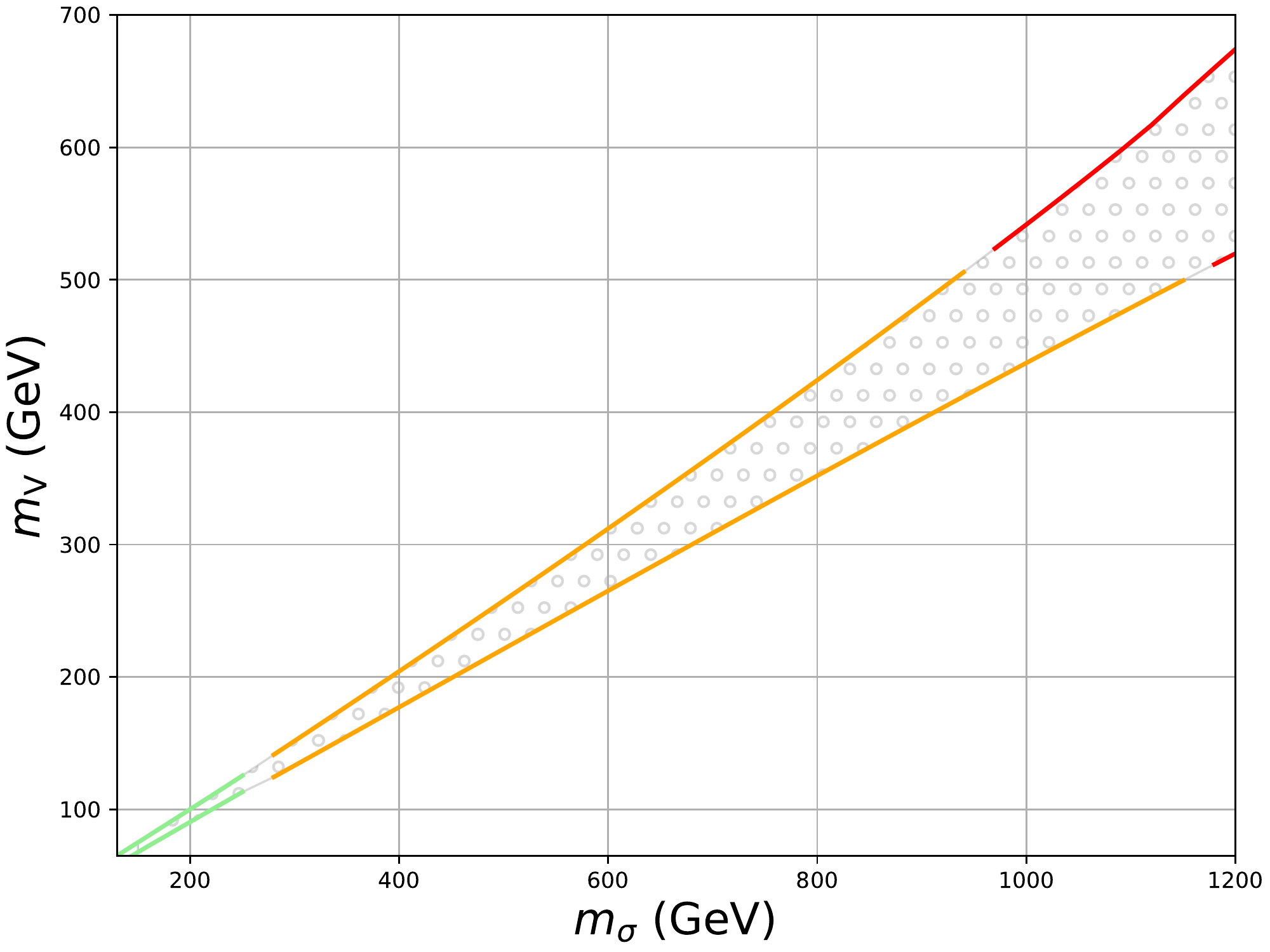}
\includegraphics[width=\columnwidth]{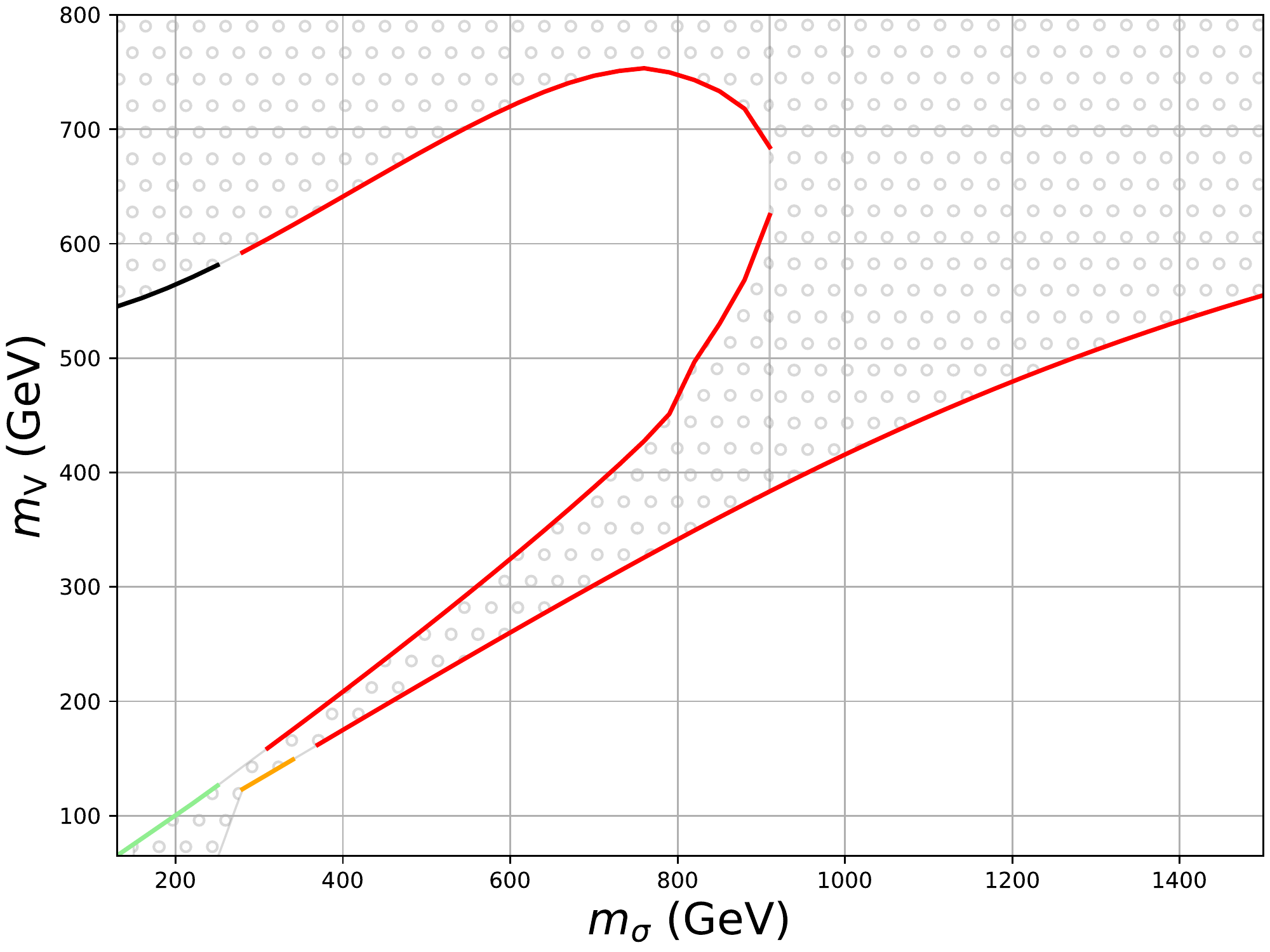}
  \caption{Combined dark matter and Higgs constraints for $\sin \alpha = 0.11$
  (upper panel) and 0.15 (lower panel) for a `gauge invariant' treatment of the
  Higgs-dilaton mixing. The description is similar as in
  figure~\ref{FIG:sin011}, except that there are no LHC or HL-LHC constraints
  from monojets or multijet searches, the cross sections being orders of
  magnitude too small.
  \label{FIG:GIDM}}
\end{figure}

On the other hand, as we discussed previously, if we make the mass mixing term gauge invariant, the decays to the SM Higgs boson then dominate, and exclude the dilaton over masses from about $300$ GeV up to the limit reached by the LHC searches (currently 3 TeV).
We therefore present the same results but for the `gauge invariant mixing' case
in figure~\ref{FIG:GIDM}. There is no dark matter parameter space available for $m_\sigma > 300$ GeV, and the LHC/HL-LHC searches are completely wiped out as the production of dark matter greatly diminishes. The `gauge invariant mixing' scenario is therefore entirely unappealing phenomenologically, and invites other model-building solutions.

\section{Future collider constraints}
\label{SEC:FCC}

We showed in the previous sections that searches for dark matter at the LHC and
at its future high-luminosity operation are not sensitive to our scenario. In
this section, we investigate instead monojet and multijet+MET collider probes
at a future 100 TeV collider. We make use of the simulation chain introduced in
section~\ref{SEC:LHC}, studying the hard-scattering process of
eq.~\eqref{monojet}. At the generator level, we impose that the transverse
momentum of the hardest jet satisfies $p_{T} > 1500$~GeV and that its
pseudo-rapidity fulfills $|\eta| < 5$. In addition, we consider as the main
backgrounds to our analysis the production of an invisible $Z$-boson with jets.

\begin{figure}
  \centering
  \includegraphics[width=\columnwidth]{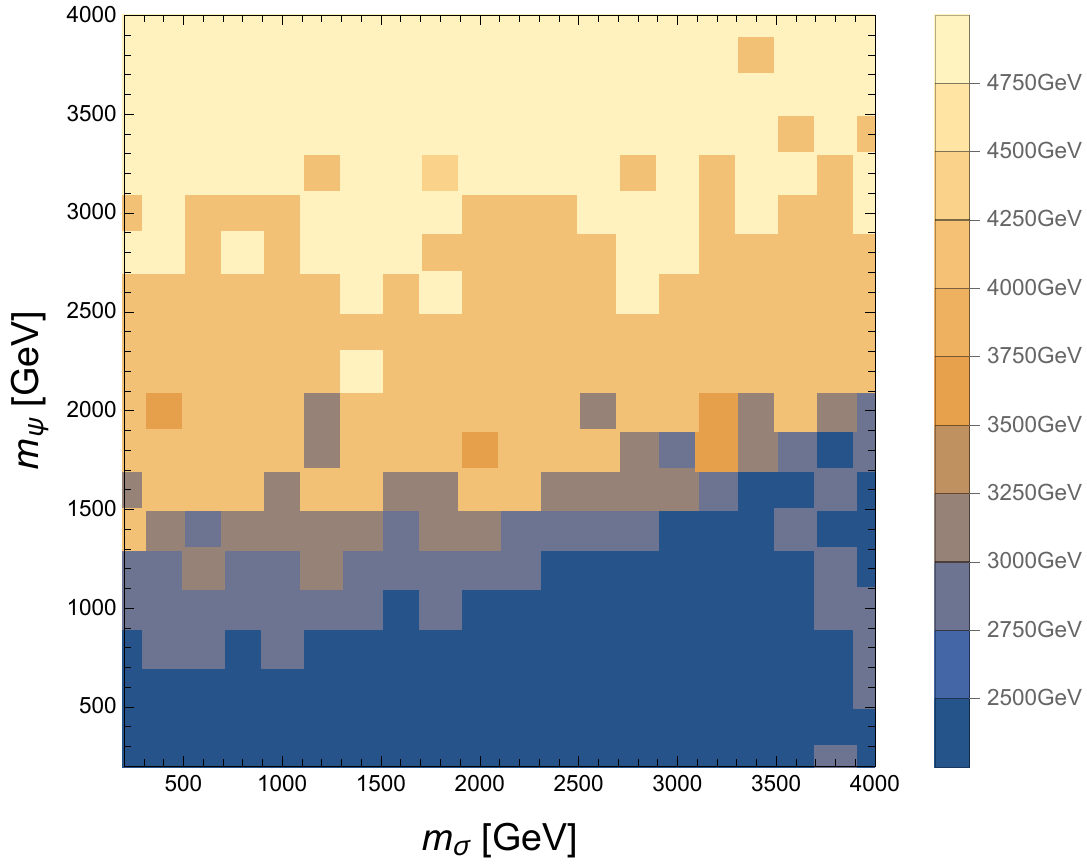}
  \includegraphics[width=\columnwidth]{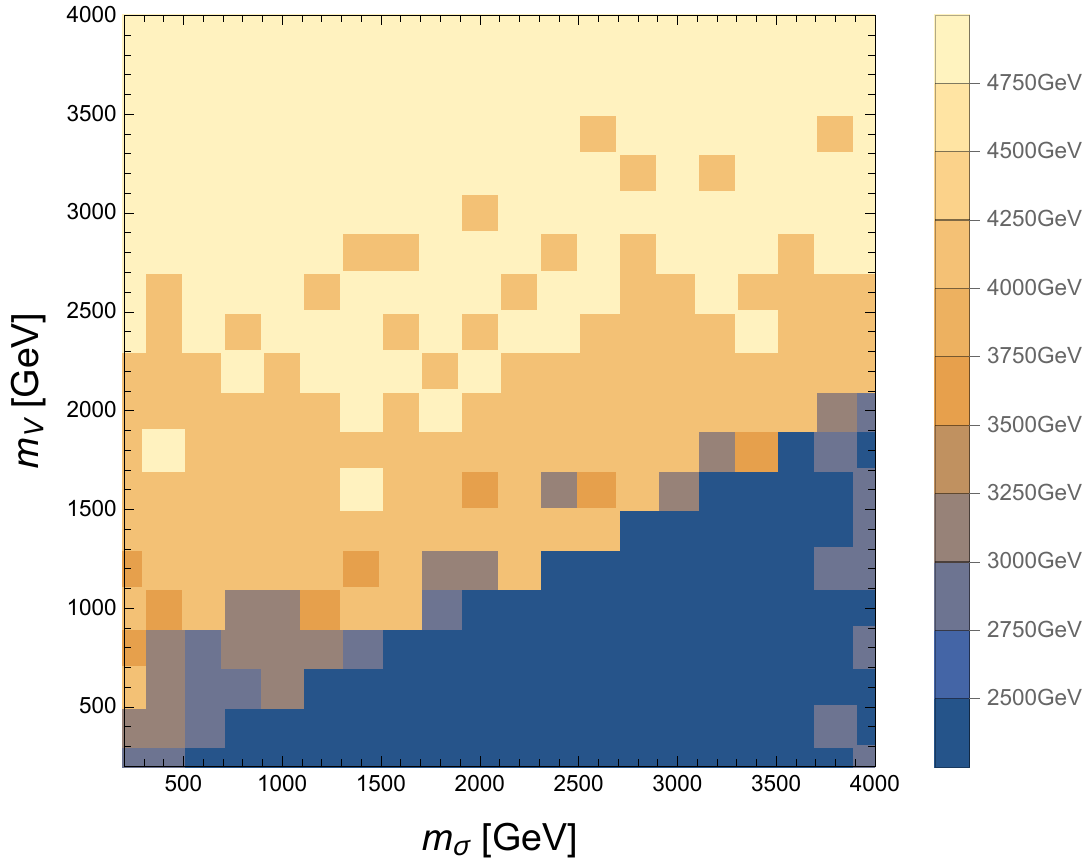}
  \caption{Missing energy selection thresholds as a function of the dilaton and
    fermionic (upper panel) and vector (lower panel) dark matter masses to
    achieve the best sensitivity at a future 100~TeV proton-proton collider. In
    order to avoid any potential unitarity issues at high energies, we impose
    that the threshold value is smaller than 5~TeV.}
  \label{FIG:METcut}
\end{figure}

In our analysis, we first veto the presence of charged leptons with a
transverse momentum $p_{T} > 20$~GeV and a pseudo-rapidity $|\eta| < 2.5$ and
2.1 for electrons and muons respectively, and then reject events featuring at
least one hadronic tau with $p_{T} > 40$~GeV and $|\eta|<2.3$. We next
require that the leading jet is central and very hard, with $p_{T}(j_{1}) >
1500$~GeV and $|\eta(j_{1})|< 2.4$, and allow for some extra hadronic activity
in the selected events. This hadronic activity is associated with the ensemble
of non-leading jets whose $p_{T} > 30$ GeV and $|\eta| < 4.5$.
The leading and all the extra jets satisfying the above requirements are further
imposed to be well separated in azimuth from the missing momentum,
\be
  \Delta\varphi(\slashed{\mathbf{p}}_{T}, j_{i}) < 0.4\,,
\ee
and additionally, the second jet is prevented from being back-to-back with the
leading jet,
\be
  \Delta\varphi(j_{1}, j_{2}) < 2.5\,.
\ee

After this preselection, our analysis relies on various signal regions to
estimate the
sensitivity of a future 100~TeV proton-proton collider to the dark matter
monojet/multijet+MET signal predicted in our model. Each signal region is
defined by a different missing transverse momentum selection,
\be
  \slashed{E}_{T} > \slashed{E}_{T}^{\rm thr.}
  \qquad\text{with}\quad \slashed{E}_{T}^{\rm thr.}\in [2-5]~{\rm TeV}\,,
\ee
so that any considered dark matter and dilaton mass configurations could be
optimally covered. In the above setup, we restrict all missing transverse energy
thresholds to be smaller than 5~TeV, which guarantees to avoid any potential
unitarity issues. The best MET threshold value for a given mass spectrum depends
on both masses, as depicted in the upper and lower panels of
figure~\ref{FIG:METcut} for the fermionic and vector dark matter cases
respectively. 

For dark matter masses much larger than half the dilaton mass (\ie\ far from any
resonant configuration), the optimal selection enforces the missing energy to be
larger than 1--3 times the dark matter mass for the two classes of models, the
MET spectrum being in general flat enough to guarantee a large signal selection
efficiency and a good rejection of the $Z$+jets background. In contrast, when
dark matter production is enhanced by the existence of a dilaton resonance
(\ie\ for $m_\Psi<m_\sigma/2$ and $m_V<m_\sigma/2$ in the fermion and vector
dark matter cases respectively), the best selection threshold is directly fixed
by the dilaton mass.

\begin{figure}
    \centering
    \includegraphics[width=\columnwidth]{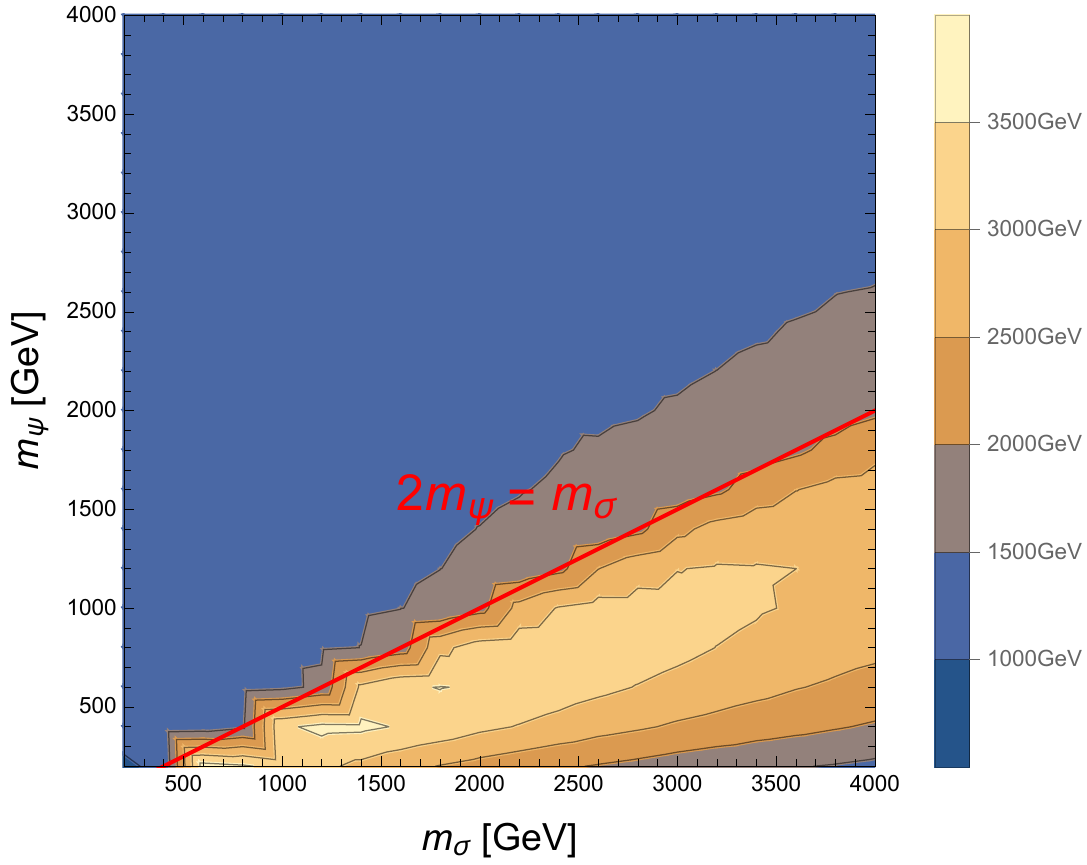}
    \includegraphics[width=\columnwidth]{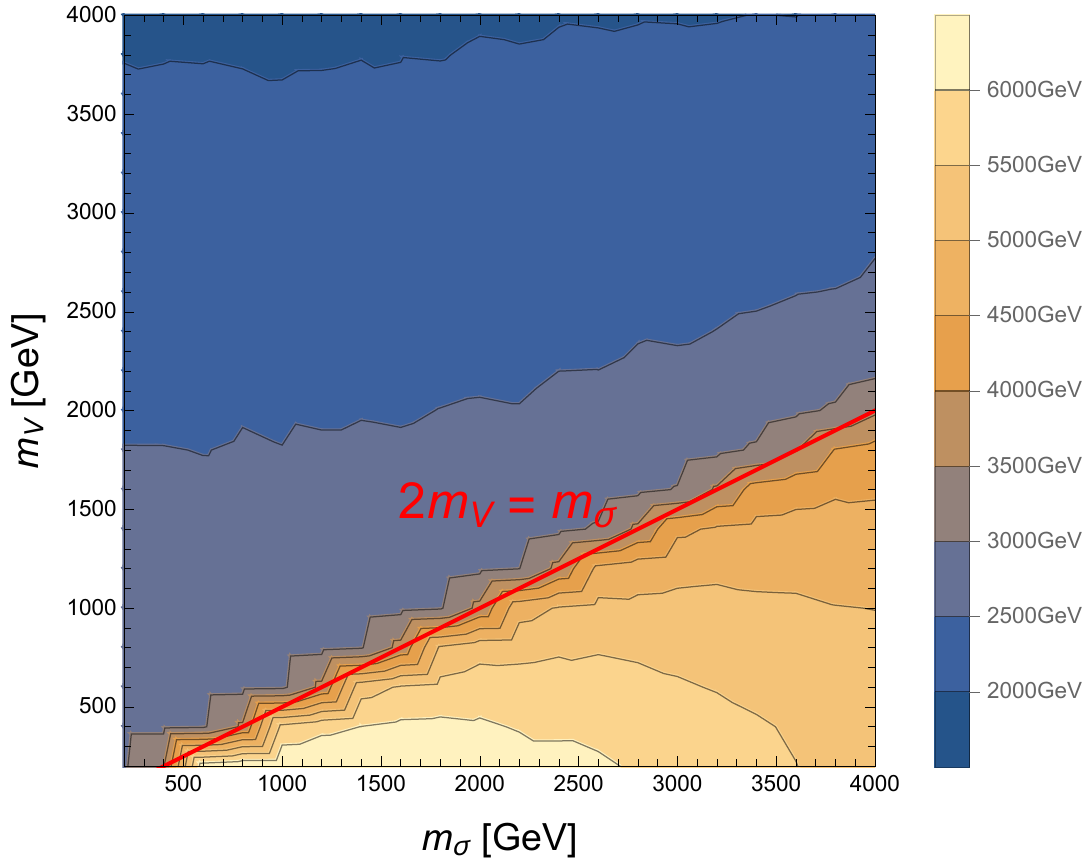}
    \caption{Sensitivity of a 100 TeV future proton-proton collider to the
      dilaton portal dark matter model considered in this work, for the case of
      Majorana (upper panel) and vector (lower panel) dark matter. The results are presented
      in the $(m_\sigma, m_{\Psi, V})$ plane and we indicate, through the colour
      coding, the expected reach on the theory cutoff scale $f$. Our findings
      correspond to integrated luminosity of 3~ab$^{-1}$.}
    \label{FIG:100TeV_result}
\end{figure}

We observe in figure~\ref{FIG:100TeV_result} that in contrast with the LHC,
3~ab$^{-1}$ of proton-proton collisions at 100~TeV are sufficient to probe
cutoff scales lying in the multi-TeV regime for both the fermion (upper panel)
and vector (lower panel) dark matter cases. The results are presented in the
$(m_\sigma, m_{\Psi, V})$ mass plane respectively, and the values of the $f$
scale that are reachable for each mass spectrum are indicated by a colour code.
For each configuration for which the dark matter can be produced from the decay
of a resonantly produced dilaton (below the red line), scales larger than 4~TeV
can be probed, providing hence complementary constraints to models allowed by
cosmological considerations.

In contrast, for configurations in which the dilaton cannot decay into a pair of
dark matter particles, the signal cross sections are smaller. This results in a
loss of sensitivity, in particular in regions favoured by cosmology.

\section{Conclusions}\label{sec:conclusion}

We have presented a comprehensive and up-to-date set of current and future constraints on the most interesting (heavy) dilaton-portal dark matter models.
While heavy scalar and unitarity constraints push the model to large masses and weak couplings, to the extent that the direct dark matter production at the LHC can only probe a light dilaton, and not reach any viable parameter space above $300$ GeV, a future collider would potentially be sensitive with the same searches. 
It would be interesting to examine future projections for heavy scalar and dark matter searches, to see whether these will be complementary.
On the other hand, the (HL) LHC reach could be enhanced if the experimental collaborations extend their published diboson limits above $3$ TeV. 
 
We discussed the fact that vector dark matter is more promising for collider searches thanks to the dilaton's much larger branching ratio into vectors compared to fermions or scalars. Allowing mixing of the dilaton with the Higgs boson then apparently leads to a way to weaken or evade heavy scalar searches (and increase the coupling of the dark matter to the visible sector via the Higgs portal) via the opening of a `magic window'. We showed that this could allow the dark matter and dilaton \emph{below} $1$ TeV in the `minimal mixing treatment'.

However, the `minimal mixing' of the dilaton with the Higgs boson is not gauge invariant. If we include operators to restore gauge invariance to the model then the `magic window' is erased, and di-Higgs bounds dominate all constraints that could be imposed on the dilaton properties. It would nevertheless be interesting to explore the high-energy origin of these Higgs-dilaton mixing terms, in the context of concrete models. Similarly, we could have formulated the theory in a gauge-invariant way by including the dilaton \emph{before electroweak symmetry breaking}, as performed in ref.~\cite{Jung:2014zga}: it would be interesting to revisit our constraints and searches in this alternative (and inequivalent) formulation of the theory.

\section*{Acknowledgments}

MDG acknowledges support from the grant
\mbox{``HiggsAutomator''} of the Agence Nationale de la Recherche
(ANR) (ANR-15-CE31-0002). He thanks his children for many stimulating conversations during the preparation of this draft. DWK is supported by a KIAS Individual Grant (Grant No. PG076201) at Korea Institute for Advanced Study. P.K. is supported in part by KIAS Individual Grant (Grant No. PG021403) at Korea Institute for Advanced Study and by National Research Foundation of Korea (NRF) Grant No. NRF-2019R1A2C3005009, funded by the Korea government (MSIT). S.J.L. acknowledge support by the Samsung Science and Technology Foundation under Project Number SSTF-BA1601-07.

\appendix

\section{Perturbative unitarity constraints}
\label{APP:UNITARITY}

Since our theory contains non-renormalisable operators, it must have a cutoff comparable to the scale $f$. We expect this to be manifest, even at tree level, in perturbative unitarity constraints on two-body scattering amplitudes.
In particular, there are some couplings that could in principle be large
compared to $f$ since they are enhanced by additional massive factors involving
the masses of the theory vector bosons (especially in the dark matter case). In
this appendix, we describe the calculation of the unitarity constraints
involving massive vectors. For simplicity, we restrict to the case of no mixing between the dilaton and the Higgs, and neglect the Higgs boson entirely. For references on the calculation of unitarity constraints, and in particular including vector bosons, see refs.~\cite{IZ,Dusedau:1984xk,Schuessler:2007av,DiLuzio:2016sur,Goodsell:2018tti}.

\subsection{Scattering to dark matter}
\label{APP:DMunit}

We first consider the scattering of vector bosons amongst themselves and into
pairs of dilatons, as well a dilaton self-scattering. The relevant terms in the
Lagrangian read
\be
\lagr_{\sigma V} = \frac{m_V^2}{f} \sigma V_\mu V^\mu + \frac{m_V^2}{2f^2} \sigma^2 V_\mu V^\mu - \xi \frac{m_\sigma^2}{f} \sigma^3 - \frac{\zeta}{24} \frac{m_{\sigma}^2}{f^2} \sigma^4,
\ee
where $\xi$ and $\zeta$ are model-dependent and usually taken to be 5/6 and 11
respectively.

Since the model has a $\mathbb{Z}_2$ symmetry, states with odd numbers of
vectors can only scatter to states with odd numbers of vectors, so that
we only need to consider the $\mathcal{M}_{\sigma \sigma \to \sigma \sigma}$,
$\mathcal{M}_{\sigma \sigma \to V V}$, $\mathcal{M}_{V V \to\sigma \sigma}$
and $\mathcal{M}_{V V \to VV }$ scattering amplitudes. These are given by
\be\bsp
  i\mathcal{M}_{\sigma\sigma\to\sigma\sigma} =&
     - 36 i \xi^2\frac{m_\sigma^4}{f^2}  \bigg[ \frac{1}{s\!-\!m_\sigma^2} +
        \frac{1}{t\!-\!m_\sigma^2} + \frac{1}{u\!-\!m_\sigma^2} \bigg]
     \\&\qquad
      - \zeta i \frac{m_\sigma^2}{f^2}
     \,, \\
  i\mathcal{M}_{VV\to\sigma\sigma} =&\
    i \frac{2m_V^2}{f^2} \epsilon_1 \!\cdot\!\epsilon_2  +
      6 i\xi \epsilon_1\!\cdot\!\epsilon_2
      \frac{m_V^2 m_\sigma^2}{f^2 (s\!-\!m_\sigma^2)}
   \\&\
    + i \left(\frac{2m_V^2}{f} \right)^2\epsilon_1^\mu \epsilon_2^\nu \bigg[ \frac{1}{t-m_V^2} \bigg( \eta^{\mu\nu} + \frac{k_1^\mu k_2^\nu}{m_V^2} \bigg)\\
  &\quad + \frac{1}{u-m_V^2} \bigg( \eta^{\mu\nu} + \frac{k_2^\mu k_1^\nu}{m_V^2} \bigg)\bigg]\,, \\
  i \mathcal{M}_{V V \rightarrow VV } =&\
  -i \frac{4m_V^4}{f^2} \epsilon_1^\mu \epsilon_2^\nu \tilde{\epsilon}_1^\rho  \tilde{\epsilon}_2^\kappa \bigg[ \frac{\eta_{\mu \nu} \eta_{\rho \kappa}}{s - m_V^2} + \frac{\eta_{\mu \rho} \eta_{\nu\kappa}}{t - m_V^2}
   \\&\qquad  + \frac{\eta_{\mu \kappa} \eta_{\nu \rho}}{u - m_V^2} \bigg]\,,
\esp\ee
where $\epsilon_{1,2}$ denote the incoming polarisation vectors and
$\tilde{\epsilon}_{1,2}$ the outgoing ones. Note that in all of the amplitudes $s/t/u$ channel poles are not possible, and so we can search over all scattering momenta up to some potential cutoff without needing to excise singular regions or submatrices as required in the general case \cite{Schuessler:2007av,Goodsell:2018tti}.

Since there are three initial polarisations possible for each vector, the scattering matrix (including the dilaton-dilaton state) is in principle of rank $10$.
However, it breaks into irreducible blocks under the Lorentz algebra, and in particular we can separate off the symmetric and antisymmetric states of $\epsilon_i^\mu \epsilon_j^\nu $ for $i \ne j$.

Typically, only the high-energy limit is considered, where we retain only longitudinal gauge bosons. We find, however, that even the transverse components can
contribute in this limit. In fact, in the low-energy regime relevant for dark
matter scattering, it is the contribution of the transverse components that
dominates. So, we can take a suitable basis of polarisation vectors, such as $(0,1,0,0)$, $(0,0,1,0)$, $(p_V/m_V,0,0,E_V/m_V)$ for a vector aligned along the third spatial component and insert them in the above amplitudes. We then extract the zeroth moment of the scattering matrix
\begin{align}
T^0_{ij} =& = \frac{1}{64\pi} \sqrt{\frac{4 |{\mathbf p}_{\rm in}|\ |{\mathbf p}_{\rm out}|}{s}} \int_{-1}^1 d (\cos \theta) \mathcal{M}_{ij}
\end{align}
where we have included appropriate symmetry factors for our incoming and
outgoing states to be identical pairs, and ${\mathbf p}_{\rm in}$ and
${\mathbf p}_{\rm out}$ are the three-momenta in the centre-of-mass frame for
the incoming and outgoing states respectively.

To find the limit from unitarity, and since the incoming and outgoing states may be related by a unitary rotation which we are not interested in, we take the square root of the eigenvalues of the
$T^0$ matrix and compare the maximum of these to $1/2$,
\begin{align}
|a_0^{\rm max}| \equiv \sqrt{\mathrm{max} \bigg[{\rm eigenvalues} \bigg(T^0 (T^0)^\dagger \bigg)\bigg] } \ < \frac{1}{2}.
\end{align}
For convenience we shall define
\begin{align}
T^0_{ij} \equiv& \frac{1}{16\pi f^2} \tilde{T}^0_{ij}.
\end{align}
The full expressions are too cumbersome to list here, but can be made available in {\tt Mathematica} format or in a {\tt c} program; we shall here identify different regions of parameter space of interest and give approximate formulae for those cases.

\subsection{Scattering at high energy}

In the limit of large $s$, we find the following for the scattering of longitudinal gauge bosons and the dilaton,
 \be\bsp
     &\tilde{T}^0_{(V_L V_L)(V_L V_L)} \rightarrow  -\frac{3}{2}m_\sigma^2, \\
   &\tilde{T}^0_{(V_LV_L) (\sigma \sigma)}  \rightarrow\frac{1}{4} \bigg[4 m_V^2 +   (6\xi -4 ) m_\sigma^2 - 8 m_V^2 \log \frac{s}{m_V^2} \bigg] ,\\
 &\tilde{T}^0_{(\sigma \sigma) (\sigma \sigma)}  \rightarrow - \frac{11}{2}  m_\sigma^2 .
\label{eq:Ttilde} \esp\ee
At this level, scattering involving the transverse modes is also relevant, and
the scattering matrix can be rotated into
 \begin{align}
  \hspace*{-.3cm}
  \tilde{T^0} \!=\!&  \left( \begin{array}{cccc} 0 & 0 & 0 & 0 \\
                                     0 & 0 & -\sqrt{2} m_V^2 & \sqrt{2} m_V^2 \\
                                     0 & -\sqrt{2} m_V^2 & \tilde{T}^0_{(V_LV_L)(V_LV_L)} & \tilde{T}^0_{(V_LV_L)( \sigma \sigma)} \\
                                     0 & \sqrt{2} m_V^2 & \tilde{T}^0_{(\sigma \sigma)( V_L V_L)} & \tilde{T}^0_{(\sigma \sigma)( \sigma \sigma)} \end{array} \right).
 \end{align}
In the limit that $m_V^2  $ is small and the scattering energy is large, the amplitude is dominated by dilaton-dilaton scattering. We find
 \begin{align}
   a_0^{\rm max} =& \frac{m_\sigma^2}{64\pi f^2} \sqrt{261+28\sqrt{65}} .
 \end{align}
This is essentially the constraint originating from dilaton self-scattering,
\begin{align}
    \frac{11m_\sigma^2}{16\pi f^2} \lesssim& 1 \rightarrow m_\sigma \lesssim 2 f,
\label{EQ:UnitarityMaxMsig} \end{align}
that is considerably stronger than the one from pure longitudinal vector
scattering.

\begin{table}
  \renewcommand{\arraystretch}{1.4}
  \setlength\tabcolsep{8pt}
  \begin{tabular}{c|c|c}
    $f$ (TeV) &  $p_{V,\mathrm{max}}$ (TeV) & Approx.~max$(m_V)$ (TeV) \\
    \hline  
     1 & 5 & 1.9 \\
     1 & 20 & 1.4 \\\hline
     2 & 10 & 3.8 \\
     2 & 20 & 3.1 \\
  \end{tabular}
  \caption{Bounds on the vector dark matter mass $m_V$ for given illustrative
   cutoff values $f$ and typical maximum centre-of-mass vector momentum 
   $p_{V,\mathrm{max}}$. \label{tab:unt}}
\end{table}

We \emph{can} use the above result in the limit that \mbox{$s\gg m_V^2$}, so
that the logarithmic term appearing in the second relation of
eq.~\eqref{eq:Ttilde} is large. For reasonable values, we find that the obtained
constraint on $m_\sigma$ is comparable or slightly stronger than the above. For
the limit $m_V \gg m_\sigma$,
\be
\frac{1}{8\pi f^2} m_V^2 \log \frac{s}{m_V^2} < 1/2\,,
\ee
which bounds $m_V$ for a given cutoff. We can directly compare this with the
values shown in figure \ref{FIG:lowf} and that we report in table~\ref{tab:unt}.
These approximate values agree with the maximum values of $m_V$ found
numerically, although the high-energy limit is generally found not to be a
very good approximation for the full scattering matrix.

Finally, while we have computed separate constraints from the
symmetric/antisymmetric scattering of the transverse and longitudinal vector
modes, in the large momentum limit these all reduce to
 \begin{align}
   \sqrt{2} \frac{m_V^2}{16\pi f^2} <& 1/2.
 \end{align}

\subsection{Scattering at low energy}

When $m_V \gg m_\sigma$ the amplitude is largest at low momenta, and dominated by the scattering of \emph{transverse} vectors of the type that obey $\epsilon \cdot k_{3,4} =0$. 
We find that the other modes reduce the scattering amplitude and mix little with the other states. The largest eigenvalue is well approximated by the $(11) \leftrightarrow (11)$ scattering, and largest when $p_V = m_\sigma$,
\begin{align}
  a_0 =& \frac{m_V^4}{16\pi f^2 E_V} \bigg[ \frac{1}{p_V} \log \left( 1 + \frac{4p_V^2}{m_\sigma^2} \right) - \frac{2p_V}{4 m_V^2 + 4 p_V^2 - m_\sigma^2} \bigg] \nn\\
  \simeq & \frac{m_V^3}{16 \pi m_\sigma f^2} \log 5.
\end{align}
This places a constraint on the minimum dilaton mass,
\begin{align}
m_\sigma \gtrsim & \frac{m_V^3}{8 \pi f^2} \log 5.
\end{align}
This can correspondingly be considered the point at which Sommerfeld enhancement of the amplitudes is significant, as clearly seen in figure~\ref{FIG:VectorDM}.

\subsection{Scattering to gluons}
\label{APP:gluonunitarity}

In this work, we are interested in vector dark matter partly because it can be
more copiously produced than fermions or scalars via the dilaton portal. It
should not be a surprise then that a unitarity limit on our theory arises from
the scattering of gluons into vector bosons via the dilaton. The corresponding
scattering amplitude is given by
\begin{align}
i\mathcal{M}_{gg\rightarrow VV} =& \left(i \frac{\alpha_s b_{3}}{4\pi f} \epsilon_1^g \cdot \epsilon_2^g s\right) \frac{i}{s-m_\sigma^2} \left( i \frac{2m_V^2}{f} \tilde{\epsilon}_1 \cdot \tilde{\epsilon}_2\right),
\end{align}
where $\epsilon_{1,2}^g$ are the gluon polarisation vectors and $b_3$ denotes
the QCD beta function.
For the longitudinal modes,
  \be \tilde{\epsilon}_1 \cdot \tilde{\epsilon}_2 = \frac{1}{m_V^2} (p^2 + E^2) = \frac{s}{2 m_V^2}\,,
\ee
so that in the $s \gg 4 m_V^2$ limit, we have
\begin{align}
i\mathcal{M}_{gg\rightarrow VV} \rightarrow& -i \frac{\alpha_s b_{3}}{4\pi f^2} \frac{s^2}{s-m_S^2}. 
\end{align}
Now $|T^0_{(gg),(V_LV_L)}| = \frac{|M|}{32\pi}$,
but there are 2 incoming spins, and 8 incoming pairs of colours that give non-zero results. So the scattering matrix looks like
\begin{align}
T^0 =\left( \begin{array}{ccc} 0 & T^0_{(gg),(V_LV_L)} &\cdots \\ T^0_{(gg),(V_LV_L)} & 0 & 0 \\ \vdots & 0 & 0 \end{array}\right)\,,
\end{align}
so
\begin{align}
T^0 (T^0)^\dagger =& \left( \begin{array}{ccc} 16 |T^0_{(gg),(V_LV_L)}|^2 & 0 &\cdots \\ 0 & 0 & 0 \\ \vdots & 0 & 0 \end{array}\right)\,.
\end{align}
Therefore, we finally obtain
\begin{align}
  a_0^{\rm max} =&      \sqrt{16} \times  \frac{\alpha_s b_{3}}{128\pi^2 f^2}s .
\end{align}
However, this is only for scattering into \emph{one} vector boson species.
After accounting for the contributions of the $Z$ and $W$ bosons (that act as
the equivalent of three individual vectors), we obtain an additional factor of two in the
limit,
\begin{align}
\sqrt{4} \times  \sqrt{16} \times  \frac{\alpha_s b_{3}}{128\pi^2 f^2} s < 1/2 \rightarrow s < \frac{8\pi^2 f^2}{7\alpha_s}.
\end{align}
Similar constraints on the maximal scattering energy via gluon fusion were found long ago in ref.~\cite{Dusedau:1984xk}.

\bibliographystyle{apsrev4-1}
\bibliography{literature}

\begin{thebibliography}{61}%
\makeatletter
\providecommand \@ifxundefined [1]{%
 \@ifx{#1\undefined}
}%
\providecommand \@ifnum [1]{%
 \ifnum #1\expandafter \@firstoftwo
 \else \expandafter \@secondoftwo
 \fi
}%
\providecommand \@ifx [1]{%
 \ifx #1\expandafter \@firstoftwo
 \else \expandafter \@secondoftwo
 \fi
}%
\providecommand \natexlab [1]{#1}%
\providecommand \enquote  [1]{``#1''}%
\providecommand \bibnamefont  [1]{#1}%
\providecommand \bibfnamefont [1]{#1}%
\providecommand \citenamefont [1]{#1}%
\providecommand \href@noop [0]{\@secondoftwo}%
\providecommand \href [0]{\begingroup \@sanitize@url \@href}%
\providecommand \@href[1]{\@@startlink{#1}\@@href}%
\providecommand \@@href[1]{\endgroup#1\@@endlink}%
\providecommand \@sanitize@url [0]{\catcode `\\12\catcode `\$12\catcode
  `\&12\catcode `\#12\catcode `\^12\catcode `\_12\catcode `\%12\relax}%
\providecommand \@@startlink[1]{}%
\providecommand \@@endlink[0]{}%
\providecommand \url  [0]{\begingroup\@sanitize@url \@url }%
\providecommand \@url [1]{\endgroup\@href {#1}{\urlprefix }}%
\providecommand \urlprefix  [0]{URL }%
\providecommand \Eprint [0]{\href }%
\providecommand \doibase [0]{http://dx.doi.org/}%
\providecommand \selectlanguage [0]{\@gobble}%
\providecommand \bibinfo  [0]{\@secondoftwo}%
\providecommand \bibfield  [0]{\@secondoftwo}%
\providecommand \translation [1]{[#1]}%
\providecommand \BibitemOpen [0]{}%
\providecommand \bibitemStop [0]{}%
\providecommand \bibitemNoStop [0]{.\EOS\space}%
\providecommand \EOS [0]{\spacefactor3000\relax}%
\providecommand \BibitemShut  [1]{\csname bibitem#1\endcsname}%
\let\auto@bib@innerbib\@empty
\bibitem [{\citenamefont {Bardeen}\ \emph {et~al.}(1986)\citenamefont
  {Bardeen}, \citenamefont {Leung},\ and\ \citenamefont
  {Love}}]{Bardeen:1985sm}%
  \BibitemOpen
  \bibfield  {author} {\bibinfo {author} {\bibfnamefont {W.~A.}\ \bibnamefont
  {Bardeen}}, \bibinfo {author} {\bibfnamefont {C.~N.}\ \bibnamefont {Leung}},
  \ and\ \bibinfo {author} {\bibfnamefont {S.~T.}\ \bibnamefont {Love}},\
  }\href {\doibase 10.1103/PhysRevLett.56.1230} {\bibfield  {journal} {\bibinfo
   {journal} {Phys. Rev. Lett.}\ }\textbf {\bibinfo {volume} {56}},\ \bibinfo
  {pages} {1230} (\bibinfo {year} {1986})}\BibitemShut {NoStop}%
\bibitem [{\citenamefont {Buchmuller}\ and\ \citenamefont
  {Dragon}(1987)}]{Buchmuller:1987uc}%
  \BibitemOpen
  \bibfield  {author} {\bibinfo {author} {\bibfnamefont {W.}~\bibnamefont
  {Buchmuller}}\ and\ \bibinfo {author} {\bibfnamefont {N.}~\bibnamefont
  {Dragon}},\ }\href {\doibase 10.1016/0370-2693(87)90041-4} {\bibfield
  {journal} {\bibinfo  {journal} {Phys. Lett. B}\ }\textbf {\bibinfo {volume}
  {195}},\ \bibinfo {pages} {417} (\bibinfo {year} {1987})}\BibitemShut
  {NoStop}%
\bibitem [{\citenamefont {Buchmuller}\ and\ \citenamefont
  {Dragon}(1989)}]{Buchmuller:1988cj}%
  \BibitemOpen
  \bibfield  {author} {\bibinfo {author} {\bibfnamefont {W.}~\bibnamefont
  {Buchmuller}}\ and\ \bibinfo {author} {\bibfnamefont {N.}~\bibnamefont
  {Dragon}},\ }\href {\doibase 10.1016/0550-3213(89)90249-6} {\bibfield
  {journal} {\bibinfo  {journal} {Nucl. Phys. B}\ }\textbf {\bibinfo {volume}
  {321}},\ \bibinfo {pages} {207} (\bibinfo {year} {1989})}\BibitemShut
  {NoStop}%
\bibitem [{\citenamefont {Rattazzi}\ and\ \citenamefont
  {Zaffaroni}(2001)}]{Rattazzi:2000hs}%
  \BibitemOpen
  \bibfield  {author} {\bibinfo {author} {\bibfnamefont {R.}~\bibnamefont
  {Rattazzi}}\ and\ \bibinfo {author} {\bibfnamefont {A.}~\bibnamefont
  {Zaffaroni}},\ }\href {\doibase 10.1088/1126-6708/2001/04/021} {\bibfield
  {journal} {\bibinfo  {journal} {JHEP}\ }\textbf {\bibinfo {volume} {04}},\
  \bibinfo {pages} {021} (\bibinfo {year} {2001})},\ \Eprint
  {http://arxiv.org/abs/hep-th/0012248} {arXiv:hep-th/0012248 [hep-th]}
  \BibitemShut {NoStop}%
\bibitem [{\citenamefont {Csaki}\ \emph {et~al.}(2001)\citenamefont {Csaki},
  \citenamefont {Graesser},\ and\ \citenamefont {Kribs}}]{Csaki:2000zn}%
  \BibitemOpen
  \bibfield  {author} {\bibinfo {author} {\bibfnamefont {C.}~\bibnamefont
  {Csaki}}, \bibinfo {author} {\bibfnamefont {M.~L.}\ \bibnamefont {Graesser}},
  \ and\ \bibinfo {author} {\bibfnamefont {G.~D.}\ \bibnamefont {Kribs}},\
  }\href {\doibase 10.1103/PhysRevD.63.065002} {\bibfield  {journal} {\bibinfo
  {journal} {Phys. Rev. D}\ }\textbf {\bibinfo {volume} {63}},\ \bibinfo
  {pages} {065002} (\bibinfo {year} {2001})},\ \Eprint
  {http://arxiv.org/abs/hep-th/0008151} {arXiv:hep-th/0008151} \BibitemShut
  {NoStop}%
\bibitem [{\citenamefont {Dominici}\ \emph {et~al.}(2003)\citenamefont
  {Dominici}, \citenamefont {Grzadkowski}, \citenamefont {Gunion},\ and\
  \citenamefont {Toharia}}]{Dominici:2002jv}%
  \BibitemOpen
  \bibfield  {author} {\bibinfo {author} {\bibfnamefont {D.}~\bibnamefont
  {Dominici}}, \bibinfo {author} {\bibfnamefont {B.}~\bibnamefont
  {Grzadkowski}}, \bibinfo {author} {\bibfnamefont {J.~F.}\ \bibnamefont
  {Gunion}}, \ and\ \bibinfo {author} {\bibfnamefont {M.}~\bibnamefont
  {Toharia}},\ }\href {\doibase 10.1016/j.nuclphysb.2003.08.020} {\bibfield
  {journal} {\bibinfo  {journal} {Nucl. Phys. B}\ }\textbf {\bibinfo {volume}
  {671}},\ \bibinfo {pages} {243} (\bibinfo {year} {2003})},\ \Eprint
  {http://arxiv.org/abs/hep-ph/0206192} {arXiv:hep-ph/0206192} \BibitemShut
  {NoStop}%
\bibitem [{\citenamefont {Csaki}\ \emph {et~al.}(2007)\citenamefont {Csaki},
  \citenamefont {Hubisz},\ and\ \citenamefont {Lee}}]{Csaki:2007ns}%
  \BibitemOpen
  \bibfield  {author} {\bibinfo {author} {\bibfnamefont {C.}~\bibnamefont
  {Csaki}}, \bibinfo {author} {\bibfnamefont {J.}~\bibnamefont {Hubisz}}, \
  and\ \bibinfo {author} {\bibfnamefont {S.~J.}\ \bibnamefont {Lee}},\ }\href
  {\doibase 10.1103/PhysRevD.76.125015} {\bibfield  {journal} {\bibinfo
  {journal} {Phys. Rev.}\ }\textbf {\bibinfo {volume} {D76}},\ \bibinfo {pages}
  {125015} (\bibinfo {year} {2007})},\ \Eprint {http://arxiv.org/abs/0705.3844}
  {arXiv:0705.3844 [hep-ph]} \BibitemShut {NoStop}%
\bibitem [{\citenamefont {Goldberger}\ \emph {et~al.}(2008)\citenamefont
  {Goldberger}, \citenamefont {Grinstein},\ and\ \citenamefont
  {Skiba}}]{Goldberger:2008zz}%
  \BibitemOpen
  \bibfield  {author} {\bibinfo {author} {\bibfnamefont {W.~D.}\ \bibnamefont
  {Goldberger}}, \bibinfo {author} {\bibfnamefont {B.}~\bibnamefont
  {Grinstein}}, \ and\ \bibinfo {author} {\bibfnamefont {W.}~\bibnamefont
  {Skiba}},\ }\href {\doibase 10.1103/PhysRevLett.100.111802} {\bibfield
  {journal} {\bibinfo  {journal} {Phys. Rev. Lett.}\ }\textbf {\bibinfo
  {volume} {100}},\ \bibinfo {pages} {111802} (\bibinfo {year} {2008})},\
  \Eprint {http://arxiv.org/abs/0708.1463} {arXiv:0708.1463 [hep-ph]}
  \BibitemShut {NoStop}%
\bibitem [{\citenamefont {Bai}\ \emph {et~al.}(2009)\citenamefont {Bai},
  \citenamefont {Carena},\ and\ \citenamefont {Lykken}}]{Bai:2009ms}%
  \BibitemOpen
  \bibfield  {author} {\bibinfo {author} {\bibfnamefont {Y.}~\bibnamefont
  {Bai}}, \bibinfo {author} {\bibfnamefont {M.}~\bibnamefont {Carena}}, \ and\
  \bibinfo {author} {\bibfnamefont {J.}~\bibnamefont {Lykken}},\ }\href
  {\doibase 10.1103/PhysRevLett.103.261803} {\bibfield  {journal} {\bibinfo
  {journal} {Phys. Rev. Lett.}\ }\textbf {\bibinfo {volume} {103}},\ \bibinfo
  {pages} {261803} (\bibinfo {year} {2009})},\ \Eprint
  {http://arxiv.org/abs/0909.1319} {arXiv:0909.1319 [hep-ph]} \BibitemShut
  {NoStop}%
\bibitem [{\citenamefont {Grzadkowski}\ \emph {et~al.}(2012)\citenamefont
  {Grzadkowski}, \citenamefont {Gunion},\ and\ \citenamefont
  {Toharia}}]{Grzadkowski:2012ng}%
  \BibitemOpen
  \bibfield  {author} {\bibinfo {author} {\bibfnamefont {B.}~\bibnamefont
  {Grzadkowski}}, \bibinfo {author} {\bibfnamefont {J.~F.}\ \bibnamefont
  {Gunion}}, \ and\ \bibinfo {author} {\bibfnamefont {M.}~\bibnamefont
  {Toharia}},\ }\href {\doibase 10.1016/j.physletb.2012.04.037} {\bibfield
  {journal} {\bibinfo  {journal} {Phys. Lett. B}\ }\textbf {\bibinfo {volume}
  {712}},\ \bibinfo {pages} {70} (\bibinfo {year} {2012})},\ \Eprint
  {http://arxiv.org/abs/1202.5017} {arXiv:1202.5017 [hep-ph]} \BibitemShut
  {NoStop}%
\bibitem [{\citenamefont {Chacko}\ and\ \citenamefont
  {Mishra}(2013)}]{Chacko:2012sy}%
  \BibitemOpen
  \bibfield  {author} {\bibinfo {author} {\bibfnamefont {Z.}~\bibnamefont
  {Chacko}}\ and\ \bibinfo {author} {\bibfnamefont {R.~K.}\ \bibnamefont
  {Mishra}},\ }\href {\doibase 10.1103/PhysRevD.87.115006} {\bibfield
  {journal} {\bibinfo  {journal} {Phys. Rev.}\ }\textbf {\bibinfo {volume}
  {D87}},\ \bibinfo {pages} {115006} (\bibinfo {year} {2013})},\ \Eprint
  {http://arxiv.org/abs/1209.3022} {arXiv:1209.3022 [hep-ph]} \BibitemShut
  {NoStop}%
\bibitem [{\citenamefont {Bellazzini}\ \emph {et~al.}(2013)\citenamefont
  {Bellazzini}, \citenamefont {Csaki}, \citenamefont {Hubisz}, \citenamefont
  {Serra},\ and\ \citenamefont {Terning}}]{Bellazzini:2012vz}%
  \BibitemOpen
  \bibfield  {author} {\bibinfo {author} {\bibfnamefont {B.}~\bibnamefont
  {Bellazzini}}, \bibinfo {author} {\bibfnamefont {C.}~\bibnamefont {Csaki}},
  \bibinfo {author} {\bibfnamefont {J.}~\bibnamefont {Hubisz}}, \bibinfo
  {author} {\bibfnamefont {J.}~\bibnamefont {Serra}}, \ and\ \bibinfo {author}
  {\bibfnamefont {J.}~\bibnamefont {Terning}},\ }\href {\doibase
  10.1140/epjc/s10052-013-2333-x} {\bibfield  {journal} {\bibinfo  {journal}
  {Eur. Phys. J.}\ }\textbf {\bibinfo {volume} {C73}},\ \bibinfo {pages} {2333}
  (\bibinfo {year} {2013})},\ \Eprint {http://arxiv.org/abs/1209.3299}
  {arXiv:1209.3299 [hep-ph]} \BibitemShut {NoStop}%
\bibitem [{\citenamefont {Ahmed}\ \emph {et~al.}(2017)\citenamefont {Ahmed},
  \citenamefont {Dillon}, \citenamefont {Grzadkowski}, \citenamefont {Gunion},\
  and\ \citenamefont {Jiang}}]{Ahmed:2015uqt}%
  \BibitemOpen
  \bibfield  {author} {\bibinfo {author} {\bibfnamefont {A.}~\bibnamefont
  {Ahmed}}, \bibinfo {author} {\bibfnamefont {B.~M.}\ \bibnamefont {Dillon}},
  \bibinfo {author} {\bibfnamefont {B.}~\bibnamefont {Grzadkowski}}, \bibinfo
  {author} {\bibfnamefont {J.~F.}\ \bibnamefont {Gunion}}, \ and\ \bibinfo
  {author} {\bibfnamefont {Y.}~\bibnamefont {Jiang}},\ }\href {\doibase
  10.1103/PhysRevD.95.095019} {\bibfield  {journal} {\bibinfo  {journal} {Phys.
  Rev. D}\ }\textbf {\bibinfo {volume} {95}},\ \bibinfo {pages} {095019}
  (\bibinfo {year} {2017})},\ \Eprint {http://arxiv.org/abs/1512.05771}
  {arXiv:1512.05771 [hep-ph]} \BibitemShut {NoStop}%
\bibitem [{\citenamefont {Ahmed}\ \emph {et~al.}(2019)\citenamefont {Ahmed},
  \citenamefont {Mariotti},\ and\ \citenamefont {Najjari}}]{Ahmed:2019csf}%
  \BibitemOpen
  \bibfield  {author} {\bibinfo {author} {\bibfnamefont {A.}~\bibnamefont
  {Ahmed}}, \bibinfo {author} {\bibfnamefont {A.}~\bibnamefont {Mariotti}}, \
  and\ \bibinfo {author} {\bibfnamefont {S.}~\bibnamefont {Najjari}},\
  }\href@noop {} {\  (\bibinfo {year} {2019})},\ \Eprint
  {http://arxiv.org/abs/1912.06645} {arXiv:1912.06645 [hep-ph]} \BibitemShut
  {NoStop}%
\bibitem [{\citenamefont {Jung}\ and\ \citenamefont {Ko}(2014)}]{Jung:2014zga}%
  \BibitemOpen
  \bibfield  {author} {\bibinfo {author} {\bibfnamefont {D.-W.}\ \bibnamefont
  {Jung}}\ and\ \bibinfo {author} {\bibfnamefont {P.}~\bibnamefont {Ko}},\
  }\href {\doibase 10.1016/j.physletb.2014.04.005} {\bibfield  {journal}
  {\bibinfo  {journal} {Phys. Lett. B}\ }\textbf {\bibinfo {volume} {732}},\
  \bibinfo {pages} {364} (\bibinfo {year} {2014})},\ \Eprint
  {http://arxiv.org/abs/1401.5586} {arXiv:1401.5586 [hep-ph]} \BibitemShut
  {NoStop}%
\bibitem [{\citenamefont {Blum}\ \emph {et~al.}(2015)\citenamefont {Blum},
  \citenamefont {Cliche}, \citenamefont {Csaki},\ and\ \citenamefont
  {Lee}}]{Blum:2014jca}%
  \BibitemOpen
  \bibfield  {author} {\bibinfo {author} {\bibfnamefont {K.}~\bibnamefont
  {Blum}}, \bibinfo {author} {\bibfnamefont {M.}~\bibnamefont {Cliche}},
  \bibinfo {author} {\bibfnamefont {C.}~\bibnamefont {Csaki}}, \ and\ \bibinfo
  {author} {\bibfnamefont {S.~J.}\ \bibnamefont {Lee}},\ }\href {\doibase
  10.1007/JHEP03(2015)099} {\bibfield  {journal} {\bibinfo  {journal} {JHEP}\
  }\textbf {\bibinfo {volume} {03}},\ \bibinfo {pages} {099} (\bibinfo {year}
  {2015})},\ \Eprint {http://arxiv.org/abs/1410.1873} {arXiv:1410.1873
  [hep-ph]} \BibitemShut {NoStop}%
\bibitem [{\citenamefont {Efrati}\ \emph {et~al.}(2015)\citenamefont {Efrati},
  \citenamefont {Kuflik}, \citenamefont {Nussinov}, \citenamefont {Soreq},\
  and\ \citenamefont {Volansky}}]{Efrati:2014aea}%
  \BibitemOpen
  \bibfield  {author} {\bibinfo {author} {\bibfnamefont {A.}~\bibnamefont
  {Efrati}}, \bibinfo {author} {\bibfnamefont {E.}~\bibnamefont {Kuflik}},
  \bibinfo {author} {\bibfnamefont {S.}~\bibnamefont {Nussinov}}, \bibinfo
  {author} {\bibfnamefont {Y.}~\bibnamefont {Soreq}}, \ and\ \bibinfo {author}
  {\bibfnamefont {T.}~\bibnamefont {Volansky}},\ }\href {\doibase
  10.1103/PhysRevD.91.055034} {\bibfield  {journal} {\bibinfo  {journal} {Phys.
  Rev.}\ }\textbf {\bibinfo {volume} {D91}},\ \bibinfo {pages} {055034}
  (\bibinfo {year} {2015})},\ \Eprint {http://arxiv.org/abs/1410.2225}
  {arXiv:1410.2225 [hep-ph]} \BibitemShut {NoStop}%
\bibitem [{\citenamefont {Kim}\ \emph {et~al.}(2016)\citenamefont {Kim},
  \citenamefont {Lee},\ and\ \citenamefont {Parolini}}]{Kim:2016jbz}%
  \BibitemOpen
  \bibfield  {author} {\bibinfo {author} {\bibfnamefont {M.}~\bibnamefont
  {Kim}}, \bibinfo {author} {\bibfnamefont {S.~J.}\ \bibnamefont {Lee}}, \ and\
  \bibinfo {author} {\bibfnamefont {A.}~\bibnamefont {Parolini}},\ }\href@noop
  {} {\  (\bibinfo {year} {2016})},\ \Eprint {http://arxiv.org/abs/1602.05590}
  {arXiv:1602.05590 [hep-ph]} \BibitemShut {NoStop}%
\bibitem [{\citenamefont {Cheng}\ and\ \citenamefont
  {Low}(2003)}]{Cheng:2003ju}%
  \BibitemOpen
  \bibfield  {author} {\bibinfo {author} {\bibfnamefont {H.-C.}\ \bibnamefont
  {Cheng}}\ and\ \bibinfo {author} {\bibfnamefont {I.}~\bibnamefont {Low}},\
  }\href {\doibase 10.1088/1126-6708/2003/09/051} {\bibfield  {journal}
  {\bibinfo  {journal} {JHEP}\ }\textbf {\bibinfo {volume} {09}},\ \bibinfo
  {pages} {051} (\bibinfo {year} {2003})},\ \Eprint
  {http://arxiv.org/abs/hep-ph/0308199} {arXiv:hep-ph/0308199 [hep-ph]}
  \BibitemShut {NoStop}%
\bibitem [{\citenamefont {Birkedal}\ \emph {et~al.}(2006)\citenamefont
  {Birkedal}, \citenamefont {Noble}, \citenamefont {Perelstein},\ and\
  \citenamefont {Spray}}]{Birkedal:2006fz}%
  \BibitemOpen
  \bibfield  {author} {\bibinfo {author} {\bibfnamefont {A.}~\bibnamefont
  {Birkedal}}, \bibinfo {author} {\bibfnamefont {A.}~\bibnamefont {Noble}},
  \bibinfo {author} {\bibfnamefont {M.}~\bibnamefont {Perelstein}}, \ and\
  \bibinfo {author} {\bibfnamefont {A.}~\bibnamefont {Spray}},\ }\href
  {\doibase 10.1103/PhysRevD.74.035002} {\bibfield  {journal} {\bibinfo
  {journal} {Phys. Rev.}\ }\textbf {\bibinfo {volume} {D74}},\ \bibinfo {pages}
  {035002} (\bibinfo {year} {2006})},\ \Eprint
  {http://arxiv.org/abs/hep-ph/0603077} {arXiv:hep-ph/0603077 [hep-ph]}
  \BibitemShut {NoStop}%
\bibitem [{\citenamefont {Lebedev}\ \emph {et~al.}(2012)\citenamefont
  {Lebedev}, \citenamefont {Lee},\ and\ \citenamefont
  {Mambrini}}]{Lebedev:2011iq}%
  \BibitemOpen
  \bibfield  {author} {\bibinfo {author} {\bibfnamefont {O.}~\bibnamefont
  {Lebedev}}, \bibinfo {author} {\bibfnamefont {H.~M.}\ \bibnamefont {Lee}}, \
  and\ \bibinfo {author} {\bibfnamefont {Y.}~\bibnamefont {Mambrini}},\ }\href
  {\doibase 10.1016/j.physletb.2012.01.029} {\bibfield  {journal} {\bibinfo
  {journal} {Phys. Lett.}\ }\textbf {\bibinfo {volume} {B707}},\ \bibinfo
  {pages} {570} (\bibinfo {year} {2012})},\ \Eprint
  {http://arxiv.org/abs/1111.4482} {arXiv:1111.4482 [hep-ph]} \BibitemShut
  {NoStop}%
\bibitem [{\citenamefont {Arcadi}\ \emph {et~al.}(2020)\citenamefont {Arcadi},
  \citenamefont {Djouadi},\ and\ \citenamefont {Kado}}]{Arcadi:2020jqf}%
  \BibitemOpen
  \bibfield  {author} {\bibinfo {author} {\bibfnamefont {G.}~\bibnamefont
  {Arcadi}}, \bibinfo {author} {\bibfnamefont {A.}~\bibnamefont {Djouadi}}, \
  and\ \bibinfo {author} {\bibfnamefont {M.}~\bibnamefont {Kado}},\ }\href@noop
  {} {\  (\bibinfo {year} {2020})},\ \Eprint {http://arxiv.org/abs/2001.10750}
  {arXiv:2001.10750 [hep-ph]} \BibitemShut {NoStop}%
\bibitem [{\citenamefont {Hambye}(2009)}]{Hambye:2008bq}%
  \BibitemOpen
  \bibfield  {author} {\bibinfo {author} {\bibfnamefont {T.}~\bibnamefont
  {Hambye}},\ }\href {\doibase 10.1088/1126-6708/2009/01/028} {\bibfield
  {journal} {\bibinfo  {journal} {JHEP}\ }\textbf {\bibinfo {volume} {01}},\
  \bibinfo {pages} {028} (\bibinfo {year} {2009})},\ \Eprint
  {http://arxiv.org/abs/0811.0172} {arXiv:0811.0172 [hep-ph]} \BibitemShut
  {NoStop}%
\bibitem [{\citenamefont {Baek}\ \emph {et~al.}(2013)\citenamefont {Baek},
  \citenamefont {Ko}, \citenamefont {Park},\ and\ \citenamefont
  {Senaha}}]{Baek:2012se}%
  \BibitemOpen
  \bibfield  {author} {\bibinfo {author} {\bibfnamefont {S.}~\bibnamefont
  {Baek}}, \bibinfo {author} {\bibfnamefont {P.}~\bibnamefont {Ko}}, \bibinfo
  {author} {\bibfnamefont {W.-I.}\ \bibnamefont {Park}}, \ and\ \bibinfo
  {author} {\bibfnamefont {E.}~\bibnamefont {Senaha}},\ }\href {\doibase
  10.1007/JHEP05(2013)036} {\bibfield  {journal} {\bibinfo  {journal} {JHEP}\
  }\textbf {\bibinfo {volume} {05}},\ \bibinfo {pages} {036} (\bibinfo {year}
  {2013})},\ \Eprint {http://arxiv.org/abs/1212.2131} {arXiv:1212.2131
  [hep-ph]} \BibitemShut {NoStop}%
\bibitem [{\citenamefont {Baek}\ \emph {et~al.}(2014)\citenamefont {Baek},
  \citenamefont {Ko},\ and\ \citenamefont {Park}}]{Baek:2014jga}%
  \BibitemOpen
  \bibfield  {author} {\bibinfo {author} {\bibfnamefont {S.}~\bibnamefont
  {Baek}}, \bibinfo {author} {\bibfnamefont {P.}~\bibnamefont {Ko}}, \ and\
  \bibinfo {author} {\bibfnamefont {W.-I.}\ \bibnamefont {Park}},\ }\href
  {\doibase 10.1103/PhysRevD.90.055014} {\bibfield  {journal} {\bibinfo
  {journal} {Phys. Rev. D}\ }\textbf {\bibinfo {volume} {90}},\ \bibinfo
  {pages} {055014} (\bibinfo {year} {2014})},\ \Eprint
  {http://arxiv.org/abs/1405.3530} {arXiv:1405.3530 [hep-ph]} \BibitemShut
  {NoStop}%
\bibitem [{\citenamefont {Ko}\ and\ \citenamefont {Yokoya}(2016)}]{Ko:2016xwd}%
  \BibitemOpen
  \bibfield  {author} {\bibinfo {author} {\bibfnamefont {P.}~\bibnamefont
  {Ko}}\ and\ \bibinfo {author} {\bibfnamefont {H.}~\bibnamefont {Yokoya}},\
  }\href {\doibase 10.1007/JHEP08(2016)109} {\bibfield  {journal} {\bibinfo
  {journal} {JHEP}\ }\textbf {\bibinfo {volume} {08}},\ \bibinfo {pages} {109}
  (\bibinfo {year} {2016})},\ \Eprint {http://arxiv.org/abs/1603.04737}
  {arXiv:1603.04737 [hep-ph]} \BibitemShut {NoStop}%
\bibitem [{\citenamefont {Kamon}\ \emph {et~al.}(2017)\citenamefont {Kamon},
  \citenamefont {Ko},\ and\ \citenamefont {Li}}]{Kamon:2017yfx}%
  \BibitemOpen
  \bibfield  {author} {\bibinfo {author} {\bibfnamefont {T.}~\bibnamefont
  {Kamon}}, \bibinfo {author} {\bibfnamefont {P.}~\bibnamefont {Ko}}, \ and\
  \bibinfo {author} {\bibfnamefont {J.}~\bibnamefont {Li}},\ }\href {\doibase
  10.1140/epjc/s10052-017-5240-8} {\bibfield  {journal} {\bibinfo  {journal}
  {Eur. Phys. J. C}\ }\textbf {\bibinfo {volume} {77}},\ \bibinfo {pages} {652}
  (\bibinfo {year} {2017})},\ \Eprint {http://arxiv.org/abs/1705.02149}
  {arXiv:1705.02149 [hep-ph]} \BibitemShut {NoStop}%
\bibitem [{\citenamefont {Dutta}\ \emph {et~al.}(2018)\citenamefont {Dutta},
  \citenamefont {Kamon}, \citenamefont {Ko},\ and\ \citenamefont
  {Li}}]{Dutta:2017sod}%
  \BibitemOpen
  \bibfield  {author} {\bibinfo {author} {\bibfnamefont {B.}~\bibnamefont
  {Dutta}}, \bibinfo {author} {\bibfnamefont {T.}~\bibnamefont {Kamon}},
  \bibinfo {author} {\bibfnamefont {P.}~\bibnamefont {Ko}}, \ and\ \bibinfo
  {author} {\bibfnamefont {J.}~\bibnamefont {Li}},\ }\href {\doibase
  10.1140/epjc/s10052-018-6071-y} {\bibfield  {journal} {\bibinfo  {journal}
  {Eur. Phys. J. C}\ }\textbf {\bibinfo {volume} {78}},\ \bibinfo {pages} {595}
  (\bibinfo {year} {2018})},\ \Eprint {http://arxiv.org/abs/1712.05123}
  {arXiv:1712.05123 [hep-ph]} \BibitemShut {NoStop}%
\bibitem [{\citenamefont {Peskin}\ and\ \citenamefont
  {Takeuchi}(1992)}]{Peskin:1991sw}%
  \BibitemOpen
  \bibfield  {author} {\bibinfo {author} {\bibfnamefont {M.~E.}\ \bibnamefont
  {Peskin}}\ and\ \bibinfo {author} {\bibfnamefont {T.}~\bibnamefont
  {Takeuchi}},\ }\href {\doibase 10.1103/PhysRevD.46.381} {\bibfield  {journal}
  {\bibinfo  {journal} {Phys. Rev. D}\ }\textbf {\bibinfo {volume} {46}},\
  \bibinfo {pages} {381} (\bibinfo {year} {1992})}\BibitemShut {NoStop}%
\bibitem [{\citenamefont {Kraml}\ \emph {et~al.}(2019)\citenamefont {Kraml},
  \citenamefont {Loc}, \citenamefont {Nhung},\ and\ \citenamefont
  {Ninh}}]{Kraml:2019sis}%
  \BibitemOpen
  \bibfield  {author} {\bibinfo {author} {\bibfnamefont {S.}~\bibnamefont
  {Kraml}}, \bibinfo {author} {\bibfnamefont {T.~Q.}\ \bibnamefont {Loc}},
  \bibinfo {author} {\bibfnamefont {D.~T.}\ \bibnamefont {Nhung}}, \ and\
  \bibinfo {author} {\bibfnamefont {L.~D.}\ \bibnamefont {Ninh}},\ }\href
  {\doibase 10.21468/SciPostPhys.7.4.052} {\bibfield  {journal} {\bibinfo
  {journal} {SciPost Phys.}\ }\textbf {\bibinfo {volume} {7}},\ \bibinfo
  {pages} {052} (\bibinfo {year} {2019})},\ \Eprint
  {http://arxiv.org/abs/1908.03952} {arXiv:1908.03952 [hep-ph]} \BibitemShut
  {NoStop}%
\bibitem [{\citenamefont {Bechtle}\ \emph {et~al.}(2014)\citenamefont
  {Bechtle}, \citenamefont {Heinemeyer}, \citenamefont {Stål}, \citenamefont
  {Stefaniak},\ and\ \citenamefont {Weiglein}}]{Bechtle:2014ewa}%
  \BibitemOpen
  \bibfield  {author} {\bibinfo {author} {\bibfnamefont {P.}~\bibnamefont
  {Bechtle}}, \bibinfo {author} {\bibfnamefont {S.}~\bibnamefont {Heinemeyer}},
  \bibinfo {author} {\bibfnamefont {O.}~\bibnamefont {Stål}}, \bibinfo
  {author} {\bibfnamefont {T.}~\bibnamefont {Stefaniak}}, \ and\ \bibinfo
  {author} {\bibfnamefont {G.}~\bibnamefont {Weiglein}},\ }\href {\doibase
  10.1007/JHEP11(2014)039} {\bibfield  {journal} {\bibinfo  {journal} {JHEP}\
  }\textbf {\bibinfo {volume} {11}},\ \bibinfo {pages} {039} (\bibinfo {year}
  {2014})},\ \Eprint {http://arxiv.org/abs/1403.1582} {arXiv:1403.1582
  [hep-ph]} \BibitemShut {NoStop}%
\bibitem [{\citenamefont {Bechtle}\ \emph {et~al.}(2020)\citenamefont
  {Bechtle}, \citenamefont {Dercks}, \citenamefont {Heinemeyer}, \citenamefont
  {Klingl}, \citenamefont {Stefaniak}, \citenamefont {Weiglein},\ and\
  \citenamefont {Wittbrodt}}]{Bechtle:2020pkv}%
  \BibitemOpen
  \bibfield  {author} {\bibinfo {author} {\bibfnamefont {P.}~\bibnamefont
  {Bechtle}}, \bibinfo {author} {\bibfnamefont {D.}~\bibnamefont {Dercks}},
  \bibinfo {author} {\bibfnamefont {S.}~\bibnamefont {Heinemeyer}}, \bibinfo
  {author} {\bibfnamefont {T.}~\bibnamefont {Klingl}}, \bibinfo {author}
  {\bibfnamefont {T.}~\bibnamefont {Stefaniak}}, \bibinfo {author}
  {\bibfnamefont {G.}~\bibnamefont {Weiglein}}, \ and\ \bibinfo {author}
  {\bibfnamefont {J.}~\bibnamefont {Wittbrodt}},\ }\href@noop {} {\  (\bibinfo
  {year} {2020})},\ \Eprint {http://arxiv.org/abs/2006.06007} {arXiv:2006.06007
  [hep-ph]} \BibitemShut {NoStop}%
\bibitem [{\citenamefont {Bélanger}\ \emph {et~al.}(2018)\citenamefont
  {Bélanger}, \citenamefont {Boudjema}, \citenamefont {Goudelis},
  \citenamefont {Pukhov},\ and\ \citenamefont {Zaldivar}}]{Belanger:2018ccd}%
  \BibitemOpen
  \bibfield  {author} {\bibinfo {author} {\bibfnamefont {G.}~\bibnamefont
  {Bélanger}}, \bibinfo {author} {\bibfnamefont {F.}~\bibnamefont {Boudjema}},
  \bibinfo {author} {\bibfnamefont {A.}~\bibnamefont {Goudelis}}, \bibinfo
  {author} {\bibfnamefont {A.}~\bibnamefont {Pukhov}}, \ and\ \bibinfo {author}
  {\bibfnamefont {B.}~\bibnamefont {Zaldivar}},\ }\href {\doibase
  10.1016/j.cpc.2018.04.027} {\bibfield  {journal} {\bibinfo  {journal}
  {Comput. Phys. Commun.}\ }\textbf {\bibinfo {volume} {231}},\ \bibinfo
  {pages} {173} (\bibinfo {year} {2018})},\ \Eprint
  {http://arxiv.org/abs/1801.03509} {arXiv:1801.03509 [hep-ph]} \BibitemShut
  {NoStop}%
\bibitem [{\citenamefont {Alloul}\ \emph {et~al.}(2014)\citenamefont {Alloul},
  \citenamefont {Christensen}, \citenamefont {Degrande}, \citenamefont {Duhr},\
  and\ \citenamefont {Fuks}}]{Alloul:2013bka}%
  \BibitemOpen
  \bibfield  {author} {\bibinfo {author} {\bibfnamefont {A.}~\bibnamefont
  {Alloul}}, \bibinfo {author} {\bibfnamefont {N.~D.}\ \bibnamefont
  {Christensen}}, \bibinfo {author} {\bibfnamefont {C.}~\bibnamefont
  {Degrande}}, \bibinfo {author} {\bibfnamefont {C.}~\bibnamefont {Duhr}}, \
  and\ \bibinfo {author} {\bibfnamefont {B.}~\bibnamefont {Fuks}},\ }\href
  {\doibase 10.1016/j.cpc.2014.04.012} {\bibfield  {journal} {\bibinfo
  {journal} {Comput. Phys. Commun.}\ }\textbf {\bibinfo {volume} {185}},\
  \bibinfo {pages} {2250} (\bibinfo {year} {2014})},\ \Eprint
  {http://arxiv.org/abs/1310.1921} {arXiv:1310.1921 [hep-ph]} \BibitemShut
  {NoStop}%
\bibitem [{\citenamefont {Haller}\ \emph {et~al.}(2018)\citenamefont {Haller},
  \citenamefont {Hoecker}, \citenamefont {Kogler}, \citenamefont {Mönig},
  \citenamefont {Peiffer},\ and\ \citenamefont {Stelzer}}]{Haller:2018nnx}%
  \BibitemOpen
  \bibfield  {author} {\bibinfo {author} {\bibfnamefont {J.}~\bibnamefont
  {Haller}}, \bibinfo {author} {\bibfnamefont {A.}~\bibnamefont {Hoecker}},
  \bibinfo {author} {\bibfnamefont {R.}~\bibnamefont {Kogler}}, \bibinfo
  {author} {\bibfnamefont {K.}~\bibnamefont {Mönig}}, \bibinfo {author}
  {\bibfnamefont {T.}~\bibnamefont {Peiffer}}, \ and\ \bibinfo {author}
  {\bibfnamefont {J.}~\bibnamefont {Stelzer}},\ }\href {\doibase
  10.1140/epjc/s10052-018-6131-3} {\bibfield  {journal} {\bibinfo  {journal}
  {Eur. Phys. J. C}\ }\textbf {\bibinfo {volume} {78}},\ \bibinfo {pages} {675}
  (\bibinfo {year} {2018})},\ \Eprint {http://arxiv.org/abs/1803.01853}
  {arXiv:1803.01853 [hep-ph]} \BibitemShut {NoStop}%
\bibitem [{\citenamefont {Bernon}\ and\ \citenamefont
  {Dumont}(2015)}]{Bernon:2015hsa}%
  \BibitemOpen
  \bibfield  {author} {\bibinfo {author} {\bibfnamefont {J.}~\bibnamefont
  {Bernon}}\ and\ \bibinfo {author} {\bibfnamefont {B.}~\bibnamefont
  {Dumont}},\ }\href {\doibase 10.1140/epjc/s10052-015-3645-9} {\bibfield
  {journal} {\bibinfo  {journal} {Eur. Phys. J. C}\ }\textbf {\bibinfo {volume}
  {75}},\ \bibinfo {pages} {440} (\bibinfo {year} {2015})},\ \Eprint
  {http://arxiv.org/abs/1502.04138} {arXiv:1502.04138 [hep-ph]} \BibitemShut
  {NoStop}%
\bibitem [{\citenamefont {Hagiwara}\ \emph {et~al.}(1994)\citenamefont
  {Hagiwara}, \citenamefont {Matsumoto}, \citenamefont {Haidt},\ and\
  \citenamefont {Kim}}]{Hagiwara:1994pw}%
  \BibitemOpen
  \bibfield  {author} {\bibinfo {author} {\bibfnamefont {K.}~\bibnamefont
  {Hagiwara}}, \bibinfo {author} {\bibfnamefont {S.}~\bibnamefont {Matsumoto}},
  \bibinfo {author} {\bibfnamefont {D.}~\bibnamefont {Haidt}}, \ and\ \bibinfo
  {author} {\bibfnamefont {C.}~\bibnamefont {Kim}},\ }\href {\doibase
  10.1007/BF01957770} {\bibfield  {journal} {\bibinfo  {journal} {Z. Phys. C}\
  }\textbf {\bibinfo {volume} {64}},\ \bibinfo {pages} {559} (\bibinfo {year}
  {1994})},\ \bibinfo {note} {[Erratum: Z.Phys.C 68, 352 (1995)]},\ \Eprint
  {http://arxiv.org/abs/hep-ph/9409380} {arXiv:hep-ph/9409380} \BibitemShut
  {NoStop}%
\bibitem [{\citenamefont {{The ATLAS collaboration}}(2019)}]{ATLAS:2019vcq}%
  \BibitemOpen
  \bibfield  {author} {\bibinfo {author} {\bibnamefont {{The ATLAS
  collaboration}}} (\bibinfo {collaboration} {ATLAS}),\ }\href@noop {} {\
  (\bibinfo {year} {2019})}\BibitemShut {NoStop}%
\bibitem [{\citenamefont {Degrande}\ \emph {et~al.}(2012)\citenamefont
  {Degrande}, \citenamefont {Duhr}, \citenamefont {Fuks}, \citenamefont
  {Grellscheid}, \citenamefont {Mattelaer},\ and\ \citenamefont
  {Reiter}}]{Degrande:2011ua}%
  \BibitemOpen
  \bibfield  {author} {\bibinfo {author} {\bibfnamefont {C.}~\bibnamefont
  {Degrande}}, \bibinfo {author} {\bibfnamefont {C.}~\bibnamefont {Duhr}},
  \bibinfo {author} {\bibfnamefont {B.}~\bibnamefont {Fuks}}, \bibinfo {author}
  {\bibfnamefont {D.}~\bibnamefont {Grellscheid}}, \bibinfo {author}
  {\bibfnamefont {O.}~\bibnamefont {Mattelaer}}, \ and\ \bibinfo {author}
  {\bibfnamefont {T.}~\bibnamefont {Reiter}},\ }\href {\doibase
  10.1016/j.cpc.2012.01.022} {\bibfield  {journal} {\bibinfo  {journal}
  {Comput. Phys. Commun.}\ }\textbf {\bibinfo {volume} {183}},\ \bibinfo
  {pages} {1201} (\bibinfo {year} {2012})},\ \Eprint
  {http://arxiv.org/abs/1108.2040} {arXiv:1108.2040 [hep-ph]} \BibitemShut
  {NoStop}%
\bibitem [{\citenamefont {Alwall}\ \emph {et~al.}(2014)\citenamefont {Alwall},
  \citenamefont {Frederix}, \citenamefont {Frixione}, \citenamefont {Hirschi},
  \citenamefont {Maltoni}, \citenamefont {Mattelaer}, \citenamefont {Shao},
  \citenamefont {Stelzer}, \citenamefont {Torrielli},\ and\ \citenamefont
  {Zaro}}]{Alwall:2014hca}%
  \BibitemOpen
  \bibfield  {author} {\bibinfo {author} {\bibfnamefont {J.}~\bibnamefont
  {Alwall}}, \bibinfo {author} {\bibfnamefont {R.}~\bibnamefont {Frederix}},
  \bibinfo {author} {\bibfnamefont {S.}~\bibnamefont {Frixione}}, \bibinfo
  {author} {\bibfnamefont {V.}~\bibnamefont {Hirschi}}, \bibinfo {author}
  {\bibfnamefont {F.}~\bibnamefont {Maltoni}}, \bibinfo {author} {\bibfnamefont
  {O.}~\bibnamefont {Mattelaer}}, \bibinfo {author} {\bibfnamefont {H.~S.}\
  \bibnamefont {Shao}}, \bibinfo {author} {\bibfnamefont {T.}~\bibnamefont
  {Stelzer}}, \bibinfo {author} {\bibfnamefont {P.}~\bibnamefont {Torrielli}},
  \ and\ \bibinfo {author} {\bibfnamefont {M.}~\bibnamefont {Zaro}},\ }\href
  {\doibase 10.1007/JHEP07(2014)079} {\bibfield  {journal} {\bibinfo  {journal}
  {JHEP}\ }\textbf {\bibinfo {volume} {07}},\ \bibinfo {pages} {079} (\bibinfo
  {year} {2014})},\ \Eprint {http://arxiv.org/abs/1405.0301} {arXiv:1405.0301
  [hep-ph]} \BibitemShut {NoStop}%
\bibitem [{\citenamefont {Ball}\ \emph {et~al.}(2015)\citenamefont {Ball} \emph
  {et~al.}}]{Ball:2014uwa}%
  \BibitemOpen
  \bibfield  {author} {\bibinfo {author} {\bibfnamefont {R.~D.}\ \bibnamefont
  {Ball}} \emph {et~al.} (\bibinfo {collaboration} {NNPDF}),\ }\href {\doibase
  10.1007/JHEP04(2015)040} {\bibfield  {journal} {\bibinfo  {journal} {JHEP}\
  }\textbf {\bibinfo {volume} {04}},\ \bibinfo {pages} {040} (\bibinfo {year}
  {2015})},\ \Eprint {http://arxiv.org/abs/1410.8849} {arXiv:1410.8849
  [hep-ph]} \BibitemShut {NoStop}%
\bibitem [{\citenamefont {Artoisenet}\ \emph {et~al.}(2013)\citenamefont
  {Artoisenet}, \citenamefont {Frederix}, \citenamefont {Mattelaer},\ and\
  \citenamefont {Rietkerk}}]{Artoisenet:2012st}%
  \BibitemOpen
  \bibfield  {author} {\bibinfo {author} {\bibfnamefont {P.}~\bibnamefont
  {Artoisenet}}, \bibinfo {author} {\bibfnamefont {R.}~\bibnamefont
  {Frederix}}, \bibinfo {author} {\bibfnamefont {O.}~\bibnamefont {Mattelaer}},
  \ and\ \bibinfo {author} {\bibfnamefont {R.}~\bibnamefont {Rietkerk}},\
  }\href {\doibase 10.1007/JHEP03(2013)015} {\bibfield  {journal} {\bibinfo
  {journal} {JHEP}\ }\textbf {\bibinfo {volume} {03}},\ \bibinfo {pages} {015}
  (\bibinfo {year} {2013})},\ \Eprint {http://arxiv.org/abs/1212.3460}
  {arXiv:1212.3460 [hep-ph]} \BibitemShut {NoStop}%
\bibitem [{\citenamefont {Alwall}\ \emph {et~al.}(2015)\citenamefont {Alwall},
  \citenamefont {Duhr}, \citenamefont {Fuks}, \citenamefont {Mattelaer},
  \citenamefont {Öztürk},\ and\ \citenamefont {Shen}}]{Alwall:2014bza}%
  \BibitemOpen
  \bibfield  {author} {\bibinfo {author} {\bibfnamefont {J.}~\bibnamefont
  {Alwall}}, \bibinfo {author} {\bibfnamefont {C.}~\bibnamefont {Duhr}},
  \bibinfo {author} {\bibfnamefont {B.}~\bibnamefont {Fuks}}, \bibinfo {author}
  {\bibfnamefont {O.}~\bibnamefont {Mattelaer}}, \bibinfo {author}
  {\bibfnamefont {D.~G.}\ \bibnamefont {Öztürk}}, \ and\ \bibinfo {author}
  {\bibfnamefont {C.-H.}\ \bibnamefont {Shen}},\ }\href {\doibase
  10.1016/j.cpc.2015.08.031} {\bibfield  {journal} {\bibinfo  {journal}
  {Comput. Phys. Commun.}\ }\textbf {\bibinfo {volume} {197}},\ \bibinfo
  {pages} {312} (\bibinfo {year} {2015})},\ \Eprint
  {http://arxiv.org/abs/1402.1178} {arXiv:1402.1178 [hep-ph]} \BibitemShut
  {NoStop}%
\bibitem [{\citenamefont {Sirunyan}\ \emph {et~al.}(2018)\citenamefont
  {Sirunyan} \emph {et~al.}}]{Sirunyan:2017jix}%
  \BibitemOpen
  \bibfield  {author} {\bibinfo {author} {\bibfnamefont {A.~M.}\ \bibnamefont
  {Sirunyan}} \emph {et~al.} (\bibinfo {collaboration} {CMS}),\ }\href
  {\doibase 10.1103/PhysRevD.97.092005} {\bibfield  {journal} {\bibinfo
  {journal} {Phys. Rev. D}\ }\textbf {\bibinfo {volume} {97}},\ \bibinfo
  {pages} {092005} (\bibinfo {year} {2018})},\ \Eprint
  {http://arxiv.org/abs/1712.02345} {arXiv:1712.02345 [hep-ex]} \BibitemShut
  {NoStop}%
\bibitem [{\citenamefont {Sjöstrand}\ \emph {et~al.}(2015)\citenamefont
  {Sjöstrand}, \citenamefont {Ask}, \citenamefont {Christiansen},
  \citenamefont {Corke}, \citenamefont {Desai}, \citenamefont {Ilten},
  \citenamefont {Mrenna}, \citenamefont {Prestel}, \citenamefont {Rasmussen},\
  and\ \citenamefont {Skands}}]{Sjostrand:2014zea}%
  \BibitemOpen
  \bibfield  {author} {\bibinfo {author} {\bibfnamefont {T.}~\bibnamefont
  {Sjöstrand}}, \bibinfo {author} {\bibfnamefont {S.}~\bibnamefont {Ask}},
  \bibinfo {author} {\bibfnamefont {J.~R.}\ \bibnamefont {Christiansen}},
  \bibinfo {author} {\bibfnamefont {R.}~\bibnamefont {Corke}}, \bibinfo
  {author} {\bibfnamefont {N.}~\bibnamefont {Desai}}, \bibinfo {author}
  {\bibfnamefont {P.}~\bibnamefont {Ilten}}, \bibinfo {author} {\bibfnamefont
  {S.}~\bibnamefont {Mrenna}}, \bibinfo {author} {\bibfnamefont
  {S.}~\bibnamefont {Prestel}}, \bibinfo {author} {\bibfnamefont {C.~O.}\
  \bibnamefont {Rasmussen}}, \ and\ \bibinfo {author} {\bibfnamefont {P.~Z.}\
  \bibnamefont {Skands}},\ }\href {\doibase 10.1016/j.cpc.2015.01.024}
  {\bibfield  {journal} {\bibinfo  {journal} {Comput. Phys. Commun.}\ }\textbf
  {\bibinfo {volume} {191}},\ \bibinfo {pages} {159} (\bibinfo {year}
  {2015})},\ \Eprint {http://arxiv.org/abs/1410.3012} {arXiv:1410.3012
  [hep-ph]} \BibitemShut {NoStop}%
\bibitem [{\citenamefont {Conte}\ \emph {et~al.}(2013)\citenamefont {Conte},
  \citenamefont {Fuks},\ and\ \citenamefont {Serret}}]{Conte:2012fm}%
  \BibitemOpen
  \bibfield  {author} {\bibinfo {author} {\bibfnamefont {E.}~\bibnamefont
  {Conte}}, \bibinfo {author} {\bibfnamefont {B.}~\bibnamefont {Fuks}}, \ and\
  \bibinfo {author} {\bibfnamefont {G.}~\bibnamefont {Serret}},\ }\href
  {\doibase 10.1016/j.cpc.2012.09.009} {\bibfield  {journal} {\bibinfo
  {journal} {Comput. Phys. Commun.}\ }\textbf {\bibinfo {volume} {184}},\
  \bibinfo {pages} {222} (\bibinfo {year} {2013})},\ \Eprint
  {http://arxiv.org/abs/1206.1599} {arXiv:1206.1599 [hep-ph]} \BibitemShut
  {NoStop}%
\bibitem [{\citenamefont {Conte}\ and\ \citenamefont
  {Fuks}(2018)}]{Conte:2018vmg}%
  \BibitemOpen
  \bibfield  {author} {\bibinfo {author} {\bibfnamefont {E.}~\bibnamefont
  {Conte}}\ and\ \bibinfo {author} {\bibfnamefont {B.}~\bibnamefont {Fuks}},\
  }\href {\doibase 10.1142/S0217751X18300272} {\bibfield  {journal} {\bibinfo
  {journal} {Int. J. Mod. Phys.}\ }\textbf {\bibinfo {volume} {A33}},\ \bibinfo
  {pages} {1830027} (\bibinfo {year} {2018})},\ \Eprint
  {http://arxiv.org/abs/1808.00480} {arXiv:1808.00480 [hep-ph]} \BibitemShut
  {NoStop}%
\bibitem [{\citenamefont {de~Favereau}\ \emph {et~al.}(2014)\citenamefont
  {de~Favereau}, \citenamefont {Delaere}, \citenamefont {Demin}, \citenamefont
  {Giammanco}, \citenamefont {Lemaître}, \citenamefont {Mertens},\ and\
  \citenamefont {Selvaggi}}]{deFavereau:2013fsa}%
  \BibitemOpen
  \bibfield  {author} {\bibinfo {author} {\bibfnamefont {J.}~\bibnamefont
  {de~Favereau}}, \bibinfo {author} {\bibfnamefont {C.}~\bibnamefont
  {Delaere}}, \bibinfo {author} {\bibfnamefont {P.}~\bibnamefont {Demin}},
  \bibinfo {author} {\bibfnamefont {A.}~\bibnamefont {Giammanco}}, \bibinfo
  {author} {\bibfnamefont {V.}~\bibnamefont {Lemaître}}, \bibinfo {author}
  {\bibfnamefont {A.}~\bibnamefont {Mertens}}, \ and\ \bibinfo {author}
  {\bibfnamefont {M.}~\bibnamefont {Selvaggi}} (\bibinfo {collaboration}
  {DELPHES 3}),\ }\href {\doibase 10.1007/JHEP02(2014)057} {\bibfield
  {journal} {\bibinfo  {journal} {JHEP}\ }\textbf {\bibinfo {volume} {02}},\
  \bibinfo {pages} {057} (\bibinfo {year} {2014})},\ \Eprint
  {http://arxiv.org/abs/1307.6346} {arXiv:1307.6346 [hep-ex]} \BibitemShut
  {NoStop}%
\bibitem [{\citenamefont {Cacciari}\ \emph {et~al.}(2012)\citenamefont
  {Cacciari}, \citenamefont {Salam},\ and\ \citenamefont
  {Soyez}}]{Cacciari:2011ma}%
  \BibitemOpen
  \bibfield  {author} {\bibinfo {author} {\bibfnamefont {M.}~\bibnamefont
  {Cacciari}}, \bibinfo {author} {\bibfnamefont {G.~P.}\ \bibnamefont {Salam}},
  \ and\ \bibinfo {author} {\bibfnamefont {G.}~\bibnamefont {Soyez}},\ }\href
  {\doibase 10.1140/epjc/s10052-012-1896-2} {\bibfield  {journal} {\bibinfo
  {journal} {Eur. Phys. J.}\ }\textbf {\bibinfo {volume} {C72}},\ \bibinfo
  {pages} {1896} (\bibinfo {year} {2012})},\ \Eprint
  {http://arxiv.org/abs/1111.6097} {arXiv:1111.6097 [hep-ph]} \BibitemShut
  {NoStop}%
\bibitem [{\citenamefont {Cacciari}\ \emph {et~al.}(2008)\citenamefont
  {Cacciari}, \citenamefont {Salam},\ and\ \citenamefont
  {Soyez}}]{Cacciari:2008gp}%
  \BibitemOpen
  \bibfield  {author} {\bibinfo {author} {\bibfnamefont {M.}~\bibnamefont
  {Cacciari}}, \bibinfo {author} {\bibfnamefont {G.~P.}\ \bibnamefont {Salam}},
  \ and\ \bibinfo {author} {\bibfnamefont {G.}~\bibnamefont {Soyez}},\ }\href
  {\doibase 10.1088/1126-6708/2008/04/063} {\bibfield  {journal} {\bibinfo
  {journal} {JHEP}\ }\textbf {\bibinfo {volume} {04}},\ \bibinfo {pages} {063}
  (\bibinfo {year} {2008})},\ \Eprint {http://arxiv.org/abs/0802.1189}
  {arXiv:0802.1189 [hep-ph]} \BibitemShut {NoStop}%
\bibitem [{\citenamefont {Dumont}\ \emph {et~al.}(2015)\citenamefont {Dumont},
  \citenamefont {Fuks}, \citenamefont {Kraml}, \citenamefont {Bein},
  \citenamefont {Chalons}, \citenamefont {Conte}, \citenamefont {Kulkarni},
  \citenamefont {Sengupta},\ and\ \citenamefont {Wymant}}]{Dumont:2014tja}%
  \BibitemOpen
  \bibfield  {author} {\bibinfo {author} {\bibfnamefont {B.}~\bibnamefont
  {Dumont}}, \bibinfo {author} {\bibfnamefont {B.}~\bibnamefont {Fuks}},
  \bibinfo {author} {\bibfnamefont {S.}~\bibnamefont {Kraml}}, \bibinfo
  {author} {\bibfnamefont {S.}~\bibnamefont {Bein}}, \bibinfo {author}
  {\bibfnamefont {G.}~\bibnamefont {Chalons}}, \bibinfo {author} {\bibfnamefont
  {E.}~\bibnamefont {Conte}}, \bibinfo {author} {\bibfnamefont
  {S.}~\bibnamefont {Kulkarni}}, \bibinfo {author} {\bibfnamefont
  {D.}~\bibnamefont {Sengupta}}, \ and\ \bibinfo {author} {\bibfnamefont
  {C.}~\bibnamefont {Wymant}},\ }\href {\doibase
  10.1140/epjc/s10052-014-3242-3} {\bibfield  {journal} {\bibinfo  {journal}
  {Eur. Phys. J.}\ }\textbf {\bibinfo {volume} {C75}},\ \bibinfo {pages} {56}
  (\bibinfo {year} {2015})},\ \Eprint {http://arxiv.org/abs/1407.3278}
  {arXiv:1407.3278 [hep-ph]} \BibitemShut {NoStop}%
\bibitem [{\citenamefont {Read}(2002)}]{Read:2002hq}%
  \BibitemOpen
  \bibfield  {author} {\bibinfo {author} {\bibfnamefont {A.~L.}\ \bibnamefont
  {Read}},\ }\bibfield  {booktitle} {\emph {\bibinfo {booktitle} {{Advanced
  Statistical Techniques in Particle Physics. Proceedings, Conference, Durham,
  UK, March 18-22, 2002}}},\ }\href {\doibase 10.1088/0954-3899/28/10/313}
  {\bibfield  {journal} {\bibinfo  {journal} {J. Phys.}\ }\textbf {\bibinfo
  {volume} {G28}},\ \bibinfo {pages} {2693} (\bibinfo {year} {2002})},\
  \bibinfo {note} {[,11(2002)]}\BibitemShut {NoStop}%
\bibitem [{\citenamefont {Araz}\ \emph {et~al.}(2020)\citenamefont {Araz},
  \citenamefont {Frank},\ and\ \citenamefont {Fuks}}]{Araz:2019otb}%
  \BibitemOpen
  \bibfield  {author} {\bibinfo {author} {\bibfnamefont {J.~Y.}\ \bibnamefont
  {Araz}}, \bibinfo {author} {\bibfnamefont {M.}~\bibnamefont {Frank}}, \ and\
  \bibinfo {author} {\bibfnamefont {B.}~\bibnamefont {Fuks}},\ }\href {\doibase
  10.1140/epjc/s10052-020-8076-6} {\bibfield  {journal} {\bibinfo  {journal}
  {Eur. Phys. J. C}\ }\textbf {\bibinfo {volume} {80}},\ \bibinfo {pages} {531}
  (\bibinfo {year} {2020})},\ \Eprint {http://arxiv.org/abs/1910.11418}
  {arXiv:1910.11418 [hep-ph]} \BibitemShut {NoStop}%
\bibitem [{\citenamefont {Banerjee}\ \emph {et~al.}(2017)\citenamefont
  {Banerjee}, \citenamefont {Barducci}, \citenamefont {Bélanger},
  \citenamefont {Fuks}, \citenamefont {Goudelis},\ and\ \citenamefont
  {Zaldivar}}]{Banerjee:2017wxi}%
  \BibitemOpen
  \bibfield  {author} {\bibinfo {author} {\bibfnamefont {S.}~\bibnamefont
  {Banerjee}}, \bibinfo {author} {\bibfnamefont {D.}~\bibnamefont {Barducci}},
  \bibinfo {author} {\bibfnamefont {G.}~\bibnamefont {Bélanger}}, \bibinfo
  {author} {\bibfnamefont {B.}~\bibnamefont {Fuks}}, \bibinfo {author}
  {\bibfnamefont {A.}~\bibnamefont {Goudelis}}, \ and\ \bibinfo {author}
  {\bibfnamefont {B.}~\bibnamefont {Zaldivar}},\ }\href {\doibase
  10.1007/JHEP07(2017)080} {\bibfield  {journal} {\bibinfo  {journal} {JHEP}\
  }\textbf {\bibinfo {volume} {07}},\ \bibinfo {pages} {080} (\bibinfo {year}
  {2017})},\ \Eprint {http://arxiv.org/abs/1705.02327} {arXiv:1705.02327
  [hep-ph]} \BibitemShut {NoStop}%
\bibitem [{\citenamefont {Aghanim}\ \emph {et~al.}(2018)\citenamefont {Aghanim}
  \emph {et~al.}}]{Aghanim:2018eyx}%
  \BibitemOpen
  \bibfield  {author} {\bibinfo {author} {\bibfnamefont {N.}~\bibnamefont
  {Aghanim}} \emph {et~al.} (\bibinfo {collaboration} {Planck}),\ }\href@noop
  {} {\  (\bibinfo {year} {2018})},\ \Eprint {http://arxiv.org/abs/1807.06209}
  {arXiv:1807.06209 [astro-ph.CO]} \BibitemShut {NoStop}%
\bibitem [{\citenamefont {Schumann}(2019)}]{Schumann:2019eaa}%
  \BibitemOpen
  \bibfield  {author} {\bibinfo {author} {\bibfnamefont {M.}~\bibnamefont
  {Schumann}},\ }\href {\doibase 10.1088/1361-6471/ab2ea5} {\bibfield
  {journal} {\bibinfo  {journal} {J. Phys. G}\ }\textbf {\bibinfo {volume}
  {46}},\ \bibinfo {pages} {103003} (\bibinfo {year} {2019})},\ \Eprint
  {http://arxiv.org/abs/1903.03026} {arXiv:1903.03026 [astro-ph.CO]}
  \BibitemShut {NoStop}%
\bibitem [{\citenamefont {Itzykson}\ and\ \citenamefont {Zuber}(1980)}]{IZ}%
  \BibitemOpen
  \bibfield  {author} {\bibinfo {author} {\bibfnamefont {C.}~\bibnamefont
  {Itzykson}}\ and\ \bibinfo {author} {\bibfnamefont {J.}~\bibnamefont
  {Zuber}},\ }\href {\doibase 10.1063/1.2916419} {\emph {\bibinfo {title}
  {Quantum Field Theory}}},\ International Series In Pure and Applied Physics\
  (\bibinfo  {publisher} {McGraw Hill},\ \bibinfo {year} {1980})\BibitemShut
  {NoStop}%
\bibitem [{\citenamefont {Dusedau}\ \emph {et~al.}(1984)\citenamefont
  {Dusedau}, \citenamefont {Lust},\ and\ \citenamefont
  {Zeppenfeld}}]{Dusedau:1984xk}%
  \BibitemOpen
  \bibfield  {author} {\bibinfo {author} {\bibfnamefont {D.}~\bibnamefont
  {Dusedau}}, \bibinfo {author} {\bibfnamefont {D.}~\bibnamefont {Lust}}, \
  and\ \bibinfo {author} {\bibfnamefont {D.}~\bibnamefont {Zeppenfeld}},\
  }\href {\doibase 10.1016/0370-2693(84)91645-9} {\bibfield  {journal}
  {\bibinfo  {journal} {Phys. Lett. B}\ }\textbf {\bibinfo {volume} {148}},\
  \bibinfo {pages} {234} (\bibinfo {year} {1984})}\BibitemShut {NoStop}%
\bibitem [{\citenamefont {Schuessler}\ and\ \citenamefont
  {Zeppenfeld}(2007)}]{Schuessler:2007av}%
  \BibitemOpen
  \bibfield  {author} {\bibinfo {author} {\bibfnamefont {A.}~\bibnamefont
  {Schuessler}}\ and\ \bibinfo {author} {\bibfnamefont {D.}~\bibnamefont
  {Zeppenfeld}},\ }in\ \href@noop {} {\emph {\bibinfo {booktitle} {{15th
  International Conference on Supersymmetry and the Unification of Fundamental
  Interactions (SUSY07)}}}}\ (\bibinfo {year} {2007})\ pp.\ \bibinfo {pages}
  {236--239},\ \Eprint {http://arxiv.org/abs/0710.5175} {arXiv:0710.5175
  [hep-ph]} \BibitemShut {NoStop}%
\bibitem [{\citenamefont {Di~Luzio}\ \emph {et~al.}(2017)\citenamefont
  {Di~Luzio}, \citenamefont {Kamenik},\ and\ \citenamefont
  {Nardecchia}}]{DiLuzio:2016sur}%
  \BibitemOpen
  \bibfield  {author} {\bibinfo {author} {\bibfnamefont {L.}~\bibnamefont
  {Di~Luzio}}, \bibinfo {author} {\bibfnamefont {J.~F.}\ \bibnamefont
  {Kamenik}}, \ and\ \bibinfo {author} {\bibfnamefont {M.}~\bibnamefont
  {Nardecchia}},\ }\href {\doibase 10.1140/epjc/s10052-017-4594-2} {\bibfield
  {journal} {\bibinfo  {journal} {Eur. Phys. J. C}\ }\textbf {\bibinfo {volume}
  {77}},\ \bibinfo {pages} {30} (\bibinfo {year} {2017})},\ \Eprint
  {http://arxiv.org/abs/1604.05746} {arXiv:1604.05746 [hep-ph]} \BibitemShut
  {NoStop}%
\bibitem [{\citenamefont {Goodsell}\ and\ \citenamefont
  {Staub}(2018)}]{Goodsell:2018tti}%
  \BibitemOpen
  \bibfield  {author} {\bibinfo {author} {\bibfnamefont {M.~D.}\ \bibnamefont
  {Goodsell}}\ and\ \bibinfo {author} {\bibfnamefont {F.}~\bibnamefont
  {Staub}},\ }\href {\doibase 10.1140/epjc/s10052-018-6127-z} {\bibfield
  {journal} {\bibinfo  {journal} {Eur. Phys. J. C}\ }\textbf {\bibinfo {volume}
  {78}},\ \bibinfo {pages} {649} (\bibinfo {year} {2018})},\ \Eprint
  {http://arxiv.org/abs/1805.07306} {arXiv:1805.07306 [hep-ph]} \BibitemShut
  {NoStop}%
\end{thebibliography}%

\end{document}